\documentclass{lmcs}


\usepackage{amssymb}
\usepackage{xcolor}
\usepackage{xspace}
\usepackage{multicol}
\usepackage{ragged2e}
\usepackage{hyperref}
\usepackage{cleveref}
\usepackage{comment}
\usepackage{array}

\hypersetup{%
	colorlinks=true,           
	allcolors=blue!70!black,   
	breaklinks=true
}

\Crefname{exa}{Example}{Examples}\crefname{exa}{Example}{Examples}
\Crefname{thm}{Theorem}{Theorems}\crefname{thm}{Theorem}{Theorems}
\Crefname{thmC}{Theorem}{Theorems}\crefname{thmC}{Theorem}{Theorems}
\Crefname{defi}{Definition}{Definitions}\crefname{defi}{Definition}{Definitions}
\Crefname{prop}{Proposition}{Propositions}\crefname{prop}{Proposition}{Propositions}
\Crefname{lem}{Lemma}{Lemmas}\crefname{lem}{Lemma}{Lemmas}
\Crefname{cor}{Corollary}{Corollaries}\crefname{cor}{Corollary}{Corollaries}
\Crefname{rem}{Remark}{Remarks}\crefname{rem}{Remark}{Remarks}
\Crefname{figure}{Figure}{Figures}\crefname{figure}{Figure}{Figures}
\Crefname{table}{Table}{Tables}\crefname{table}{Table}{Tables}
\Crefname{section}{Section}{Sections}\crefname{section}{Section}{Sections}

\usepackage{todonotes}

\usepackage{tikz}
\usetikzlibrary{backgrounds, positioning, arrows, automata, calc}
\tikzstyle{state}=[thick,minimum size=18pt, circle,draw]
\tikzstyle{transition}=[->,thick,>=stealth,shorten >=1pt,shorten <=1pt, font=\small]
\tikzstyle{loop above right}=[out=60,in=30, min distance=5mm, looseness=8]
\tikzstyle{loop above left}=[out=150,in=120, min distance=5mm, looseness=8]
\tikzstyle{loop below left}=[out=-120,in=-150, min distance=5mm, looseness=8]
\tikzstyle{loop below right}=[out=-30,in=-60, min distance=5mm, looseness=8]
\newcommand{\zokiInitialLength}{12pt}
\newcommand{\zokiInitialAngle}{180}
\newcommand{\zokiInitialPos}{left}
\newcommand{\zokiInitialText}{}
\tikzoption{initial length}{\tikzaddafternodepathoption{\renewcommand{\zokiInitialLength}{#1}}}
\tikzoption{initial angle}{\tikzaddafternodepathoption{\renewcommand{\zokiInitialAngle}{#1}}}
\tikzoption{initial text}{\tikzaddafternodepathoption{\renewcommand{\zokiInitialText}{#1}}}
\tikzoption{initial pos}{\tikzaddafternodepathoption{\renewcommand{\zokiInitialPos}{#1}}}
\tikzstyle{initial}=[after node path={{%
		[to path={[transition] (\tikztostart) -- (\tikztotarget) \tikztonodes}]%
		(\tikzlastnode.\zokiInitialAngle)++(\zokiInitialAngle:\zokiInitialLength) edge[at start] node[\zokiInitialPos, inner sep=0pt, outer sep=1pt] {\zokiInitialText} (\tikzlastnode)%
}}]
\tikzstyle{final}=[after node path={ node[state, scale=.8] at (\tikzlastnode) {} }]
\tikzset{
	bg/.default={},
	bg/.style={execute at end picture={
		\begin{scope}[on background layer]
			\node[xshift=0mm, yshift=0mm] (ss) at (current bounding box.south) {};
			\node[xshift=0mm, yshift=0mm] (sw) at (current bounding box.south west) {};
			\node[xshift=1mm, yshift=1mm] (ne) at (current bounding box.north east) {};
			\node[xshift=1mm, yshift=-1mm] (nw) at (current bounding box.north west) {};

			\ifx&#1&\else
			\node[anchor=north, xshift=2pt] at (ss) {#1};
			\fi
		\end{scope}
	}},
}

\newcommand{\DD}{\ensuremath{\mathbb{D}}\xspace}
\newcommand{\NN}{\ensuremath{\mathbb{N}}\xspace}
\newcommand{\QQ}{\ensuremath{\mathbb{Q}}\xspace}
\newcommand{\RR}{\ensuremath{\mathbb{R}}\xspace}

\renewcommand{\AA}{\ensuremath{\mathbb{A}}\xspace}
\newcommand{\BB}{\ensuremath{\mathbb{B}}\xspace}

\newcommand{\A}{\mathcal{A}\xspace}
\newcommand{\B}{\mathcal{B}\xspace}
\newcommand{\C}{\mathcal{C}\xspace}
\newcommand{\D}{\mathcal{D}\xspace}
\newcommand{\T}{\mathcal{T}\xspace}
\newcommand{\M}{\mathcal{M}\xspace}
\newcommand{\N}{\mathcal{N}\xspace}

\newcommand{\trans}[3]{#1\xrightarrow[]{#2}#3}

\newcommand{\printValueFunction}[1]{\mathsf{#1}}
\newcommand{\Val}{\printValueFunction{Val}}
\newcommand{\Inf}{\printValueFunction{Inf}}
\newcommand{\Sup}{\printValueFunction{Sup}}
\newcommand{\DSum}{\printValueFunction{DSum}}
\newcommand{\LimInf}{\printValueFunction{LimInf}}
\newcommand{\LimSup}{\printValueFunction{LimSup}}
\newcommand{\LimInfAvg}{\printValueFunction{LimInfAvg}}
\newcommand{\LimSupAvg}{\printValueFunction{LimSupAvg}}

\newcommand{\E}{\mathbb{E}\xspace}

\providecommand{\st}{}
\renewcommand{\st}{\;\ifnum\currentgrouptype=16 \middle\fi|\;}
\newcommand{\compare}{\mathrel{\triangleright}}

\newcommand{\CompClass}[1]{{\textsc{#1}}}
\newcommand{\NLogSpace}{\CompClass{NLogSpace}\xspace}
\newcommand{\PTime}{\CompClass{PTime}\xspace}

\newcommand{\PSpace}{\CompClass{PSpace}\xspace}
\newcommand{\NPSpace}{\CompClass{NPSpace}\xspace}
\newcommand{\PSpaceC}{\CompClass{PSpace}\text{-complete}\xspace}
\newcommand{\PSpaceH}{\CompClass{PSpace}\text{-hard}\xspace}
\newcommand{\coPSpace}{\text{co-}\CompClass{PSpace}\xspace}

\newcommand{\Entry}[1]{\begin{tabular}[c]{@{}c@{}} #1\end{tabular}}
\newcommand{\RefEntry}[2]{\Entry{#1\\#2}}

\newif\ifhideproofs
\hideproofsfalse
\ifhideproofs
	\excludecomment{proof}
\fi

\newcommand{\hidehyperproof}[1]{}

\begin{document}

\title[Quantitative Language Automata]{Quantitative Language Automata}

\thanks{This work was supported by the European Research Council (ERC) Grants VAMOS (No. 101020093) and HYPER (No. 101055412).}

\author[T.\,A.~Henzinger]{Thomas A. {Henzinger}\lmcsorcid{0000-0002-2985-7724}}[a]
\author[P.~Kebis]{Pavol {Kebis}\lmcsorcid{0009-0004-0379-5766}}[a]
\author[N.~Mazzocchi]{Nicolas {Mazzocchi}\lmcsorcid{0000-0001-6425-5369}}[b]
\author[N.\,E.~Sara\c{c}]{N. Ege {Sara\c{c}}\lmcsorcid{0009-0000-2866-8078}}[c]

\address{Institute of Science and Technology Austria (ISTA), Austria}
\email{tah@ista.ac.at, pavol.kebis@ista.ac.at}

\address{Slovak University of Technology in Bratislava, Slovak Republic}
\email{nicolas.mazzocchi@stuba.sk}

\address{CISPA Helmholtz Center for Information Security, Germany}
\email{ege.sarac@cispa.de}

\begin{abstract}
	A {\em quantitative word automaton} (QWA) defines a function from infinite words to values.
	For example, every infinite run of a limit-average QWA $\mathcal{A}$ obtains a mean payoff, and every word $w \in \Sigma^\omega$ is assigned the supremum of the mean-payoff values of nondeterministic runs of $\mathcal{A}$ over~$w$.
	We introduce {\em quantitative language automata} (QLAs) that define functions from language generators, which represent implementations, to values, where a language generator can be nonprobabilistic, inducing a set of infinite words, or probabilistic, inducing a probability measure over infinite words.
	A QLA consists of a QWA together with a language aggregator.
	For example, given a QWA $\mathcal{A}$, the infimum aggregator maps each language $L \subseteq \Sigma^\omega$ to the greatest lower bound assigned by $\mathcal{A}$ to any word in~$L$.
	Over the boolean domain, QWAs capture trace properties, and QLAs capture hyperproperties.
	For more general value sets, QLAs serve as a specification language for a generalization of hyperproperties, called {\em quantitative hyperproperties}.
	A nonprobabilistic (resp.\ probabilistic) quantitative hyperproperty assigns a value to each set (resp.\ distribution) $G$ of traces, e.g., the minimal (resp.\ expected) average response time exhibited by the traces in~$G$ (resp.\ by traces sampled according to~$G$).
	We give several examples of quantitative hyperproperties and investigate three paradigmatic problems for QLAs: evaluation, nonemptiness, and universality.
	In the {\em evaluation} problem, given a QLA $\mathbb{A}$ and an implementation~$G$, we ask for the value that $\mathbb{A}$ assigns to $G$.
	In the {\em nonemptiness} (resp.\ {\em universality}) problem, given a QLA $\mathbb{A}$, a threshold~$k$, and a comparison relation ${\mathrel{\triangleright}} \in \{{>},{\geq}\}$, we ask whether $\mathbb{A}$ assigns a value ${\mathrel{\triangleright}} k$ to some (resp.\ every) language.
	We provide a comprehensive picture of decidability and complexity for these problems for QLAs with common aggregators as well as their restrictions to $\omega$-regular languages and trace distributions generated by finite-state Markov chains.
\end{abstract}

\maketitle

\section{Introduction}\label{sec:introduction}

The specification and verification of system properties have traditionally taken a boolean view.
While this view is natural for many correctness properties, it does not capture quantitative aspects of system behavior such as performance or robustness.
Quantitative trace properties and quantitative word automata~\cite{DBLP:journals/tocl/ChatterjeeDH10} address this limitation.
Rather than partitioning traces into correct and incorrect ones, they assign to each execution a value from a richer domain, such as the real numbers.
This makes it possible to specify, for example, the maximal or average response time along a server execution, or the degree to which a system behavior violates a desired boolean property~\cite{DBLP:conf/concur/HenzingerO13}.

Many interesting system properties are inherently multi-trace, especially in security, where one often needs to relate several executions of the same system.
In the boolean setting, such properties are captured by hyperproperties~\cite{DBLP:journals/jcs/ClarksonS10}, which are interpreted over sets of traces rather than over individual traces.
A standard example is observational determinism, which requires that any two executions with matching observable inputs also have matching observable outputs.
Thus, trace properties classify individual executions, whereas hyperproperties classify sets of executions and, through them, system implementations.
In a similar vein, while quantitative trace properties and quantitative word automata describe quantitative aspects of individual executions, \emph{quantitative hyperproperties} are needed to capture quantitative aspects of system-wide behavior.
For example, they can express worst-case or best-case response time across the executions of an implementation, or quantify how severely an implementation violates a desired boolean property.
In this paper, we introduce \emph{quantitative language automata} (QLAs), an automata-theoretic model for the specification and verification of quantitative hyperproperties.
Unlike quantitative word automata, which evaluate individual executions, quantitative language automata evaluate sets of executions and, in the probabilistic setting, distributions over executions.

Quantitative word automata (QWAs) extend boolean $\omega$-automata with weighted transitions.
Each infinite run induces an infinite sequence of weights, which is evaluated by a \emph{run aggregator} (also called a value function).
Typical run aggregators include $\Sup$ (the maximal weight occurring along a run), $\LimSup$ (the largest weight occurring infinitely often), and $\LimSupAvg$ (the long-run average of the weight sequence).
When a word admits multiple runs, as is generally the case for nondeterministic automata, the resulting run values are combined by a \emph{word aggregator}.
The most common word aggregator is $\Sup$, which takes the least upper bound of the values achievable by resolving the nondeterministic choices and thus generalizes the usual acceptance semantics, where one accepting run suffices.
Other word aggregators are also natural.
For example, $\LimSup$ assigns to a word $w$ the largest value realized by infinitely many runs over $w$.
In the probabilistic setting, the word aggregator $\E$ instead assigns to $w$ the expected value over the induced distribution on runs.

Quantitative language automata extend QWAs with a third layer of aggregation, called the \emph{language aggregator}.
Intuitively, after a QWA assigns a value to each word, the language aggregator combines these word values across all words generated by the system under consideration.
A language generator can be nonprobabilistic, in which case it defines a set of infinite words, or probabilistic, in which case it defines a probability measure over infinite words.
Accordingly, a QLA $\AA = (h,\A)$ consists of a QWA $\A$ together with a language aggregator $h$, and maps a language generator $G$ to the aggregated value $h_{w \sim G}\A(w)$.
For nonprobabilistic generators, aggregators such as $\Inf$ and $\Sup$ capture worst-case and best-case values over all generated executions.
Their limit variants $\LimInf$ and $\LimSup$ refine this view by focusing on values that occur infinitely often across generated words.
For probabilistic generators, the language aggregator $\E$ captures average-case behavior with respect to the induced probability measure.
Combining run, word, and language aggregators in this way allows QLAs to express nested quantitative views, such as the expected value, over generated executions, of the best value realizable by a nondeterministic specification.

The standard automata-theoretic decision problems extend naturally to QLAs.
Consider a quantitative language automaton $\AA$ and a rational threshold $k$.
In the evaluation problem, the input is a finite-state language generator: an $\omega$-regular language in the nonprobabilistic case, and a finite-state Markov chain in the probabilistic case, and the task is to compute the value assigned by $\AA$ to that input.
In the nonemptiness and universality problems, we ask whether $\AA$ maps some, respectively every, (nonempty) language generator to a value that is at least $k$ or greater than $k$.
We study these problems both for unrestricted language generators and for natural restricted classes, most notably finite-state ones.

\setlength{\tabcolsep}{1.5pt}
\begin{table}[h!]
	\centering
	\renewcommand{\arraystretch}{1}

	\begin{center}
		\scalebox{1}{\begin{tabular}{c||c|c||c||c||c}
				& \multicolumn{2}{c||}{$h,g \in \{\Inf, \Sup\}$}
				& $h = \E$
				& $h \in \{\Inf, \Sup\}$
				& $h = \E$         \\
				\cline{2-6}
				$f$
				& $h = g$
				& $h \neq g$
				& $g \in \{\Inf, \Sup\}$
				& $g = \E$
				& $g = \E$         \\
				\hline\hline
				$\left\{\!\begin{smallmatrix}\Inf,\Sup,\\\LimInf,\LimSup\end{smallmatrix}\!\right\}$
				& \RefEntry{\PTime}{\scriptsize\Cref{evaluation:classic}}
				& \RefEntry{\PSpaceC}{\scriptsize\Cref{evaluation:classic}}
				& \RefEntry{\PSpaceC}{\scriptsize\Cref{stochastic:classic:evaluation}}
				& \RefEntry{Undecidable}{\scriptsize\Cref{nondet:stochastic:evaluation}}
				& \RefEntry{\PTime}{\scriptsize\Cref{stochastic:stochastic:evaluation}}
				\\
				\hline
				$\left\{\!\begin{smallmatrix}\LimInfAvg,\\\LimSupAvg\end{smallmatrix}\!\right\}$
				& \RefEntry{\PTime}{\scriptsize\Cref{evaluation:avg}}
				& \RefEntry{Undecidable}{\scriptsize\Cref{evaluation:avg}}
				& \RefEntry{Undecidable}{\scriptsize\Cref{stochastic:avg:evaluation}}
				& \RefEntry{Undecidable}{\scriptsize\Cref{nondet:stochastic:evaluation}}
				& \RefEntry{\PTime}{\scriptsize\Cref{stochastic:stochastic:evaluation}}
				\\
				\hline
				$\DSum$
				& \RefEntry{\PTime}{\scriptsize\Cref{evaluation:dsum}}
				& \RefEntry{Open $^\dagger$}{\scriptsize\Cref{evaluation:dsum}}
				& \RefEntry{Open $^\dagger$}{\scriptsize\Cref{stochastic:nondet:dsum}}
				& \RefEntry{Open}{}
				& \RefEntry{\PTime}{\scriptsize\Cref{stochastic:stochastic:evaluation}}
				\\
			\end{tabular}}\\[2pt]
		\textbf{(a)} Evaluation for word and language aggregators $g, h \in \{\Inf, \Sup, \E\}$ (\Cref{sec:evaluation}).
	\end{center}

	\vspace{8pt}

	\begin{center}
		\scalebox{1}{\begin{tabular}{c||c|c|c}
				& \multicolumn{3}{c}{$h \in \{\Inf, \Sup, \E\}$}            \\
				\cline{2-4}
				$f$
				& $g = \Sup$
				& $g = \Inf$
				& $g = \E$                                                   \\
				\hline\hline
				$\left\{\!\begin{smallmatrix}\Inf,\Sup,\\\LimInf,\LimSup\end{smallmatrix}\!\right\}$
				& \RefEntry{\PTime}{\scriptsize\Cref{ptime:nonemptiness}}
				& \RefEntry{\PSpaceC}{\scriptsize\Cref{emptiness:classic}}
				& \RefEntry{Undecidable}{\scriptsize\Cref{emptiness:mixe}}
				\\
				\hline
				$\left\{\!\begin{smallmatrix}\LimInfAvg,\\\LimSupAvg\end{smallmatrix}\!\right\}$
				& \RefEntry{\PTime}{\scriptsize\Cref{ptime:nonemptiness}}
				& \RefEntry{Undecidable / Open $^\ast$}{\scriptsize\Cref{emptiness:avg}}
				& \RefEntry{Undecidable}{\scriptsize\Cref{emptiness:mixe}}
				\\
				\hline
				$\DSum$
				& \RefEntry{\PTime}{\scriptsize\Cref{ptime:nonemptiness}}
				& \RefEntry{Open $^\dagger$}{\scriptsize\Cref{emptiness:dsum}}
				& \RefEntry{Undecidable / Open $^\ast$}{\scriptsize\Cref{emptiness:mixe}}
				\\
			\end{tabular}}\\[2pt]
		\textbf{(b)} Nonemptiness for word and language aggregators $g, h \in \{\Inf, \Sup, \E\}$ (\Cref{sec:emptiness}).
	\end{center}

	\vspace{8pt}

	\begin{center}
		\scalebox{1}{\begin{tabular}{c||>{\centering\arraybackslash}p{3.2cm}||>{\centering\arraybackslash}p{3.2cm}}
				\multicolumn{3}{>{\centering\arraybackslash}p{7.5cm}}{$f,h \in \{\Inf,\Sup,\LimInf,\LimSup\}$,}\\
				\multicolumn{3}{>{\centering\arraybackslash}p{7.5cm}}{at least one of $g,h$ in $\{\LimInf,\LimSup\}$}\\
				\hline
				$g$
				& Evaluation
				& Nonemptiness                                               \\
				\hline\hline
				$\Sup$
				& \RefEntry{\PSpaceC}{\scriptsize\Cref{cor:PSpace:complete:evaluation:limit}}
				& \RefEntry{\PTime}{\scriptsize\Cref{top:PTIME}}
				\\
				\hline
				$\LimSup$
				& \RefEntry{\PSpaceC}{\scriptsize\Cref{cor:PSpace:complete:evaluation:limit}}
				& \RefEntry{\PSpaceC}{\scriptsize\Cref{top:PSpace}}
				\\
				\hline
				$\Inf$
				& \RefEntry{\PSpaceC}{\scriptsize\Cref{cor:PSpace:complete:evaluation:limit}}
				& \RefEntry{\PSpaceC}{\scriptsize\Cref{thm:limit:emptiness:hard}}
				\\
				\hline
				$\LimInf$
				& \RefEntry{\PSpaceC}{\scriptsize\Cref{cor:PSpace:complete:evaluation:limit}}
				& \RefEntry{\PSpaceC}{\scriptsize\Cref{thm:limit:emptiness:hard}}
				\\
			\end{tabular}}\\[2pt]
		\textbf{(c)} Evaluation and nonemptiness for all aggregators $f,g,h \in \{\Inf,\Sup,\LimInf,\LimSup\}$ with at least one of $g,h$ in $\{\LimInf,\LimSup\}$ (\Cref{sec:limit}).
	\end{center}

	\caption{
		Complexity of evaluation and nonemptiness for QLAs $\AA = (h,(g,f,\T))$.
		Universality is dual to nonemptiness (see \Cref{rem:duality}), e.g., in \PTime for QLAs with $f \in \{\Inf$, $\Sup$, $\LimInf$, $\LimSup\}$, $g = \Inf$, and $h \in \{\Inf, \Sup, \E\}$.
		Except for the cell marked with $\ast$, all results hold also for the finite-state restriction of the problem, where the language generators are restricted to $\omega$-regular languages or trace distributions generated by finite-state Markov chains.
		For the cell marked with $\ast$, we note also that the strict version of the problem is undecidable while the non-strict one is open.
		For the cells marked with $\dagger$, the problem is at least as hard as the universality problem for nondeterministic $\DSum$ QWAs, which is a major open problem~\cite{DBLP:conf/lics/BokerHO15}.
		\label{tbl:complexity}
	}
\end{table}

\paragraph{Contribution and Overview.}
Our main contribution is the introduction and systematic study of quantitative language automata (QLAs), based on three layers of aggregation: run, word, and language aggregation.
This framework supports the specification and verification of quantitative hyperproperties over both nonprobabilistic and probabilistic language generators.

\cref{sec:preliminaries} introduces the formal framework, and \cref{sec:applications} illustrates it through examples of quantitative hyperproperties specified by QLAs.
\cref{sec:problems} defines the evaluation, nonemptiness, and universality problems studied in the paper.
\cref{sec:evaluation} analyzes evaluation for QLAs with word and language aggregators in $\{\Inf,\Sup,\E\}$, and \cref{sec:emptiness} studies the corresponding nonemptiness and universality problems.
\cref{sec:limit} turns to the limit-aggregator fragment, namely QLAs whose word or language aggregator is $\LimInf$ or $\LimSup$.
\cref{sec:conclusion} concludes with directions for future work.
We summarize our main decidability and complexity results in \cref{tbl:complexity}.

To present a comprehensive picture, we overcome several technical challenges:

First, for evaluation (\cref{sec:evaluation}), we show that even for $\omega$-regular input languages, the optimal value need not be attained in a simple way: in the limit-average setting, it may fail to be achieved by any lasso word in the language, and in the discounted-sum setting, it may fail to be achieved by any word of the language at all.
For the polynomial-time solvable nonprobabilistic cases with matching word and language aggregators, we overcome these obstacles by proving that the value can still be computed effectively:
in the limit-average case by analyzing strongly connected components of the product automaton, and in the discounted-sum case by passing to the safety closure, which preserves the value and realizes it.
We complement these upper bounds with matching hardness results, including undecidability for limit-average when the word and language aggregators differ, and hardness for discounted-sum via reductions from universality problems of the underlying QWAs.

Second, for nonemptiness and universality (\cref{sec:emptiness}), we study extremal values of the underlying QWAs, in particular their bottom values.
We show that the bottom value of nondeterministic discounted-sum QWAs is always achieved by some word, and that the bottom value of nondeterministic $\LimSupAvg$ QWAs can be approximated arbitrarily closely by lasso words, in sharp contrast to the $\LimInfAvg$ case~\cite{DBLP:journals/tocl/ChatterjeeDH10}.
These results allow us to derive hardness results and transfer principles for nonemptiness and universality of several QLA classes.
Moreover, for many of the classes we analyze, the unrestricted versions of nonemptiness and universality coincide with their finite-state restrictions, yielding finite-model properties.

Third, in the limit-aggregator fragment (\cref{sec:limit}), where all aggregators belong to $\{\Inf,\Sup,\LimInf,\LimSup\}$ and at least one of the word or language aggregators is a limit aggregator, we study which values occur infinitely often, both across runs of a fixed word and across words of a language.
To do so, we combine a finite combinatorial pattern witnessing infinitely many words with automata-theoretic constructions that isolate exact-value runs and detect when such runs occur infinitely often.
These tools yield expressive-power results and complexity bounds for this fragment, including a uniform upper bound for evaluation and comparison, and a precise picture for nonemptiness and universality.

This article extends the conference version~\cite{DBLP:conf/concur/HenzingerKMS25}.
The main new material concerns QLAs with limit aggregators, developed in \cref{sec:limit}, where we introduce new automata-theoretic techniques for reasoning about values realized infinitely often and use them to obtain expressive-power results as well as tight complexity bounds for evaluation, comparison, nonemptiness, and universality in this fragment.
Beyond the limit-aggregator setting, we also sharpen the remainder of the paper by completing the complexity picture for the previously studied cases and tightening several of the bounds.
We also provide the full proofs omitted from the conference version due to space constraints, and we have polished the exposition throughout for clarity and readability.

\paragraph{Related Work.}
Our work builds on quantitative languages and quantitative word automata~\cite{DBLP:journals/tocl/ChatterjeeDH10,DBLP:conf/fct/ChatterjeeDH09}, as well as on games with quantitative objectives~\cite{DBLP:conf/csl/DegorreDGRT10}.
A particularly relevant line on multi-trace reasoning over quantitative signals is HyperSTL~\cite{DBLP:conf/memocode/NguyenKJDJ17}, which extends signal temporal logic with trace quantifiers to specify hyperproperties of real-valued signals.
HyperSTL also equips a fragment with robustness-based quantitative semantics.
This is close in motivation to our interest in quantitative system-wide properties, but different in formalization: HyperSTL measures robustness of satisfaction of a temporal formula, whereas QLAs aggregate quantitative word-automaton values across runs and across words.

There have also been other notions of quantitative hyperproperties~\cite{DBLP:conf/cav/FinkbeinerHT18,DBLP:conf/cav/SahaiS020}.
In these works, the quantitative aspect is based on counting how many traces participate in a given relation.
As a result, they capture counting- and cardinality-based specifications such as quantitative information-flow bounds, but not system-wide extremal or average-value properties such as the worst-case server uptime from \cref{sec:applications}.
Conversely, QLAs with the aggregators considered here do not express such counting constraints.

Closer in spirit to our semantic view is the predicate-transformer approach of~\cite{DBLP:journals/pacmpl/ZhangZK024}, which reasons about quantitative hyperproperties of finite program executions via functions from sets or distributions of final states to quantitative values.
It can express, for example, extremal quantities over final-state values and statistical quantities such as variance.
However, its setting is finite executions of imperative programs, and it does not target long-run word properties such as $\LimSup$, $\LimInfAvg$, or $\DSum$ over infinite traces.
Conversely, QLAs do support such long-run quantitative objectives, but with the current aggregators they do not express quantities such as variance that combine multiple expectations.

To the best of our knowledge, our work provides the first systematic framework for quantitative hyperproperties understood as functions from sets or distributions of infinite words to quantitative values, and the first systematic study of the specification and verification of such properties through automata.

\section{Definitional Framework}\label{sec:preliminaries}

Let $\Sigma$ be a finite \emph{alphabet} of letters.
A \emph{word} (or \emph{trace}) over $\Sigma$ is a finite or infinite sequence of letters from $\Sigma$.
We denote by $|w|$ the length of a finite word $w$.
We denote by $\Sigma^*$ (resp. $\Sigma^\omega$) the set of all finite (resp. infinite) words over $\Sigma$.
An infinite word $w$ is \emph{ultimately periodic} (a.k.a. \emph{lasso}) iff $w = u v^\omega$ for some $u,v \in \Sigma^*$ such that $|v| \geq 1$.
A \emph{language} is a set of infinite words.
Given a language $L \subseteq \Sigma^\omega$, we denote by $L^\complement$ its \emph{complement} $\Sigma^\omega \setminus L$.
Given $u \in \Sigma^*$ and $w \in \Sigma^\omega$, we write $u \prec w$ when $u$ is a prefix of $w$.

We denote by $\NN$ the set of natural numbers (including 0), $\QQ$ the set of rational numbers, $\RR$ the set of real numbers.
We further let $\overline{\NN} = \NN \cup \{\infty\}$ and $\overline{\RR} = \RR \cup \{-\infty, +\infty\}$.
Consider a set $S$.
An $S$-multiset is a function $M : S \to \overline{\NN}$ that maps each element of $S$ to a value denoting its multiplicity.
The support of a multiset $M$ is the set $\textit{Supp}(M) = \{ x \in S \st M(x) \geq 1\}$ of distinct elements in $M$, and its infinite support is the set $\textit{InfSupp}(M) = \{ x \in S \st M(x) = \infty\}$ of elements with infinite multiplicities.

A \emph{value domain} $\DD$ is a nontrivial complete lattice.
A \emph{quantitative property} is a total function $\varPhi : \Sigma^\omega \to \DD$.

\subsection{Quantitative Word Automata}
A \emph{weighted labeled transition system} is a tuple $\T = (\Sigma, Q, q_I, \delta, d)$, where:
$\Sigma$ is a finite alphabet, $Q$ is a finite nonempty set of states, $q_I \in Q$ is the initial state, $\delta \colon Q \times \Sigma \to 2^{\QQ \times Q}$ is a finite transition function over weight--state pairs, and $d \colon Q \times \Sigma \times \QQ \times Q \to [0,1]$ is a probability distribution satisfying, for all $q \in Q$ and $a \in \Sigma$:
\begin{enumerate}[(i)]
	\item $d(q,a,x,q') > 0$ iff $(x,q') \in \delta(q,a)$, and
	\item $\sum_{(x,q') \in \delta(q,a)} d(q,a,x,q') = 1$.
\end{enumerate}

A \emph{transition} is a tuple $t = (q,a,x,q')$ with $(x,q') \in \delta(q,a)$, written $\trans{q}{a:x}{q'}$.
Its weight is $\gamma(t) = x$, and we write $d(t)$ for $d(q,a,x,q')$.
Two transitions $(q,a,x,p)$ and $(q',a',x',p')$ are \emph{parallel} if $q = q'$, $a = a'$, $p = p'$, and $x \neq x'$. 
We say that $\T$ is \emph{deterministic} iff $|\delta(q,a)| = 1$ for every $q \in Q$ and $a \in \Sigma$, and \emph{complete} (a.k.a.\ \emph{total}) iff $|\delta(q,a)| \geq 1$ for every $q \in Q$ and $a \in \Sigma$.
Throughout the paper, we assume that weighted labeled transition systems are complete and have no parallel transitions. 

Although weighted labeled transition systems are probabilistic by definition, they can be viewed as \emph{nondeterministic} by considering only the support of $d$, i.e., treating all transitions with positive probability as nondeterministic choices.

The \emph{dual} $\hat{\T}$ of $\T$ is obtained by negating all weights:
$(x,q') \in \delta(q,a)$ becomes $(-x,q') \in \hat{\delta}(q,a)$, while probabilities are preserved: $\hat{d}(q,a,-x,q') := d(q,a,x,q')$.

For a word $w = a_0 a_1 \ldots \in \Sigma^\omega$, a \emph{run of $\T$ on $w$} is an infinite sequence
$\rho = \trans{q_0}{a_0:x_0}{q_1}\, \trans{q_1}{a_1:x_1}{q_2}\, \ldots$ of transitions such that $q_0 = q_I$ and $(x_i, q_{i+1}) \in \delta(q_i, a_i)$ for all $i \geq 0$.
We write $\gamma(\rho) = x_0 x_1 \ldots \in \QQ^\omega$ for the induced weight sequence and $R(\T,w)$ for the set of all runs of $\T$ on $w$.
For a finite run prefix $r = t_0 t_1 \ldots t_n$, define $\Pr(r) = \prod_{i=0}^{n} d(t_i)$.
The family of cylinder probabilities $\Pr(\cdot)$ induces a unique probability measure $\mu_{\T,w}$ on the Borel $\sigma$-algebra over $R(\T,w)$.

\paragraph{Run Aggregators.}
A \emph{run aggregator} (a.k.a.\ value function) is a function $f \colon \QQ^\omega \to \overline{\RR}$ that accumulates an infinite sequence of weights into a single value.
We consider the following run aggregators over a sequence $x = x_0 x_1 \ldots$ of rational weights and a discount factor
$\lambda \in \QQ \cap (0,1)$:

\hfill\break
\noindent\begin{minipage}{\linewidth}\medskip
	\begin{minipage}[t]{.45\linewidth}
		\begin{itemize}
			\item $\Inf(x) = \inf\{x_n \st n \in \NN\}$
			\item $\Sup(x) = \sup\{x_n \st n \in \NN\}$
			\item $\LimInf(x) = \liminf_{n\to\infty} x_n$
			\item $\LimSup(x) = \limsup_{n\to\infty} x_n$
		\end{itemize}
	\end{minipage}%
	\begin{minipage}[t]{.55\linewidth}
		\begin{itemize}
			\item $\LimInfAvg(x) = \liminf_{n\to\infty} \bigl(\frac{1}{n} \sum_{i=0}^{n-1} x_i\bigr)$
			\item $\LimSupAvg(x) = \limsup_{n\to\infty} \bigl(\frac{1}{n} \sum_{i=0}^{n-1} x_i\bigr)$
			\item $\DSum_{\lambda}(x) = \sum_{i \geq 0} \lambda^i\, x_i$
		\end{itemize}
	\end{minipage}%
	\medskip\end{minipage}
\indent

We write $\DSum$ when the discount factor $\lambda$ is left unspecified.
The run aggregators $\Inf$ and $\Sup$ (resp., $\LimInf$ and $\LimSup$, $\LimInfAvg$ and $\LimSupAvg$) are duals; $\DSum_\lambda$ is self-dual.

\paragraph{Word Aggregators.}
Given $w$, $\T$, and $f$, we define the \emph{multiset of run values} $M_{w,\T,f} \colon \overline{\RR} \to \overline{\NN}$ by
\[
	M_{w,\T,f}(x) =
	\begin{cases}
		k      & \text{if } \bigl|\{\rho \in R(\T,w) \mid f(\gamma(\rho)) = x\}\bigr| = k \in \NN, \\
		\infty & \text{if } \{\rho \in R(\T,w) \mid f(\gamma(\rho)) = x\} \text{ is infinite.}
	\end{cases}
\]
Let $\textit{Supp}(M) = \{x \mid M(x) \geq 1\}$ and $\textit{InfSupp}(M) = \{x \mid M(x) = \infty\}$.
Moreover, define the \emph{pushforward distribution} $\nu_{w,\T,f}$ on $\overline{\RR}$ by
\[
	\nu_{w,\T,f}(B) = \mu_{\T,w}\bigl(\{\rho \mid f(\gamma(\rho)) \in B\}\bigr)
\]
for every Borel set $B \subseteq \overline{\RR}$.

A \emph{word aggregator} accumulates the values of runs into a single value.
The aggregators $\Inf^\tau,\Sup^\beta,\LimInf^\tau,\LimSup^\beta$ act on $\overline{\RR}$-multisets $M$, while $\E$ acts on probability measures $\nu$ on $\overline{\RR}$.
We consider:
\begin{align*}
	\Inf^\tau(M) &=
	\begin{cases}
		\inf\, \textit{Supp}(M) & \text{if } \textit{Supp}(M) \neq \emptyset, \\
		\tau & \text{otherwise,}
	\end{cases} \\[4pt]
	\Sup^\beta(M) &=
	\begin{cases}
		\sup\, \textit{Supp}(M) & \text{if } \textit{Supp}(M) \neq \emptyset, \\
		\beta & \text{otherwise,}
	\end{cases} \\[4pt]
	\LimInf^\tau(M) &=
	\begin{cases}
		\inf\, \textit{InfSupp}(M) & \text{if } \textit{InfSupp}(M) \neq \emptyset, \\
		\tau & \text{otherwise,}
	\end{cases} \\[4pt]
	\LimSup^\beta(M) &=
	\begin{cases}
		\sup\, \textit{InfSupp}(M) & \text{if } \textit{InfSupp}(M) \neq \emptyset, \\
		\beta & \text{otherwise,}
	\end{cases} \\[4pt]
	\E(\nu) &= \int_{\overline{\RR}} x\, d\nu(x).
\end{align*}
Here $\tau,\beta \in \overline{\RR}$ are constants we will fix below.
The word aggregators $\Inf^\tau$ and $\Sup^\beta$ (resp.\ $\LimInf^\tau$ and $\LimSup^\beta$) are duals; $\E$ is self-dual.

\paragraph{QWA Syntax and Semantics.}
A \emph{quantitative word automaton} (QWA)~\cite{DBLP:journals/tocl/ChatterjeeDH10} is a tuple $\A = (g, f, \T)$ where $\T$ is a weighted labeled transition system, $f$ is a run aggregator, and $g$ is a word aggregator.
For $w \in \Sigma^\omega$, define
\[
	\A(w) =
	\begin{cases}
		g(M_{w,\T,f}) & \text{if } g \in \{\Inf, \Sup, \LimInf^\tau, \LimSup^\beta\}, \\
		\E(\nu_{w,\T,f}) & \text{if } g = \E.
	\end{cases}
\]

The \emph{top value} of $\A$ is $\top_{\A} = \sup_{w \in \Sigma^\omega} \A(w)$ and its \emph{bottom value} is $\bot_{\A} = \inf_{w \in \Sigma^\omega} \A(w)$.
We say that $\top_{\A}$ (resp.\ $\bot_{\A}$) is \emph{achievable} iff there exists a word $w$ with $\A(w) = \top_{\A}$ (resp.\ $\A(w) = \bot_{\A}$).

Since $\T$ is complete, every word admits at least one run, so $\textit{Supp}(M_{w,\T,f}) \neq \emptyset$.
Hence, $\Inf^\tau$ and $\Sup^\beta$ are always well-defined on $M_{w,\T,f}$, and $\E$ is well-defined on $\nu_{w,\T,f}$.
In contrast, for $g \in \{\LimInf^\tau, \LimSup^\beta\}$ it may happen that $\textit{InfSupp}(M_{w,\T,f}) = \emptyset$; the constants $\tau,\beta$ handle this degenerate case.
To make $\LimInf^\tau$ and $\LimSup^\beta$ total in a canonical way, we set:
given $\A^{(g \gets g')} = (g', f, \T)$, let
$\tau = \top_{\A^{(g \gets \Inf)}}$ if $g = \LimInf^\tau$, and
$\beta = \bot_{\A^{(g \gets \Sup)}}$ if $g = \LimSup^\beta$.
Henceforth, we leave $\tau$ and $\beta$ implicit and simply write $\Inf$, $\Sup$, $\LimInf$, and $\LimSup$ for the word aggregators of a QWA.

\paragraph{Duals and Special Cases.}
A QWA is \emph{nondeterministic} when its word aggregator is $\Sup$, and
\emph{universal} when $\Inf$.
The \emph{dual} of a QWA $\A = (g, f, \T)$ is
$\hat{\A} = (\hat{g}, \hat{f}, \hat{\T})$, where $\hat{g}$, $\hat{f}$, and
$\hat{\T}$ are the duals of $g$, $f$, and $\T$, respectively.

boolean word automata are the special case of QWAs with weights in $\{0,1\}$,
$g \in \{\Inf, \Sup, \LimInf, \LimSup\}$, and
$f \in \{\Inf, \Sup, \LimInf, \LimSup\}$.
In particular, a nondeterministic B\"uchi automaton (NBA) is a boolean word automaton
with $g = \Sup$ and $f = \LimSup$.
Given a boolean word automaton $\B$, we write
$L(\B) = \{w \in \Sigma^\omega \mid \B(w) = 1\}$ for its language.

\subsection{Quantitative Language Automata}

A \emph{language generator} is a function $G \colon 2^{\Sigma^\omega} \to [0,1]$.
A language generator $G$ is \emph{nonprobabilistic} iff there is a language $L_G \subseteq \Sigma^\omega$ such that $G(L_G) = 1$ and $G(L) = 0$ for every $L \neq L_G$;
it is \emph{nonempty} iff $L_G \neq \emptyset$.
A language generator $G$ is \emph{probabilistic} iff its restriction to the Borel $\sigma$-algebra over $\Sigma^\omega$ is a probability measure, which we denote by $\mu_G$.
A \emph{nonprobabilistic quantitative hyperproperty} (resp.\ \emph{probabilistic quantitative hyperproperty}) is a total function from the set of all nonprobabilistic (resp.\ probabilistic) language generators to a value domain $\DD$.

\paragraph{Language Aggregators.}
A \emph{language aggregator} accumulates the values of words into a single value.
The aggregators $\Inf^\tau, \Sup^\beta, \LimInf^\tau, \LimSup^\beta$ act on $\overline{\RR}$-multisets, while $\E$ acts on probability measures on $\overline{\RR}$.
We use the same definitions as at the word level.
Here $\tau, \beta \in \overline{\RR}$ are constants fixed below.
The duality relationships are the same: $\Inf^\tau$ and $\Sup^\beta$ (resp., $\LimInf^\tau$ and $\LimSup^\beta$) are duals; $\E$ is self-dual.

\paragraph{QLA Syntax and Semantics.}
A \emph{quantitative language automaton} (QLA) is a pair $\AA = (h, \A)$ where $\A = (g, f, \T)$ is a QWA over $\Sigma$ and $h$ is a language aggregator.

Below, we assume the input to a QLA is a nonprobabilistic (resp.\ probabilistic) language generator when $h \neq \E$ (resp.\ $h = \E$).
We write $L$ to denote a nonprobabilistic language generator, and $\mu$ to denote a probabilistic one.

For the \emph{nonprobabilistic case}, consider a nonempty language generator $G$ with language $L_G$ and a QLA $\AA = (h, \A)$.
We define the $\overline{\RR}$-multiset $M_{G,\A} \colon \overline{\RR} \to \overline{\NN}$ by
\[
	M_{G,\A}(x) =
	\begin{cases}
		k      & \text{if } \bigl|\{\, w \in L_G \mid \A(w) = x \,\}\bigr| = k \in \NN, \\
		\infty & \text{if } \{\, w \in L_G \mid \A(w) = x \,\} \text{ is infinite.}
	\end{cases}
\]
and set
\[
	\AA(G) = h(M_{G,\A}) = h_{w \in L_G} \A(w).
\]

For the \emph{probabilistic case} (i.e., $h = \E$), consider a probabilistic language generator $G$.
We define the pushforward distribution $\nu_{G,\A}$ on $\overline{\RR}$ by
\[
	\nu_{G,\A}(B) = \mu_G\bigl(\{\, w \in \Sigma^\omega \mid \A(w) \in B \,\}\bigr)
\]
for every Borel set $B \subseteq \overline{\RR}$, and let
\[
	\AA(G) = \E(\nu_{G,\A})
		= \int_{\overline{\RR}} x\, d\nu_{G,\A}(x)
		= \int_{w \in \Sigma^\omega} \A(w)\, d\mu_G(w).
\]

The \emph{top value} of a QLA $\AA$, denoted $\top_{\AA}$, is the supremum of $\AA(G)$ over all admissible inputs $G$, and its \emph{bottom value}, denoted $\bot_{\AA}$, is the corresponding infimum.
Here, the admissible inputs are all nonempty nonprobabilistic language generators when $h \neq \E$, and all probabilistic language generators when $h = \E$.

\begin{prop}\label{all:top:bot}\label{all:top:bottom}
	Consider a QLA $\AA = (h, \A)$ with $h \in \{\Inf, \Sup, \E\}$.
	Then, $\top_{\AA} = \top_{\A}$ and $\bot_{\AA} = \bot_{\A}$.
\end{prop}
\begin{proof}
	Let $\AA = (h, \A)$ be a language automaton as in the statement.
	Assume $h = \Sup$.
	First, observe that for every language $L_1$ and $L_2$ with $L_1 \subseteq L_2$ we have $\AA(L_1) \leq \AA(L_2)$.
	We immediately obtain $\top_{\AA} = \AA(\Sigma^\omega)$.
	By definition, $\AA(\Sigma^\omega) = \sup_{w \in \Sigma^\omega} \A(w) = \top_{\A}$.
	For the case of bottom values, we have $\bot_{\AA} = \inf_{w \in \Sigma^\omega} \AA(\{w\})$ by the observation above, which implies that for every word $w$ and language $L$ with $w \in L$ we have $\AA(\{w\}) \leq \AA(L)$.
	Then, as $\AA(\{w\}) = \A(w)$, we conclude $\bot_{\AA} = \bot_{\A}$.
	The case of $h = \Inf$ is similar.
	
	Now, assume $h = \E$.
	Consider a sequence of words $w_1, w_2, \ldots$ such that $\lim_{i \to \infty} \A(w_i) = \top_{\A}$.
	Then, there is a sequence of probabilistic language generators $G_1, G_2, \ldots$ such that for each $i \geq 1$ the measure $\mu_i$ of $G_i$ assign probability 1 to $\{w_i\}$, which implies $\top_{\AA} \geq \top_{\A}$.
	Moreover, thanks to monotonicity of expected value, we have $\top_{\AA} \leq \top_{\A}$, which results in $\top_{\AA} = \top_{\A}$.
	The case of bottom values is similar.
\end{proof}

\paragraph{Degenerate Cases.}
For a nonempty nonprobabilistic language generator, every multiset $M_{G,\A}$ has nonempty support, so $\Inf^\tau$ and $\Sup^\beta$ are well-defined regardless of $\tau$ and $\beta$.
To handle the case of empty language ($L_G = \emptyset$), we set $\tau = \top_\A$ and $\beta = \bot_\A$.

For $h \in \{\LimInf^\tau, \LimSup^\beta\}$, the set $\textit{InfSupp}(M_{G,\A})$ may be empty even for nonempty language generators.
To make $\LimInf^\tau$ and $\LimSup^\beta$ total in a canonical way, given $\AA^{(h \gets h')} = (h', \A)$, let
$\tau = \top_{\AA^{(h \gets \Inf)}}$ if $h = \LimInf^\tau$, and
$\beta = \bot_{\AA^{(h \gets \Sup)}}$ if $h = \LimSup^\beta$.
Henceforth, we leave $\tau$ and $\beta$ implicit and simply write $\Inf$, $\Sup$, $\LimInf$, and $\LimSup$ for the language aggregators of a QLA.

\paragraph{Duals.}
Given a QLA $\AA = (h, \A)$, its \emph{dual} is $\hat{\AA} = (\hat{h}, \hat{\A})$ where $\hat{\A}$ is the dual of $\A$ and $\hat{h}$ is the dual of $h$.

\begin{prop}\label{all:dual}
	Consider a QLA $\AA$ and its dual $\hat{\AA}$.
	Then, $\AA(G) = -\hat{\AA}(G)$ for every language generator $G$.
\end{prop}
\begin{proof}
	Follows from the fact that $\inf X = - \sup -X$, $\liminf X = - \limsup -X$ and the self duality of $\E$ and $\DSum$.
\end{proof}

\section{Applications of Language Automata}\label{sec:applications}

\begin{figure}[t]
	\noindent\begin{minipage}[b]{.12\linewidth}\centering
		\scalebox{0.8}{
			\begin{tikzpicture}[bg={\scalebox{1}{$\T_{\text{up}}$}}, node distance =1.8cm]
				\node[state, initial] (0) {};
				\path[transition]
				(0) edge[loop above] node[align=center] {\small $\texttt{on}:1$, $\texttt{off}:0$} (0)
				(0) edge[opacity=0, loop below] (0)
				;
			\end{tikzpicture}
		}
	\end{minipage}%
	\noindent\begin{minipage}[b]{.30\linewidth}\centering
		\scalebox{0.8}{
			\begin{tikzpicture}[bg={\scalebox{1}{$\T_{\text{sim}}$}}, node distance =1.8cm]
				\node[state, initial] (0) {};
				\node[state, right of = 0] (1) {};
				\node[state, right of = 1] (2) {};

				\path[transition]
				(0) edge[loop above] node {\small $\texttt{idle},\texttt{gra}:0$} (0)
				(0) edge[bend right] node[below] {\small $\texttt{req}:0$} (1)
				(1) edge[bend right] node[above] {\small $\texttt{gra}:0$} (0)
				(1) edge[loop above] node {\small $\texttt{idle}:0$} (1)
				(1) edge node[above] {\small $\texttt{req}:0$} (2)
				(2) edge[loop above] node {\small $\Sigma:1$} (2)
				;
			\end{tikzpicture}
		}
	\end{minipage}%
	\noindent\begin{minipage}[b]{.30\linewidth}\centering
		\scalebox{0.8}{
			\begin{tikzpicture}[bg={\scalebox{1}{$\M$}}, node distance =1.8cm]
				\node[state, initial] (0) {};
				\node[state, right of = 0] (1) {};
				\node[state, right of = 1] (2) {};

				\path[transition]
				(0) edge[loop above] node[] {\small $\texttt{idle}:1-p$} (0)
				(0) edge[bend right] node[below] {\small $\texttt{req}:p$} (1)
				(1) edge[bend right] node[above] {\hspace{1em}\small $\texttt{gra}:1-q$} (0)
				(1) edge node[above] {\small $\texttt{req}:q$} (2)
				(2) edge[loop above] node {\small $\Sigma:1$} (2)
				;
			\end{tikzpicture}
		}
	\end{minipage}%
	\noindent\begin{minipage}[b]{.28\linewidth}\centering
		\scalebox{0.8}{
			\begin{tikzpicture}[bg={\scalebox{1}{$\T_{\text{com}}$}}, node distance =1.8cm]
				\node[state, initial] (0) {};
				\node[state, right of = 0] (1) {};
				\node[state, left of = 0] (2) {};

				\path[transition]
				(0) edge[bend right] node[below] {\small \phantom{kk}$\texttt{send}:\frac{9}{10}:1$} (1)
				(1) edge[bend right] node[above] {\small $\texttt{ack}:1$} (0)
				(0) edge[bend left] node[below] {\small $\texttt{send}:\frac{1}{10}:1$\phantom{kk}} (2)
				(2) edge[bend left] node[above] {\small $\texttt{ack}:5$} (0)
				(0) edge[loop above] node {\small $\texttt{ack}:1$} (0)
				(1) edge[loop above] node {\small $\texttt{send}:1$} (1)
				(2) edge[loop above] node {\small $\texttt{send}:1$} (2)

				;
			\end{tikzpicture}
		}
	\end{minipage}

	\centering
	\scalebox{0.8}{
		\begin{tikzpicture}[bg={\scalebox{1.2}{$\T_{\text{sta}}$}}, node distance =1.5cm]
			\node[state, initial] (0) {};
			\node[state, right of = 0, initial] (1) {};
			\node[state, right of = 1] (2) {};
			\node[state, right of = 2] (3) {};
			\node[state, left of = 0, initial] (-1) {};
			\node[state, left of = -1] (-2) {};
			\node[state, left of = -2] (-3) {};

			\path[transition]
			(0) edge[bend right] node[below] {\small $\texttt{r}:1$} (1)
			(0) edge[bend right] node[above] {\small $\texttt{l}:1$} (-1)
			(1) edge[bend right] node[below] {\small $\texttt{r}:2$} (2)
			(1) edge[bend right] node[above] {\small $\texttt{l}:0$} (0)
			(2) edge[bend right] node[below] {\small $\texttt{r}:3$} (3)
			(2) edge[bend right] node[above] {\small $\texttt{l}:1$} (1)
			(3) edge[bend right] node[above] {\small $\texttt{l}:2$} (2)
			(3) edge[loop above] node {\small $\texttt{r},\texttt{i}:3$} (3)
			(-1) edge[bend right] node[above] {\small $\texttt{l}:2$} (-2)
			(-1) edge[bend right] node[below] {\small $\texttt{r}:0$} (0)
			(-2) edge[bend right] node[above] {\small $\texttt{l}:3$} (-3)
			(-2) edge[bend right] node[below] {\small $\texttt{r}:1$} (-1)
			(-3) edge[bend right] node[below] {\small $\texttt{r}:2$} (-2)
			(-3) edge[loop above] node {\small $\texttt{l},\texttt{i}:3$} (-3)
			(-2) edge[loop above] node {\small $\texttt{i}:2$} (-2)
			(-1) edge[loop above] node {\small $\texttt{i}:1$} (-1)
			(0) edge[loop above] node {\small $\texttt{i}:0$} (0)
			(2) edge[loop above] node {\small $\texttt{i}:2$} (2)
			(1) edge[loop above] node {\small $\texttt{i}:1$} (1)
			;
		\end{tikzpicture}
	}
	\caption{%
		Weighted labeled transition systems $\T_{\text{up}}$, $\T_{\text{sim}}$, $\T_{\text{com}}$, and $\T_{\text{sta}}$ can be used to specify QWAs and QLAs with various aggregators, as demonstrated in \Cref{sec:applications}.
		Transitions are marked with $\sigma : p : x$, where $\sigma$ denotes a letter, $p$ the transition probability (dropped when $p=1$), and $x$ the transition weight.
		The Markov chain $\M$ is a language generator with parameters $0 \leq p,q \leq 1$ (its transitions are not weighted, and those with probability 0 are not shown).
		In $\T_{\text{sta}}$, the three initial arrows are shorthand for the three admissible initial positions.
	}
	\label{fig:examples}
\end{figure}

QWAs describe quantitative specifications \emph{per execution}, whereas QLAs describe \emph{system-wide} quantitative specifications.
Below we illustrate several applications of QLAs for specifying quantitative hyperproperties.
Notice that QLAs are an \emph{operational} specification language: the underlying word automaton is part of the specification itself.

\paragraph{Server Uptime.}
QWAs can model performance metrics of individual executions.
Let $\Sigma = \{\texttt{on}, \texttt{off}\}$ be a finite alphabet of observations modeling a server's activity.
Consider the weighted labeled transition system $\T_{\text{up}}$ from \Cref{fig:examples}, and let $\A_{\text{up}} = (g, f, \T_{\text{up}})$ be a QWA with $g = \Sup$ and $f = \LimInfAvg$.
The automaton $\A_{\text{up}}$ maps each infinite word $w$ to the lower long-run average frequency of $\texttt{on}$, i.e., the uptime of the execution modeled by $w$.
For example, if $w = \texttt{on} \cdot (\texttt{on} \cdot \texttt{off})^\omega$, then $\A_{\text{up}}(w) = 0.5$.

QLAs can model \emph{system-wide} performance metrics.
Let $\AA_{\text{up}} = (h, \A_{\text{up}})$.
Setting $h = \Inf$ makes $\AA_{\text{up}}$ map a language $L$, representing a server implementation, to the worst-case uptime over all executions in $L$, i.e., $\AA_{\text{up}}(L) = \inf_{w \in L} \A_{\text{up}}(w)$.
For example, let $U = \{\texttt{on}^n \texttt{off}^k \st n > k \geq 0\}$ and consider the language $L = \{u_1 u_2 \ldots \st \forall i \geq 1 : u_i \in U\}$ of server executions in which each block of $\texttt{on}$'s is followed by a strictly shorter block of $\texttt{off}$'s.
Every prefix of every word in $L$ contains strictly more $\texttt{on}$'s than $\texttt{off}$'s, so every word in $L$ has uptime at least $0.5$.
On the other hand, the word $w = s_1 s_2 \ldots \in L$, where $s_i = \texttt{on}^i \texttt{off}^{i-1} \in U$ for each $i \geq 1$, has uptime $0.5$.
Hence, $\AA_{\text{up}}(L) = 0.5$.
In particular, the value of $\AA_{\text{up}}(L)$ is not achieved by any ultimately periodic word in $L$; we later show that this can happen even for $\omega$-regular languages (\Cref{ultim:avg}).
Setting $h = \LimInf$ instead makes $\AA_{\text{up}}$ express the ``almost'' worst-case uptime, i.e., it considers only uptime values realized by infinitely many executions.
For example, let $L = \Sigma^* \texttt{on}^\omega \cup \{(\texttt{on}^n \texttt{off})^\omega \st n \geq 0\}$.
The set $\Sigma^* \texttt{on}^\omega$ contributes infinitely many executions with uptime $1$, while each word $(\texttt{on}^n \texttt{off})^\omega$ contributes the value $\frac{n}{n+1}$ only once.
Thus, although the infimum over $L$ is $0$, the only uptime value realized infinitely often is $1$, so $\AA_{\text{up}}(L) = 1$.

\paragraph{Implementation Distance.}
QWAs can specify the distance of individual executions from a desired boolean trace property.
Let $\Sigma = \{\texttt{req}, \texttt{gra}, \texttt{idle}\}$ be a finite alphabet of observations modeling a server receiving requests and issuing grants, and consider the boolean safety property $P$ requiring that no two requests are simultaneously open, i.e., there is a $\texttt{gra}$ between every two $\texttt{req}$'s.
Consider the transition system $\T_{\text{sim}}$ from \Cref{fig:examples}, and let $\A_{\text{sim}} = (g, f, \T_{\text{sim}})$ be a QWA where $f = \DSum_{0.5}$ and $g \in \{\Inf,\Sup\}$ (the choice of $g$ does not matter because $\T_{\text{sim}}$ is deterministic).
The automaton $\A_{\text{sim}}$ maps each infinite word $w$ to its Cantor distance from $P$, i.e., $\A_{\text{sim}}(w) = \inf_{w' \in P} d(w, w')$.
For example, if $w = \texttt{req} \cdot \texttt{idle} \cdot \texttt{req}^\omega$, then $\A_{\text{sim}}(w) = 0.25$.

QLAs can specify the distance of \emph{systems} from a desired boolean trace property.
Let $\AA_{\text{sim}} = (h, \A_{\text{sim}})$.
Setting $h = \Sup$ makes $\AA_{\text{sim}}$ map a language $L$, representing a server implementation, to the worst-case distance from $L$ to $P$.
Equivalently, $\AA_{\text{sim}}(L) = \sup_{w \in L} \inf_{w' \in P} d(w,w')$, where $d$ denotes the Cantor distance.
For example, if $L = \Sigma^* \texttt{req}^2 \Sigma^\omega$, then $\AA_{\text{sim}}(L) = 0.5$.
Setting $h = \Inf$ makes $\AA_{\text{sim}}$ map each language to its best-case distance from $P$.
For the same language $L$, we get $\AA_{\text{sim}}(L) = 0$ even though every word in $L$ violates $P$.
Indeed, for every $i \geq 0$, the word $w_i = \texttt{idle}^i \texttt{req}^2 \texttt{idle}^\omega$ belongs to $L$ and $\lim_{i \to \infty} \A_{\text{sim}}(w_i) = 0$.
This illustrates a challenge in evaluating discounted-sum QLAs: the value of $L$ need not be realized by any word in $L$, even when $L$ is $\omega$-regular.
Setting $h = \E$ makes $\AA_{\text{sim}}$ map a probabilistic language generator to the average-case distance to $P$.
For example, let $\mu_{\M}$ be the trace distribution induced by the Markov chain $\M$ in \Cref{fig:examples}.
The expected value can be computed by solving the corresponding system of linear equations, yielding $\AA_{\text{sim}}(\mu_{\M}) = \frac{2pq}{2+p(1+q)}$.

\paragraph{Robot Stability.}
QWAs can express stability constraints of individual executions.
Let $\Sigma = \{\texttt{l}, \texttt{r}, \texttt{i}\}$ be a finite alphabet representing robot movements (\emph{left}, \emph{right}, and \emph{idle}) on a one-dimensional finite grid.
An execution is $\varepsilon$-stable, for some $\varepsilon \geq 0$, if there exists $0 \leq \delta \leq \varepsilon$ such that whenever the system starts within a $\delta$-ball around the origin, it remains forever within an $\varepsilon$-ball around the origin.
For a fixed $\delta \geq 0$, QWAs can express the least admissible $\varepsilon$ associated with each execution.
Consider the weighted labeled transition system $\T_{\text{sta}}$ from \Cref{fig:examples}, and let $\A_{\text{sta}} = (\Sup, \Sup, \T_{\text{sta}})$.
Here the three initial arrows are shorthand for the three admissible starting positions $-1$, $0$, and $1$; equivalently, one can encode them using a fresh initial state.
The automaton captures the case $\delta = 1$, meaning that the robot starts at distance at most $1$ from the origin.
Transition weights record distances from the origin, and a run value is the maximal distance reached.
Thus, the automaton's value on a word is the worst-case required $\varepsilon$ over all admissible initial positions.
For instance, the word $w = (\texttt{l} \cdot \texttt{r})^\omega$ has three runs with values $1$, $1$, and $2$, so $\A_{\text{sta}}(w) = 2$.

QLAs can express stability constraints of \emph{systems}.
Let $\AA_{\text{sta}} = (h, \A_{\text{sta}})$.
If $h = \Sup$, then for a language $L$ modeling the robot's behavior, the automaton $\AA_{\text{sta}}$ maps $L$ to the least upper bound of the per-execution $\varepsilon$-values induced by $L$ and $\A_{\text{sta}}$.
That is, $\AA_{\text{sta}}(L) = \sup_{w \in L} \A_{\text{sta}}(w)$, which is the smallest $\varepsilon$ ensuring that every execution in $L$ is $\varepsilon$-stable with $\delta = 1$.
If $h = \LimSup$, then $\AA_{\text{sta}}(L)$ captures the almost worst-case required $\varepsilon$: it is the largest $\varepsilon$-value realized by infinitely many executions, while values achieved only by finitely many executions are ignored.

\paragraph{Communication Channel Cost.}
Probabilistic QWAs can specify the expected cost of individual executions.
Let $\Sigma = \{\texttt{send}, \texttt{ack}\}$ be an alphabet modeling a communication channel.
Consider the transition system $\T_{\text{com}}$ from \Cref{fig:examples}, and let $\A_{\text{com}} = (\E, \LimSup, \T_{\text{com}})$ be a probabilistic QWA.
Each run is assigned its long-term maximal cost, i.e., the $\LimSup$ of its weight sequence, and each infinite word is assigned the expected value over the induced distribution of runs.
For example, for $w = (\texttt{send} \cdot \texttt{ack})^\omega$ we obtain $\A_{\text{com}}(w) = 5$.
Indeed, at each occurrence of $\texttt{send}$, the high-cost branch is taken with probability $\frac{1}{10}$, so it is taken infinitely often with probability $1$, and therefore the run value is almost surely $5$.

QLAs can specify the aggregate cost of a communication channel.
Let $\AA_{\text{com}} = (h, \A_{\text{com}})$.
If $h = \E$, then $\AA_{\text{com}}$ maps a probabilistic language generator to the expected long-run maximal cost of the channel.
Consider a Markov chain defining the uniform probability measure $\mu$ over the alphabet $\Sigma = \{\texttt{send}, \texttt{ack}\}$.
Then, $\AA_{\text{com}}(\mu) = 5$.
Indeed, in the product of $\mu$ and $\A_{\text{com}}$, every state is visited infinitely often with probability $1$.
In particular, the high-cost state (on the left in \Cref{fig:examples}) is visited infinitely often almost surely, and from this state the transition with weight 5 is taken with probability $\frac{1}{2}$.
Hence, this weight-$5$ transition is taken infinitely often with probability $1$.
Since all other transitions have weight $1$, the $\LimSup$ of the weight sequence is almost surely $5$, and therefore the expected value is $5$.

\section{Problems on Language Automata}\label{sec:problems}

Let $\AA = (h, \A)$ be a QLA, and let ${\compare} \in \{{>}, {\geq}\}$.
We study the following problems.

\begin{description}
	\item[Nonprobabilistic Evaluation ($h \neq \E$)]
	Given a B\"uchi automaton $\B$ with $L(\B) \neq \emptyset$, compute $\AA(L(\B))$.
	\item[Probabilistic Evaluation ($h = \E$)]
	Given a finite-state Markov chain $\M$, compute $\AA(\mu_\M)$, where $\mu_\M$ is the induced probability measure on $\Sigma^\omega$.
	\item[${\compare}$-Nonemptiness]
	Given $k \in \QQ$, is $\AA(G) \compare k$ for some nonempty language generator $G$?
	\item[${\compare}$-Universality]
	Given $k \in \QQ$, is $\AA(G) \compare k$ for every nonempty language generator $G$?
\end{description}

In the sequel, we identify a nonprobabilistic language generator with its language and a probabilistic one with its induced measure.

\begin{rem}
	Since evaluation may return a real value, our complexity statements are formulated for the associated threshold problem: given a language generator $G$ and a rational threshold $k \in \QQ$, decide whether $\AA(G) \geq k$.
	The complexity is measured in the size of the representation of $G$.
	This decision problem trivially polynomial-time reduces to exact evaluation, so every hardness result for the threshold version also applies to exact evaluation.
	Conversely, when $h,g \in \{\Inf,\Sup,\LimInf,\LimSup\}$ and $f \in \{\Inf,\Sup,\LimInf,\LimSup\}$, every run value, word value, and language value belongs to the finite set of distinct transition weights of the underlying weighted transition system.
	Hence, exact evaluation reduces to polynomially many threshold queries, one for each distinct transition weight.
	Therefore, for the complexity classes considered in this paper, the same upper bounds apply to exact evaluation in these cases.
	No analogous finite-valued reduction is available in general when $h = \E$ or $g = \E$, or when $f \in \{\LimInfAvg,\LimSupAvg,\DSum\}$, since the resulting value need not belong to the finite set of transition weights.
	Accordingly, in those cases we state our complexity results for the threshold version.
\end{rem}

For nonemptiness and universality, we also consider the following variants with restricted quantification over language generators.

\begin{description}
	\item[Borel] Nonprobabilistic generators $G$ with $L_G$ Borel in $\Sigma^\omega$ under the Cantor topology.
	\item[Markov] Probabilistic generators $G$ with $\mu_G$ Markovian.
	\item[Finite-state] Nonprobabilistic generators $G$ with $L_G$ $\omega$-regular; probabilistic generators $G$ with $\mu_G$ finite-state Markovian.
\end{description}

We will repeatedly relate the above questions for QLAs to the corresponding questions for their underlying QWAs.
For completeness, we spell out these word-level notions here.
Consider a QWA $\A$, a rational $k \in \QQ$, and ${\compare} \in \{{>}, {\geq}\}$.
Then, $\A$ is ${\compare}$-nonempty (resp.\ ${\compare}$-universal) for $k$ iff $\A(w) \compare k$ for some (resp.\ all) $w \in \Sigma^\omega$.
Similarly, the \emph{lasso-word restriction} requires quantification over lasso words instead of all words.

Moreover, $\A$ is \emph{approximate-nonempty} for $k$ iff $\top_{\A} \geq k$.
Equivalently, either there exists $w \in \Sigma^\omega$ such that $\A(w) \geq k$, or for every $\varepsilon > 0$ there exists $w$ such that $\A(w) > k - \varepsilon$.
Hence, if $\top_{\A}$ is achievable, then approximate-nonemptiness and $\geq$-nonemptiness coincide.
Note also that $\top_{\A} > k$ iff $\A(w) > k$ for some word $w$, independently of whether $\top_{\A}$ is achievable.
Dually, $\A$ is \emph{approximate-universal} for $k$ iff $\bot_{\A} > k$.

For the classes of QLAs considered in this paper, the Borel (resp.\ Markov) restrictions coincide with the unrestricted problems.
Thus, for decision problems we focus on the unrestricted and finite-state cases.

\begin{prop}\label{Borel:reduction}
	Consider a QLA $\AA = (h, \A)$ with $h \in \{\Inf,\Sup,\LimInf,\LimSup\}$ (resp.\ $h = \E$), a rational $k \in \QQ$, and ${\compare} \in \{{>}, {\geq}\}$.
	Then, $\AA$ is ${\compare}$-nonempty for $k$ iff $\AA$ is Borel ${\compare}$-nonempty (resp.\ Markov ${\compare}$-nonempty) for $k$.
	The same holds for universality.
\end{prop}
\begin{proof}
	We prove the claim for $\geq$-nonemptiness; the strict case is analogous.
	The implication from the restricted variant to the unrestricted one is immediate.

	Assume first that $h \in \{\Sup,\LimSup\}$ and that there exists a nonempty language $L \subseteq \Sigma^\omega$ with $\AA(L) \geq k$.
	For both aggregators, the map $L \mapsto \AA(L)$ is monotone with respect to inclusion.
	Hence $\AA(\Sigma^\omega) \geq \AA(L) \geq k$.
	Since $\Sigma^\omega$ is Borel, this yields a Borel witness.

	Assume next that $h = \Inf$ and that $\AA(L) = \inf_{w \in L}\A(w) \geq k$ for some nonempty language $L$.
	Then, every word in $L$ has value at least $k$, so for any $w \in L$ we have $\AA(\{w\}) = \A(w) \geq k$.
	Thus, a singleton, hence a closed set, witnesses Borel nonemptiness.

	Now let $h = \LimInf$, and suppose that $\AA(L) \geq k$ for some nonempty language $L$.
	If $\textit{InfSupp}(M_{L,\A}) = \emptyset$, then $\AA(L)$ is the default value of $\LimInf$.
	In this case every singleton language has the same value, so any singleton witnesses Borel nonemptiness.
	Otherwise, choose any value $v \in \textit{InfSupp}(M_{L,\A})$.
	Since $\AA(L) = \inf \textit{InfSupp}(M_{L,\A}) \geq k$, we have $v \geq k$.
	The set $L_v = \{ w \in L \st \A(w) = v \}$ is infinite by definition of $\textit{InfSupp}$, so it contains a countably infinite subset $L' \subseteq L_v$.
	As a countable union of singletons, $L'$ is Borel, and $\AA(L') = v \geq k$.

	Finally, let $h = \E$, and suppose that $\AA(\mu) \geq k$ for some probability measure $\mu$.
	Then, there must exist a word $w$ with $\A(w) \geq k$; otherwise $\A(w) < k$ for all $w$, and the expectation would be $< k$.
	A Markov chain generating $w$ with probability $1$ witnesses Markov nonemptiness.

	The universality claims follow from the corresponding nonemptiness claims by duality.
\end{proof}

For QLAs with language aggregators $\Inf$, $\Sup$, and $\E$, the decision problems often reduce to the corresponding problems for the underlying QWA.
Moreover, several of these classes enjoy a finite-model property.

\begin{prop}\label{bigreduction}\label{emptiness:reduction}
	Consider a QLA $\AA = (h, \A)$ with $h \in \{\Inf,\Sup,\E\}$, a rational $k \in \QQ$, and ${\compare} \in \{{>}, {\geq}\}$.
	\begin{enumerate}[(a)]
		\item Unrestricted variant
		\begin{enumerate}[(i)]\itemindent=-1.6em
			\item If $h \neq \Sup$ or ${\compare} = {>}$, then $\AA$ is $\compare$-nonempty for $k$ iff $\A$ is ${\compare}$-nonempty for $k$.
			\item If $h = \Sup$, then $\AA$ is $\geq$-nonempty for $k$ iff $\A$ is approximate-nonempty for $k$.
		\end{enumerate}
		\item Finite-state variant
		\begin{enumerate}[(i)]\itemindent=-1.6em
			\item If $h = \Sup$, then $\AA$ is finite-state ${\compare}$-nonempty for $k$ iff $\AA$ is ${\compare}$-nonempty for $k$.
			\item If $h = \Inf$, then $\AA$ is finite-state ${\compare}$-nonempty for $k$ iff $\A$ is lasso-word ${\compare}$-nonempty for $k$.
			\item If $\top_{\A}$ is achievable by a lasso word, then $\AA$ is finite-state ${\compare}$-nonempty for $k$ iff $\AA$ is ${\compare}$-nonempty for $k$ iff $\A$ is lasso-word ${\compare}$-nonempty for $k$.
		\end{enumerate}
	\end{enumerate}
	For universality, we obtain the dual statements by exchanging $\Inf$ with $\Sup$, $>$ with $\geq$, nonempty with universal, approximate-nonempty with approximate-universal, and $\top_{\A}$ with $\bot_{\A}$.
\end{prop}
\begin{proof}
	We prove the items one by one.

	(a)(i) Suppose first that $h = \Inf$.
	If $\AA(L) = \inf_{w \in L} \A(w) \compare k$ for some nonempty language $L$, then every word in $L$ satisfies $\A(w) \compare k$.
	Conversely, if $\A(w) \compare k$ for some word $w$, then $\AA(\{w\}) = \A(w) \compare k$.

	Suppose next that $h = \E$.
	If $\AA(\mu) \compare k$ for some probability measure $\mu$, then there exists a word $w$ with $\A(w) \compare k$; otherwise every word would have value failing ${\compare}\,k$, and so would the expectation.
	Conversely, if $\A(w) \compare k$ for some word $w$, then the Dirac measure concentrated on $w$ witnesses ${\compare}$-nonemptiness of $\AA$.

	Finally, suppose that $h = \Sup$ and ${\compare} = {>}$.
	If $\AA(L) = \sup_{w \in L} \A(w) > k$ for some nonempty language $L$, then there exists $w \in L$ with $\A(w) > k$.
	Conversely, if $\A(w) > k$ for some word $w$, then $\AA(\{w\}) > k$.

	(a)(ii) Assume $h = \Sup$.
	For every language $L \subseteq \Sigma^\omega$ we have $\AA(L) \leq \AA(\Sigma^\omega)$.
	Hence there exists $L \subseteq \Sigma^\omega$ with $\AA(L) \geq k$ iff $\AA(\Sigma^\omega) \geq k$.
	By \Cref{all:top:bot}, this is equivalent to $\top_{\AA} \geq k$, hence to $\top_{\A} \geq k$, i.e., to approximate-nonemptiness of $\A$.

	(b)(i) If $h = \Sup$, then there exists $L \subseteq \Sigma^\omega$ with $\AA(L) \compare k$ iff $\AA(\Sigma^\omega) \compare k$.
	Since $\Sigma^\omega$ is $\omega$-regular, the unrestricted and finite-state variants coincide.

	(b)(ii) Assume $h = \Inf$.
	If $\AA(L) = \inf_{w \in L} \A(w) \compare k$ for some nonempty $\omega$-regular language $L$, then every word in $L$ has value ${\compare}\,k$.
	Since every nonempty $\omega$-regular language contains a lasso word, there exists a lasso word $w \in L$ with $\A(w) \compare k$.
	Conversely, if $\A(w) \compare k$ for some lasso word $w$, then $\AA(\{w\}) = \A(w) \compare k$, and $\{w\}$ is $\omega$-regular.

	(b)(iii) Suppose first that $h = \E$.
	For every measure $\mu$ we have $\AA(\mu) \leq \top_{\A}$.
	Hence, if $\AA(\mu) \compare k$ for some $\mu$, then $\top_{\A} \compare k$.
	By assumption, there exists a lasso word $w$ with $\A(w) = \top_{\A}$, and thus $\A(w) \compare k$.
	Conversely, if $\top_{\A} \compare k$, then the same lasso word witnesses it, and the Dirac measure concentrated on $w$ is a finite-state and unrestricted witness for $\AA$.

	Suppose next that $h = \Inf$.
	By (a)(i), unrestricted ${\compare}$-nonemptiness of $\AA$ is equivalent to the existence of a word $w$ with $\A(w) \compare k$.
	This is in turn equivalent to $\top_{\A} \compare k$.
	Since $\top_{\A}$ is achieved by a lasso word, the latter is equivalent to the existence of a lasso word $w$ with $\A(w) \compare k$.
	Combining this with (b)(ii) gives the claim.

	Finally, suppose that $h = \Sup$.
	For ${\compare} = {>}$, unrestricted nonemptiness is equivalent to the existence of a word of value $> k$ by (a)(i), hence to $\top_{\A} > k$, and therefore to the existence of a lasso word of value $> k$ by the lasso-achievability assumption.
	For ${\compare} = {\geq}$, unrestricted nonemptiness is equivalent to approximate-nonemptiness of $\A$ by (a)(ii), that is, to $\top_{\A} \geq k$.
	Again, since $\top_{\A}$ is achieved by a lasso word, this is equivalent to lasso-word $\geq$-nonemptiness of $\A$.
	Using (b)(i), we conclude that the finite-state and unrestricted variants coincide and are both equivalent to lasso-word ${\compare}$-nonemptiness of $\A$.

	The universality statements are the corresponding duals.
\end{proof}

The next proposition isolates a useful sufficient condition for reducing unrestricted nonemptiness and universality to their finite-state variants.

\begin{prop}\label{approximability-reduction}
	Consider a QLA $\AA = (h, \A)$ with $h \in \{\Inf,\Sup,\E\}$ and $k \in \QQ$.
	If for every $\varepsilon > 0$ there is a lasso word $w$ such that $\A(w) \geq \top_{\A} - \varepsilon$, then $\AA$ is $>$-nonempty for $k$ iff $\AA$ is finite-state $>$-nonempty for $k$.
	Dually, if for every $\varepsilon > 0$ there is a lasso word $w$ such that $\A(w) \leq \bot_{\A} + \varepsilon$, then $\AA$ is $\geq$-universal for $k$ iff $\AA$ is finite-state $\geq$-universal for $k$.
\end{prop}
\begin{proof}
	If $h = \Sup$, the claim is immediate from \Cref{bigreduction}(b)(i).

	If $h = \Inf$, then by \Cref{bigreduction}(a)(i), $\AA$ is $>$-nonempty for $k$ iff there exists a word $w$ with $\A(w) > k$.
	This is equivalent to $\top_{\A} > k$.
	Choose $\varepsilon = (\top_{\A} - k)/2 > 0$.
	By assumption, there exists a lasso word $w$ with $\A(w) \geq \top_{\A} - \varepsilon > k$.
	Hence unrestricted $>$-nonemptiness is equivalent to lasso-word $>$-nonemptiness of $\A$, and now \Cref{bigreduction}(b)(ii) yields the finite-state equivalence.

	If $h = \E$, then by \Cref{bigreduction}(a)(i), $\AA$ is $>$-nonempty for $k$ iff there exists a word $w$ with $\A(w) > k$, equivalently iff $\top_{\A} > k$.
	With the same choice of $\varepsilon$, the assumption yields a lasso word $w'$ such that $\A(w') > k$.
	A finite-state Markov chain generating $w'$ with probability $1$ therefore witnesses finite-state $>$-nonemptiness.

	The dual statement follows by the same argument.
\end{proof}

\section{Solving Evaluation}\label{sec:evaluation}

In this section, we investigate the evaluation problem for QLAs with language and word aggregators $h,g \in \{\Inf, \Sup, \E\}$, for which we provide a full picture of complexity results.
First, in \cref{sec:nonprobeval}, we focus on the nonprobabilistic evaluation problem (where $h \in \{\Inf, \Sup\}$) and then, in \cref{sec:probeval}, on the probabilistic one (where $h = \E$).

\subsection{Nonprobabilistic Evaluation}\label{sec:nonprobeval}

First, we consider the evaluation problem for QLAs with $h \in \{\Inf,\Sup\}$.
We start with QLAs whose underlying word automata are universal or nondeterministic (i.e., $g \in \{\Inf,\Sup\}$).
We study various run aggregators $f$ separately and show that the problem is in \PTime when the word aggregator $g$ and the language aggregator $h$ coincide.
When they differ, the problem becomes harder: while it remains algorithmically solvable in \PSpace for the ``standard'' run aggregators (i.e., $f \in \{\Inf,\Sup,\LimInf,\LimSup\}$), we show that it is not computable for limit-average and at least as hard as a long-standing open problem for discounted-sum.
Finally, for QLAs whose underlying word automata are probabilistic (i.e., $g = \E$), we show that the problem is not computable.
At their core, the easiness results rely on analyzing the extreme values of the underlying word automata.
Similarly, we establish the hardness results by reductions from word automata problems.

\paragraph{QLAs with Standard QWAs.}
For QLAs with run aggregators $f\in\{\Inf$, $\Sup$, $\LimInf$, $\LimSup\}$, the nonprobabilistic evaluation problem can be solved by reasoning on lasso words since both top and bottom values of the underlying word automata are realized by lasso words.

\begin{thm}\label{evaluation:classic}
	Consider a QLA $\AA = (h,(g,f,\T))$ with $f\in\{\Inf$, $\Sup$, $\LimInf$, $\LimSup\}$ and $g,h \in \{\Inf$, $\Sup\}$.
	The evaluation of $\AA$ is in \PTime when $g = h$ and it is \PSpaceC when $g \neq h$.
\end{thm}
\begin{proof}
	Let $S = L(\B) \neq \emptyset$ be the input language, and let $\A = (g,f,\T)$ be the underlying QWA of $\AA$.
	Let $X$ be the finite set of weights of $\T$.
	Since $f \in \{\Inf$, $\Sup$, $\LimInf$, $\LimSup\}$, every run value belongs to $X$; hence every word value $\A(w)$ and language value $\AA(S)$ also belong to $X$.
	Therefore, exact evaluation reduces to deciding, for each $x \in X$, whether $\AA(S) \geq x$.

	We first assume $g = \Sup$.
	For every $x \in X$, a B\"uchi automaton recognizing $L(\A_{\geq x}) = \{ w \in \Sigma^\omega \st \A(w) \geq x \}$ can be constructed in linear time~\cite{DBLP:journals/tocl/ChatterjeeDH10}.

	If $h = \Sup$, then for every $x \in X$ we have $\AA(S) \geq x$ iff $\A$ is ${\geq}$-nonempty for $x$ iff $L(\B) \cap L(\A_{\geq x}) \neq \emptyset$.
	Hence, $\AA(S)$ is the largest $x \in X$ such that $L(\B) \cap L(\A_{\geq x}) \neq \emptyset$.
	Since emptiness of B\"uchi automata is decidable in polynomial time, evaluation is in \PTime.

	If $h = \Inf$, then for every $x \in X$ we have $\AA(S) \geq x$ iff $\A$ is ${\geq}$-universal for $x$ iff $L(\B) \subseteq L(\A_{\geq x})$.
	Hence, $\AA(S)$ is the largest $x \in X$ such that $L(\B) \subseteq L(\A_{\geq x})$.
	Since inclusion of $\omega$-regular languages is decidable in \PSpace, evaluation is in \PSpace.
	The cases with $g = \Inf$ follow by duality: by \Cref{all:dual}, $\AA(S) = -\hat{\AA}(S)$, where $\hat{\AA}$ has $\hat{g} = \Sup$.
	Thus $(g,h) = (\Inf,\Inf)$ reduces to $(\Sup,\Sup)$, and $(g,h) = (\Inf,\Sup)$ reduces to $(\Sup,\Inf)$.

	We now prove \PSpace-hardness for the $g \neq h$ cases.
	Suppose $g = \Sup$ and $h = \Inf$.
	We reduce the ${\geq}$-universality problem for nondeterministic QWAs $\A = (\Sup,f,\T)$ with $f \in \{\Inf$, $\Sup$, $\LimInf$, $\LimSup\}$ to evaluation of QLAs by letting $\AA = (\Inf,\A)$ and $S = \Sigma^\omega$.
	Indeed, $\AA(\Sigma^\omega) = \inf_{w \in \Sigma^\omega} \A(w)$, so $\AA(\Sigma^\omega) \geq k$ iff $\A$ is ${\geq}$-universal for $k$.
	Since this universality problem is \PSpaceH~\cite{DBLP:journals/tocl/ChatterjeeDH10,DBLP:conf/vmcai/KupfermanL07}, so is evaluation.
	The case $g = \Inf$ and $h = \Sup$ follows by duality.
\end{proof}

\paragraph{QLAs with Limit-Average QWAs.}
QLAs with run aggregators $f \in \{\LimInfAvg, \LimSupAvg\}$ differ from the previous case in the sense that it is not sufficient to consider only the lasso words in a given $\omega$-regular language, even when the underlying word automaton is deterministic.
To witness this, consider the best-case average uptime QLA
$\AA = (\Sup, (\Sup, \LimInfAvg, \T_{\text{up}}))$ as in \Cref{sec:applications,fig:examples}.
Having $L = (\Sigma^* \texttt{off})^\omega$, we get $\AA(L) = 1$ as there is a word in $L$ with infinitely many $\texttt{off}$'s but longer and longer blocks of $\texttt{on}$'s, but no lasso word in $L$ has an average uptime of 1.

\begin{prop}\label{ultim:avg}
	There is a QLA $\AA$ with a run aggregator $f\in\{\LimInfAvg, \LimSupAvg\}$ and an $\omega$-regular language $S$ such that no lasso word in $S$ achieves the value $\AA(S)$.
\end{prop}
\begin{proof}
	Consider a language automaton $\AA = (\Sup,\A)$ where $\A = (g,f,\T)$ and $\T$ consists of a single state that loops over $a$ and $b$ with respectively weight 1 and 0.
	Since $\T$ is deterministic, the choice of $g \in \{\Inf,\Sup,\E\}$ is not relevant.
	Let $f \in \{\LimInfAvg, \LimSupAvg\}$ and $L = (\Sigma^* b)^\omega$.
	Observe that a lasso word $w$ fulfills $\A(w) = 1$ iff it has finitely many letters $b$.
	We define $w' = b a b a^2 b a^4 b a^8\dots$.
	Since $w' \in L$ and $\A(w') = 1$, we have $\AA(L) = \Sup_{w \in L} \A(w) = 1$.
	Hence, although the value $\AA(L)$ is achieved by some word, it cannot be achieved by an ultimately periodic word in $L$.
\end{proof}

Nonetheless, for QLAs with matching word and language aggregators, we show that the value of an $\omega$-regular language given by a B\"uchi automaton $\B$ is computable by analyzing the strongly connected components (SCCs) of the underlying word automaton's product with $\B$ as follows:
Among all SCCs that are reachable from the initial state, we find the ones that contain at least one state whose B\"uchi component is accepting.
Then, in each such SCC, we compute the maximum mean weight of its cycles by Karp's algorithm \cite{DBLP:journals/dm/Karp78}.
The largest among these mean values is exactly the value of the given language.
Even though such a cycle may not involve an accepting state of $\B$, we can construct a run of the product that visits an accepting state infinitely often while going over this cycle with increasing frequency (hence the long-run average converges to the cycle's mean).
When these aggregators differ, the problem is undecidable by reduction from the universality of limit-average QWAs~\cite{DBLP:conf/csl/DegorreDGRT10,DBLP:conf/concur/ChatterjeeDEHR10,DBLP:journals/tcs/HunterPPR18}.

\begin{thm}\label{evaluation:avg}
	Consider a QLA $\AA = (h,(g,f,\T))$ with $f\in\{\LimInfAvg$, $\LimSupAvg\}$ and $g,h \in \{\Inf$, $\Sup\}$.
	The evaluation of $\AA$ is in \PTime when $g = h$ and undecidable when $g \neq h$.
\end{thm}
\begin{proof}
	Let $S \subseteq \Sigma^\omega$ be an $\omega$-regular language given by a B\"uchi automaton~$\B$.
	Let $\A$ be the underlying QWA of $\AA$.
	First, suppose $h = g = \Sup$.
	Let $\B$ be the input B\"uchi automaton.
	Construct the product automaton $\A \times \B$.
	Among all SCCs of $\A \times \B$ that are reachable from the initial state, find the ones that contain at least one state whose $\B$-component is accepting in $\B$.
	In each such SCC, compute the maximum mean of its cycles by Karp's algorithm \cite{DBLP:journals/dm/Karp78}.
	The largest among these mean values is exactly $\AA(S)$.
	Even though such a cycle $C$ may not involve an accepting state of $\B$, we can construct a run of the product that visits an accepting state infinitely often while going over $C$ with increasing frequency.
	Overall, $\AA(S)$ is computable in \PTime.
	The case of $h = g =\Inf$ is dual thanks to \cref{all:dual}.
	Now, suppose $h = \Inf$ and $g = \Sup$.
	Let $S = \Sigma^\omega$ and notice that $\AA(\Sigma^\omega) = \bot_{\A}$.
	An algorithm to compute $\bot_{\A}$ would imply the decidability of the universality problem of nondeterministic limit-average automata since $\A(w) \geq k$ for every word $w$ iff $\bot_{\A} \geq k$.
	Since this universality problem is undecidable \cite{DBLP:conf/csl/DegorreDGRT10,DBLP:conf/concur/ChatterjeeDEHR10}, the value $\AA(S)$ is not computable.
	The case of $h = \Sup$ and $g = \Inf$ is dual thanks to \cref{all:dual}.
\end{proof}

\paragraph{QLAs with Discounted-Sum QWAs.}
QLAs with the run aggregator $f=\DSum$ have the particular behavior that the value assigned to an $\omega$-regular language $L$ may be not achievable by any word in $L$, even when the underlying word automaton is deterministic.
To witness this, consider $\AA = (\Sup, \A)$ with $\A = (\Sup, \DSum_{0.5}, \T_{\text{up}})$ as in \Cref{fig:examples}.
We have $\AA((\Sigma^* \texttt{off})^\omega) = 2$ since $\lim_{n \to \infty} \A(\texttt{on}^n \texttt{off}^\omega) = 2$.
However, only for $w = \texttt{on}^\omega \notin (\Sigma^* \texttt{off})^\omega$ we have $\A(w) = 2$.

\begin{prop}\label{ultim:dsum}
	There is a QLA $\AA$ with the run aggregator $f=\DSum$ and an $\omega$-regular language $S$ such that no word in $S$ achieves the value $\AA(S)$.
\end{prop}
\begin{proof}
	Consider $L = (\Sigma^* b)^\omega$ and the automata $\AA$ and $\A$ from \Cref{ultim:avg} with $f = \DSum_{0.5}$.
	Observe that, $a^\omega$ is the only word achieving the value $2$, i.e., $\A(w) < 2$ for all $w \neq a^\omega$.
	For each $i \geq 1$ we define $w_i = a^i b^\omega \in L$.
	Since $\lim_{i \to \infty} \A(w_i) = 2$ and $\top_{\AA} = 2$, we have $\AA(L) = 2$.
	Therefore, the value $\AA(L)$ is not achievable by any word in $L$.
\end{proof}

We establish that such a behavior is not possible when the input language includes all its limit points, i.e., it is safe in the boolean sense~\cite{DBLP:journals/tse/Lamport77,DBLP:journals/ipl/AlpernS85}:
Consider a sequence of words in the safety language whose values approach the supremum.
We build by a diagonalization argument an infinite word $w$ whose every finite prefix already appears in the language, so $w$ is in the safety language.
Applying the same construction to the corresponding optimal runs yields an infinite run on $w$.
This run's value equals the supremum since the contribution of the remaining tail is bounded by a vanishing geometric series due to discounting.

\begin{prop}\label{dsum:safe:in}
	Consider a QLA $\AA = (\Sup, (\Sup, \DSum, \T))$.
	For every nonempty safety language $S \subseteq \Sigma^\omega$, the value $\AA(S)$ is achievable by some run of a word in $S$.
\end{prop}
\begin{proof}
	Let $\AA$ be as in the statement and $\A$ be its underlying QWA.
	Since $\AA(S) = \sup_{w \in S} \A(w)$, there exists a sequence of words $\{w_i\}_{i\in\NN}$ all coming from $S$ such that $\lim_{i \to \infty} \A(w_i) = \AA(S)$.
	As in the proof of \cite[Thm. 9]{DBLP:conf/fsttcs/BokerL21}, for each $i \in \NN$, we can choose a run $\pi_i$ on $w_i$ satisfying $\DSum(\gamma(\pi_i)) = \A(w_i)$, thus $\lim_{i \to \infty} \DSum(\gamma(\pi_i)) = \AA(S)$.
	Since the alphabet $\Sigma$ is finite, for each $j \in \NN$ there exists at least one finite word $u_j \in \Sigma^j$ (of length $j$) that occurs as a prefix of infinitely many $w_i$'s.
	By a diagonalization argument, we extract an infinite subsequence $(w'_j)_{j \in \NN}$ (of $(w_i)_{i \in \NN}$) such that $u_j \prec w'_j$ for each $j$, and $\lim_{j \to \infty} \A(w'_j) = \AA(S)$.

	Similarly, consider the corresponding runs $\rho'_j$ on $w'_j$ (with $\DSum(\gamma(\rho'_j)) = \A(w'_j)$).
	For each $j \in \NN$ there exists a finite run fragment $\theta_j$ of length $j$ that is a prefix of infinitely many $\rho'_j$'s.
	Thus, we can extract an infinite subsequence (which we continue to index by $j$) such that $\theta_j \prec \rho'_j$ and $\theta_j \prec \theta_k$ for all $j < k$.
	Denote by $\rho_j$ the run $\rho'_j$ from the extracted subsequence.
	Let $w = \lim_{j \to \infty} u_j$.
	Since each $u_j$ is a prefix of some word in $S$ and $S$ is a safety language, we have $w \in S$.
	Next, define the infinite run $\rho$ on $w$ by letting the $j$th transition of $\rho$ equal the $j$th transition of $\rho_j$.
	The run $\rho$ is well defined because $\theta_j \prec \theta_k$ for every $j,k \in \NN$ with $j < k$.

	For each $j \in \NN$, note that the difference between the values $\DSum(\gamma(\rho))$ and $\DSum(\gamma(\rho_j))$ comes from the tail of the run: $\left|\DSum(\gamma(\rho)) - \DSum(\gamma(\rho_j))\right| \leq \sum_{i=j}^\infty \lambda^i \ell = \frac{\ell \lambda^j}{1-\lambda}$ where $\ell$ is the maximal transition weight in $\A$ and $\lambda$ the discount factor.
	Since $\lim_{j\to\infty} \frac{\ell \lambda^j}{1-\lambda} = 0$, we have $\lim_{j \to \infty} \DSum(\gamma(\rho_j)) = \DSum(\gamma(\rho))$.
	Moreover, since $\lim_{j \to \infty} \DSum(\gamma(\rho_j)) = \lim_{j \to \infty} \DSum(\gamma(\rho'_j)) = \AA(S)$, we conclude $\DSum(\gamma(\rho)) = \A(w) = \AA(S)$.
\end{proof}

The value of a language $S$ matches that of its safety closure $S'$ (i.e., the smallest safety language containing it) because every word in the safety closure can be approximated arbitrarily closely by words from the original language:
If $S'$ achieves a value on a word $w$, we can isolate a prefix of $w$ whose contribution is close enough to the value of $w$.
By construction, the same prefix also occurs in a word of $S$, and completing the run along this word in $S$ can change the total value by at most an arbitrarily small amount due to discounting.
Hence the maximal value in $S'$ can be approximated arbitrarily closely by words in $S$, and the two suprema coincide.

\begin{prop} \label{dsum:safe:equal}
	Consider a QLA $\AA = (\Sup, (\Sup, \DSum, \T))$.
	For every language $S \subseteq \Sigma^\omega$ we have $\AA(S) = \AA(S')$ where $S'$ is the safety closure of $S$.
\end{prop}
\begin{proof}
	Let $\AA$ be as in the statement and $\A$ be its underlying QWA.
	Since $S \subseteq S'$ by definition, we immediately have $\AA(S) \leq \AA(S')$.
	We prove below $\AA(S) \geq \AA(S')$.

	Suppose towards contradiction that $\AA(S) < \AA(S')$ and let $\varepsilon = \AA(S') - \AA(S) > 0$.
	Since $S'$ is safe, thanks to \cref{dsum:safe:in} there exists a word $w \in S'$ and a run $\rho$ of $\A$ over $w$ such that $\A(w) = \DSum(\gamma(\rho)) = \AA(S')$.
	Moreover, since $\A$ is discounting, there exists $n \in \NN$ such that $\sum_{i = n}^{\infty} \ell \lambda^i < \frac{\varepsilon}{2}$, where $\ell$ is the maximal transition weight in $\A$ and $\lambda$ the discount factor.

	Let $u$ be the prefix of $w$ of length $n$ and $\theta$ the corresponding finite prefix of $\rho$.
	By definition of $S'$, there exists $w' \in S$ with $u \prec w'$.
	Let $\rho'$ be a run on $w'$ with the finite prefix $\theta$.
	By the discounting property mentioned above, we have $\left|\DSum(\gamma(\rho)) - \DSum(\gamma(\rho'))\right| \leq \sum_{i=n}^{\infty} \ell \lambda^i < \frac{\varepsilon}{2}$.
	Therefore, we obtain $\DSum(\gamma(\rho')) \geq \DSum(\gamma(\rho)) - \frac{\varepsilon}{2} = \AA(S') - \frac{\varepsilon}{2}$.
	Since $\rho'$ is a run of a word in $S$, we also have $\AA(S) \geq \DSum(\gamma(\rho'))$.
	It implies that $\AA(S) \geq \AA(S') - \frac{\varepsilon}{2}$, which contradicts the initial assumption that $\AA(S') - \AA(S) = \varepsilon > 0$.
\end{proof}

The observation above helps us provide a \PTime algorithm when the word and language aggregators match:
We first construct the Büchi automaton's safety closure, so the optimal value is achieved by a run that never reaches the rejecting sink.
Then, we compute the product of the underlying word automaton and the safety closure automaton.
Computing the best (or worst) discounted sum over all sink-avoiding paths in the product can be done by solving a one-player discounted-payoff game~\cite{esslli2006-students-And}.
When the two aggregators differ, the evaluation problem is at least as hard as the universality problem for nondeterministic discounted-sum automata, which is a long-standing open problem~\cite{DBLP:journals/tocl/ChatterjeeDH10,DBLP:conf/lics/BokerHO15}.

\begin{thm}\label{evaluation:dsum}
	Consider a QLA $\AA = (h,(g,\DSum,\T))$ with $g,h \in \{\Inf$, $\Sup\}$.
	The evaluation of $\AA$ is in \PTime when $g = h$ and at least as hard as the ${\geq}$-universality of nondeterministic discounted-sum word automata when $g \neq h$.
\end{thm}
\begin{proof}
	Let $S \subseteq \Sigma^\omega$ be an $\omega$-regular language given by a B\"uchi automaton~$\B$.
	We start with the case with $h=g=\Sup$.
	Let $\B$ be the B\"uchi automaton recognizing $S$.
	The B\"uchi automaton $\B'$ recognizing the safety closure of $S$ is computable in \PTime thanks to~\cite{DBLP:journals/dc/AlpernS87}.
	By safety condition, we can assume without loss of generality that $\B'$ is complete and has only one rejecting state which is a sink, called $p$ hereafter.
	We construct the discounted-sum word automaton $\C$ as the cross-product of $\B'$ and $\A$ where the transition weights are taken from $\A$.
	By \Cref{dsum:safe:equal}, we have $\AA(L(\B)) = \AA(L(\B'))$, and by \Cref{dsum:safe:in}, the value $\AA(L(\B'))$  is achievable by some run of a word of $L(\B')$.
	This implies that the value $\AA(L(\B))$ is achieved by a run that never visits a pair with $p$.
	Hence, we can compute this value in polynomial time thanks to the algorithm solving the one-player discounted payoff objectives game of~\cite{esslli2006-students-And} on the arena defined by $\C$ where all pairs containing $p$ are removed.
	The case of $g = h = \Inf$ is dual thanks to~\Cref{all:dual}.

	Suppose $h = \Inf$ and $g = \Sup$.
	Considering $S = \Sigma^\omega$, we have $\AA(S) = \bot_{\A}$.
	Notice that $\A$ is ${\geq}$-universal for $k$ iff $\bot_{\A} \geq k$.
	Therefore, the evaluation problem for this class of QLAs is at least as hard as the universality problem of nondeterministic discounted-sum word automata.
	The case of $h = \Sup$ and $g = \Inf$ is dual.
\end{proof}

\paragraph{QLAs with Probabilistic QWAs.}
When the underlying word automaton is probabilistic, i.e., has the word aggregator $g = \E$, the nonprobabilistic evaluation problem has no algorithmic solution due to inapproximability of their top values~\cite{DBLP:journals/ai/MadaniHC03}. We leave the case $f=\DSum$ open.

\begin{thm}\label{nondet:stochastic:evaluation}
	Consider a QLA $\AA = (h,(\E,f,\T))$ with $f \in \{\Inf,$ $\Sup$, $\LimInf$, $\LimSup$, $\LimInfAvg$, $\LimSupAvg \}$ and $h \in \{\Inf$, $\Sup\}$.
	The evaluation of $\AA$ is undecidable.
\end{thm}
\begin{proof}
	For $h = \Sup$, we sketch a reduction from the approximate-nonemptiness problem of QWA with $f$ and $g$ as in the statement, which is undecidable as proved in~\cref{stochastic:undecidability:rest}.
	Evaluating $\AA(\Sigma^{\omega})$ is equivalent to evaluating $\sup_{w \in \Sigma^{\omega}}\A(w)$, which is equal to $\top_{\A}$.
	Recall that deciding approximate-nonemptiness for $k$ is equivalent to deciding whether $\top_{\A} \geq k$.
	For $h = \Inf$, the proof goes by a reduction to the approximate-universality problem, which is undecidable (see ~\cref{stochastic:undecidability:rest}).
	Recall that deciding approximate-universality for $k$ is equivalent to deciding whether $\bot_{\A} > k$.
\end{proof}

\subsection{Probabilistic Evaluation}\label{sec:probeval}

Now, we consider the evaluation problem for QLAs with $h = \E$ and follow the same structure as in \cref{sec:nonprobeval}: we start with the cases of $g \in \{\Inf,\Sup\}$ and study various run aggregators $f$ separately, and then look at the case of $g = \E$.

\paragraph{QLAs with Standard QWAs.}
For QLAs with standard run aggregators $f \in \{\Inf$, $\Sup$, $\LimInf$, $\LimSup\}$, probabilistic evaluation is already \PSpaceC.
Our proof avoids the exponential-time determinization route of \cite[Thm.~8]{DBLP:journals/jcss/MichaliszynO20}.
The key observation is that since every word value belongs to the finite set of transition weights of the underlying weighted transition system, we can express the expected value as a telescoping linear combination of probabilities of threshold languages of the form $\{w \in \Sigma^\omega \mid \A(w) \geq x\}$ or $\{w \in \Sigma^\omega \mid \A(w) < x\}$.
Each such threshold language is recognized by nondeterministic B\"uchi automaton whose size is linear in the size of $\A$, and the required acceptance probabilities against a finite-state Markov chain can be computed in polynomial space by \cite[Cor.~43]{DBLP:conf/dcfs/KieferW21}.
The matching lower bound follows from a reduction from universality of nondeterministic finite automata (NFAs) over finite words.

\begin{thm}\label{stochastic:classic:evaluation}
	Consider a QLA $\AA = (\E,(g,f,\T))$ with $g \in \{\Inf,\Sup\}$ and	$f \in \{\Inf$, $\Sup$, $\LimInf$, $\LimSup\}$.
	The evaluation of $\AA$ is \PSpaceC.
\end{thm}
\begin{proof}
	Let $\A = (g,f,\T)$ be the underlying QWA of $\AA$, and let $X = \{x_1, \dots, x_n\}$ be the set of distinct transition weights of $\T$, ordered so that $x_1 < \dots < x_n$.
	Since weighted labeled transition systems are complete, every word admits at least one run.
	As $f \in \{\Inf, \Sup, \LimInf, \LimSup\}$, every run value belongs to $X$.
	Hence, because $g \in \{\Inf, \Sup\}$, every word value $\A(w)$ also belongs to $X$.

	We first show the \PSpace membership.
	If $n = 1$, then $\A(w) = x_1$ for every word $w$, so $\AA(\M) = x_1$ and the threshold test is trivial.
	Assume henceforth that $n \geq 2$.
	Define $D = x_n - x_1 > 0$ and $\alpha_i = (x_i - x_{i-1}) / D$ for $2 \leq i \leq n$.
	Then, each $\alpha_i$ is a nonnegative rational and $\sum_{i=2}^n \alpha_i = 1$.

	For $x \in X$, let $T_{\geq x}$ and $T_{< x}$ be the sets of transitions of $\T$ with weight at least $x$ and below $x$, respectively.
	For each $i \in \{2, \dots, n\}$, define
	$L_i = \{w \in \Sigma^\omega \st \A(w) \geq x_i\}$ if $g = \Sup$, and
	$L_i = \{w \in \Sigma^\omega \st \A(w) < x_i\}$ if $g = \Inf$.

	Each language $L_i$ is recognized by an NBA of size $O(|\T|)$.
	We spell this out only for $g = \Sup$; the case $g = \Inf$ is obtained symmetrically by replacing $T_{\geq x_i}$ with $T_{< x_i}$ and dualizing the condition.
	If $g = \Sup$, then $w \in L_i$ iff $\T$ has a run over $w$ whose transition sequence, relative to $T_{\geq x_i}$, is always in $T_{\geq x_i}$ when $f = \Inf$, hits $T_{\geq x_i}$ at least once when $f = \Sup$, is eventually always in $T_{\geq x_i}$ when $f = \LimInf$, and hits $T_{\geq x_i}$ infinitely often when $f = \LimSup$.
	These are the standard safety, reachability, co-B\"uchi, and B\"uchi conditions, respectively, and are all recognized by linear-size NBAs.
	Let $\B_i$ be such an NBA for $L_i$.

	Introduce fresh letters $\#_2, \dots, \#_n \notin \Sigma$, and let $\Sigma' = \Sigma \cup \{\#_2, \dots, \#_n\}$.
	Construct an NBA $\B$ over $\Sigma'$ such that $L(\B) = \bigcup_{i=2}^n \#_i L(\B_i) = \bigcup_{i=2}^n \{\#_i w \mid w \in L_i\}$.
	This is done by a fresh initial state that reads one marker $\#_i$ and then enters a copy of $\B_i$.
	Since $n \leq |\T|$, the automaton $\B$ has polynomial size.

	Next, construct a finite-state Markov chain $\M'$ over $\Sigma'$ that, in its first step, chooses the letter $\#_i$ with probability $\alpha_i$ and thereafter simulates $\M$.
	Then $P_{u \sim \mu_{\M'}}[u \in L(\B)] = \sum_{i=2}^n \alpha_i \cdot P_{w \sim \mu_{\M}}[w \in L_i]$.

	Given a language $L$ and a word $w$, we let $\chi(L, w) = 1$ if $w \in L$, and $\chi(L, w) = 0$ otherwise.
	If $g = \Sup$, then for every word $w$ we have $\A(w) = x_1 + \sum_{i=2}^n (x_i - x_{i-1}) \cdot \chi(L_i, w)$, because both sides evaluate to $x_j$ whenever $\A(w) = x_j$.
	Taking expectation with respect to $\mu_\M$ and using the definition of the $\alpha_i$ yields $\AA(\M) = x_1 + D \cdot P_{u \sim \mu_{\M'}}[u \in L(\B)]$.
	Hence $\AA(\M) \geq k$ if and only if $P_{u \sim \mu_{\M'}}[u \in L(\B)] \geq (k - x_1) / D$.
	If $g = \Inf$, then for every word $w$ we have $\A(w) = x_n - \sum_{i=2}^n (x_i - x_{i-1}) \cdot \chi(L_i, w)$, again because both sides equal $x_j$ whenever $\A(w) = x_j$.
	Therefore $\AA(\M) = x_n - D \cdot P_{u \sim \mu_{\M'}}[u \in L(\B)]$, and so $\AA(\M) \geq k$ if and only if $P_{u \sim \mu_{\M'}}[u \in L(\B)] \leq (x_n - k) / D$.

	By~\cite[Cor.~43]{DBLP:conf/dcfs/KieferW21}, for every NBA $\B$ and every finite-state Markov chain $\M'$, the acceptance probability $P_{u \sim \mu_{\M'}}[u \in L(\B)]$ can be computed in \PSpace.
	The above reduction is polynomial-time, and the final threshold comparison is over rationals of polynomial size.
	Thus, the problem is in \PSpace.

	Now, we show \PSpace-hardness.
	We reduce from universality of NFAs over finite words, which is \PSpace-hard.
	Let $\C = (Q,\Sigma,\delta,q_I,F)$ be such an NFA.
	By adding a rejecting sink if necessary, we may assume that $\C$ is complete.

	Introduce a fresh letter $\#$ and let $\Sigma' = \Sigma \cup \{\#\}$.
	Let $\M_\#$ be the two-state Markov chain over $\Sigma'$ that, from its initial state, emits each letter $a \in \Sigma$ with probability $1/(2|\Sigma|)$ and stays in the same state, emits $\#$ with probability $1/2$ and moves to an absorbing state, and from there emits only $\#$ forever.
	Then, every word of the form $u\#^\omega$ with $u \in \Sigma^*$ has positive probability, and every other word has probability $0$.

	We first treat $g = \Sup$ and $f = \Sup$.
	Construct a weighted labeled transition system $\T^{\Sup}_\C$ with state set $Q \cup \{t_0,t_1\}$ as follows.
	For every transition of $\C$ on a letter from $\Sigma$, keep the same transition and assign it weight $0$.
	For each $q \in F$, add a transition $q \xrightarrow{\#} t_1$ of weight $1$.
	For each $q \in Q \setminus F$, add a transition $q \xrightarrow{\#} t_0$ of weight $0$.
	Finally, for each $b \in \{0,1\}$, let $t_b$ loop on every letter of $\Sigma'$ with weight $b$.
	Let $\A^{\Sup}_\C = (\Sup,\Sup,\T^{\Sup}_\C)$ and $\AA^{\Sup}_\C = (\E,\A^{\Sup}_\C)$.
	For every $u \in \Sigma^*$, we have $\A^{\Sup}_\C(u\#^\omega) = 1$ iff $u \in L(\C)$:
	if $u$ is accepted, some run reaches a state in $F$ after reading $u$ and then has weight sequence $0^{|u|}1^\omega$;
	if $u$ is rejected, every run has weight sequence $0^\omega$.
	Hence, $\AA^{\Sup}_\C(\M_\#) = 1$ iff $L(\C) = \Sigma^*$.
	So, the evaluation problem is \PSpace-hard for $(g,f) = (\Sup,\Sup)$.
	The same construction also yields hardness for $(g,f) = (\Sup,\LimSup)$, since every run is eventually trapped in either $t_0$ or $t_1$, and therefore $\Sup$ and $\LimSup$ coincide on its weight sequence.

	Next, let us treat $g = \Sup$ and $f = \Inf$.
	Construct $\T^{\Inf}_\C$ exactly as above, except that every transition on a letter from $\Sigma$ now has weight $1$.
	Let $\A^{\Inf}_\C = (\Sup,\Inf,\T^{\Inf}_\C)$ and $\AA^{\Inf}_\C = (\E,\A^{\Inf}_\C)$.
	Then, for every $u \in \Sigma^*$, we again have $\A^{\Inf}_\C(u\#^\omega) = 1$ iff $u \in L(\C)$:
	accepted words admit a run of weight sequence $1^\omega$, whereas rejected words admit only runs of the form $1^{|u|}0^\omega$.
	Thus, $\AA^{\Inf}_\C(\M_\#) = 1$ iff $L(\C) = \Sigma^*$, so the evaluation problem is \PSpace-hard for $(g,f) = (\Sup,\Inf)$.
	Again, the same construction also yields hardness for $(g,f) = (\Sup,\LimInf)$, because every run is eventually constant.

	Finally, the cases $g = \Inf$ are obtained by the dual constructions; thus, the first construction also yields hardness for $(g,f) = (\Inf,\Inf)$ and $(g,f) = (\Inf,\LimInf)$, and the second for $(g,f) = (\Inf,\Sup)$ and $(g,f) = (\Inf,\LimSup)$.
\end{proof}

\paragraph{QLAs with Limit-Average QWAs.}
Undecidability of probabilistic evaluation for limit-average QLAs was shown in~\cite[Thm.~7]{DBLP:journals/jcss/MichaliszynO20} by a reduction from the universality problem of quantitative automata on finite words with the summation run aggregator and weights in $\{-1,0,1\}$, a.k.a. weighted automata over the tropical semiring of integers~\cite{DBLP:journals/ijac/Krob94,DBLP:journals/iandc/AlmagorBK22}.

\begin{thmC}[\cite{DBLP:journals/jcss/MichaliszynO20}]\label{stochastic:avg:evaluation}
	Consider a QLA $\AA = (\E,(g,f,\T))$ with $f \in \{\LimInfAvg,$ $\LimSupAvg\}$ and $g \in \{\Inf$, $\Sup\}$.
	The evaluation of $\AA$ is undecidable.
\end{thmC}

\paragraph{QLAs with Discounted-Sum QWAs.}
Next, we show the hardness of probabilistic evaluation for $\DSum$ QLAs.
As in the nonprobabilistic case, we provide a reduction from the universality problem of nondeterministic discounted-sum QWAs.

\begin{thm}\label{stochastic:nondet:dsum}
	Consider a QLA $\AA = (\E,(g,\DSum,\T))$ with $g \in \{\Inf$, $\Sup\}$.
	The evaluation of $\AA$ is at least as hard as the ${\geq}$-universality of nondeterministic discounted-sum word automata.
\end{thm}
\begin{proof}
	Given a $\DSum$ word automaton $\A$ and a threshold $k$, we can decide whether $\A(w) \geq k$ holds for every $w\in\Sigma^\omega$ by reducing the evaluation of some language automaton $\AA' = (\E, \A')$ where $\A$ has $g = \Sup$, and $f = \DSum$.
	We construct a word automaton $\A'$ such that  $\A'(w) = \max(k, \A(w))$, for all $w\in\Sigma^\omega$.
	This is possible by introducing a new run from the initial state, for which the first transition is weighted by $k$ and all other transition is weighted by $0$.
	The construction thus consists in adding a new state with self-loop weighed $0$ that is reachable only from the initial state with weight $k$.
	Let $\AA = (\E, \A)$ and $\AA' = (\E, \A')$ be two language automata, and $\M$ be the uniform Markov chain (i.e., single state with a uniform self loop).
	We show that $\AA(\M) = \AA'(\M)$ if and only if $\A(w) \geq k$ for every $w\in\Sigma^\omega$.
	If $\A(w) \geq k$ holds for every $w\in\Sigma^\omega$, then both values are trivially equal.
	Otherwise, there exists $w\in\Sigma^\omega$ such that $\A(w) < k$, which by discounting implies that there exists a finite prefix $u \prec w$ such that $\A(uw') < k$ for all $w' \in \Sigma^{\omega}$.
	Since $\M$ is the uniform Markov chain, the measure of the set $u\Sigma^\omega$ is nonzero, and thus $\AA(\M) \neq \AA'(\M)$.
\end{proof}

\paragraph{QLAs with Probabilistic QWAs.}
Finally, consider QLAs with $h = g = \E$.
The randomness of the input Markov chain and the randomness of the QWA can be merged into one finite product Markov chain.
Its states record the current generator state, the current automaton state, and the last weight produced by the automaton.
Reading that stored weight as the reward gives exactly the same random weight sequence as in the semantics of the QLA.
So evaluation reduces to computing the expected payoff of a finite weighted Markov chain.
For the objectives considered here, this is polynomial-time.

\begin{thm}\label{stochastic:stochastic:evaluation}
	Consider a QLA $\AA = (\E, (\E, f, \T))$ with $f \in \{\Inf$, $\Sup$, $\LimInf$, $\LimSup$, $\LimInfAvg$, $\LimSupAvg$, $\DSum\}$.
	The evaluation of $\AA$ is in \PTime.
\end{thm}
\begin{proof}
	Let $\M$ be the input Markov chain, with state space $Q_\M$ and initial state $m_I$.
	Let $Q_\T$ be the state space of $\T$, and let $X$ be the finite set of transition weights of $\T$.
	Define a Markov chain $\widehat{\C}$ with state space $\widehat Q = Q_\M \times Q_\T \times X$.
	\begin{itemize}
		\item Its initial distribution is $\alpha(m',q',x) = \sum_{a \in \Sigma} P_\M(m_I,a,m') d(q_I,a,x,q')$,
		\item transition matrix is $\widehat P((m,q,x),(m',q',x')) = \sum_{a \in \Sigma} P_\M(m,a,m') d(q,a,x',q')$,
		\item and reward function is $r(m,q,x) = x$.
	\end{itemize}

	If we sample a path $m_0 \xrightarrow{a_0} m_1 \xrightarrow{a_1} \cdots$ of $\M$ and, along the generated word, a run $q_I \xrightarrow{a_0:x_0} q_1 \xrightarrow{a_1:x_1} \cdots$ of $\T$, then $X_n = (m_{n+1}, q_{n+1}, x_n)$ is exactly a run of $\widehat{\C}$ and $r(X_n) = x_n$ for all $n \geq 0$.
	Hence, $\AA(\M)$ is the expected value of $f$ on the reward sequence of $\widehat{\C}$.
	It remains to evaluate a finite weighted Markov chain.

	Let $V = \{\nu_1 < \cdots < \nu_m\}$ be the set of rewards of $\widehat{\C}$.
	For $f = \Inf$, let $T_i = \{s \in \widehat Q \mid r(s) \leq \nu_i\}$ and $p_i = \Pr_\alpha[\Diamond T_i]$.
	Then $\Pr_\alpha[\Inf = \nu_i] = p_i - p_{i-1}$, with $p_0 = 0$, so $\E_\alpha[\Inf] = \sum_{i = 1}^m \nu_i (p_i - p_{i-1})$.
	Each $p_i$ is a reachability probability, so it is computable in polynomial time.
	The case $f = \Sup$ is symmetric, using $T_i = \{s \in \widehat Q \mid r(s) \geq \nu_i\}$.

	For $f \in \{\LimInf, \LimSup\}$, the chain almost surely reaches a bottom strongly connected component and then stays there.
	Inside such a component every state is visited infinitely often almost surely.
	Hence, on a component $B$, the value of $\LimInf$ is $\min_{s \in B} r(s)$ and the value of $\LimSup$ is $\max_{s \in B} r(s)$ almost surely.
	So, $\E_\alpha[f]$ is the sum of these component values weighted by the probabilities of reaching the corresponding components.
	This is again polynomial-time.

	For $f \in \{\LimInfAvg, \LimSupAvg\}$, the same decomposition applies.
	If $B$ is a bottom strongly connected component and $\pi_B$ is its stationary distribution, then the averages $\frac{1}{n} \sum_{k = 0}^{n-1} r(X_k)$ converge almost surely to $\sum_{s \in B} \pi_B(s) r(s)$ on the event of eventually staying in $B$.
	So, $\LimInfAvg$ and $\LimSupAvg$ coincide almost surely on that event, and their expectation is obtained by weighting this mean payoff of each component by its reachability probability.
	The distributions $\pi_B$ are computed by linear equations, so this case is also polynomial-time.

	Finally, if $f = \DSum_\lambda$, let $u(s)$ be the expected discounted payoff from state $s$.
	Then $u = r + \lambda \widehat P u$, or equivalently $(I - \lambda \widehat P)u = r$, where $I$ is the identity matrix.
	Since $\lambda \in (0,1)$, the matrix $I - \lambda \widehat P$ is invertible, so $u$ is computable in polynomial time.
	The required value is $\AA(\M) = \sum_{s \in \widehat Q} \alpha(s) u(s)$.
\end{proof}

\section{Deciding Nonemptiness and Universality}\label{sec:emptiness}

In this section, we study the nonemptiness and universality problems for QLAs with language and word aggregators $h,g \in \{\Inf, \Sup, \E\}$.
This complements \cref{sec:evaluation} for the same fragment, while the cases in which $g$ or $h$ is a limit aggregator (i.e., in $\{\LimInf, \LimSup\}$) are studied in \cref{sec:limit}.
We establish the main decidability and complexity results for this fragment and indicate the remaining open cases explicitly.
For many classes, the unrestricted and finite-state variants coincide; where they do not, we state separately what is known for the finite-state restriction.

Universality is handled by duality.
The precise correspondence is stated in \cref{rem:duality}.
Accordingly, in the remainder of the section, we focus on nonemptiness, and the corresponding universality results follow by duality.

\begin{rem}\label{rem:duality}
	Consider a QLA $\AA = (h, \A)$ and its dual $\hat{\AA}$.
	Let $k \in \QQ$.
	Thanks to~\cref{all:dual}, the QLA $\AA$ is $\geq$-nonempty (resp. $\geq$-universal) for $k$ iff $\hat{\AA}$ is not $>$-universal (resp. $>$-nonempty) for $-k$.
	The statement holds also for the finite-state restriction.
\end{rem}

We organize the remainder of the section by the word aggregator of the underlying QWA.
We begin with nondeterministic QWAs, that is, the case $g = \Sup$, for which nonemptiness is in \PTime.
We then turn to universal QWAs, that is, the case $g = \Inf$.
For $f \in \{\Inf, \Sup, \LimInf, \LimSup\}$ we obtain a \PSpace-complete characterization.
For $f \in \{\LimInfAvg, \LimSupAvg\}$, the strict version is undecidable and several non-strict cases remain open.
For $f = \DSum$, we obtain hardness via the universality problem for nondeterministic discounted-sum QWAs.
Finally, we consider probabilistic QWAs, that is, the case $g = \E$, and show that nonemptiness is undecidable there as well.

\paragraph{QLAs with Nondeterministic QWAs.}
We begin with QLAs whose underlying word automata are nondeterministic, that is, with $g = \Sup$.
In this case, nonemptiness reduces to the corresponding lasso-word nonemptiness problem for the underlying QWA via \cref{bigreduction}.
Since for the run aggregators considered here the top value of a nondeterministic QWA is achieved by a lasso word, this yields a polynomial-time algorithm.
The same argument also applies in the finite-state case.

\begin{thm}\label{ptime:nonemptiness}
	Consider a QLA $\AA = (h,(\Sup,f,\T))$ with $f \in \{\Inf,$ $\Sup$, $\LimInf$, $\LimSup$, $\LimInfAvg$, $\LimSupAvg$, $\DSum \}$ and $h \in \{\Inf$, $\Sup$, $\E\}$.
	Let ${\compare}\in\{{>}, {\geq}\}$.
	The ${\compare}$-nonemptiness of $\AA$ is in \PTime.
	The statement holds also for the finite-state restriction.
\end{thm}
\begin{proof}
	Let $\A$ be the underlying QWA of $\AA$.
	As $g = \Sup$, the value $\top_{\A}$ is achievable by some lasso word~\cite[Thm. 3]{DBLP:journals/tocl/ChatterjeeDH10}.
	Thus, using \cref{bigreduction}(b)(iii), we can reduce whether $\AA$ is (finite-state) ${\compare}$-nonempty for $k$ to whether $\A$ is (lasso-word) ${\compare}$-nonempty for $k$, which is decidable in \PTime.
\end{proof}

\paragraph{QLAs with Universal Standard QWAs.}
We now turn to QLAs whose underlying word automata are universal, that is, with $g = \Inf$.
For the standard run aggregators $f \in \{\Inf, \Sup, \LimInf, \LimSup\}$, the relevant extremal values of the underlying QWA are attained by lasso words.
Combined with \cref{bigreduction}, this reduces both unrestricted and finite-state nonemptiness to lasso-word nonemptiness of the underlying QWA.
This yields a \PSpace-complete classification.

\begin{thm}\label{emptiness:classic}\label{emptiness:classic:stochastic}
	Consider a QLA $\AA = (h,(\Inf,f,\T))$ with $f\in\{\Inf$, $\Sup$, $\LimInf$, $\LimSup\}$ and $h \in \{\Inf,\Sup,\E\}$.
	Let ${\compare}\in\{{>}, {\geq}\}$.
	The ${\compare}$-nonemptiness of $\AA$ is \PSpaceC.
	The statement holds also for the finite-state restriction.
\end{thm}
\begin{proof}
	Let $\A = (\Inf,f,\T)$ be the underlying QWA of $\AA$.

	For any QWA $\C = (g,f,\T')$ with $f\in\{\Inf$, $\Sup$, $\LimInf$, $\LimSup\}$, the values $\top_{\C}$ and $\bot_{\C}$ are both achievable by some lasso words.
	This is because every word value of $\C$ equals one of the weights in $\C$, and for each weight $x$ the language $\C_{=x}$ is $\omega$-regular.
	Thus, using \cref{bigreduction}(b)(iii), we obtain that both unrestricted and finite-state ${\compare}$-nonemptiness of $\AA$ reduce to lasso-word ${\compare}$-nonemptiness of $\A$.
	Since $\top_{\A}$ is achieved by a lasso word, lasso-word and unrestricted ${\compare}$-nonemptiness coincide for $\A$.
	Hence both problems are in \PSpace by~\cite[Thm.~3]{DBLP:conf/fct/ChatterjeeDH09}.

	For hardness, let $\hat{\compare} \in \{{>},{\geq}\} \setminus \{{\compare}\}$ be the dual comparison, and let $\hat{f}$ be the dual of $f$.
	We reduce from lasso-word ${\hat{\compare}}$-universality of nondeterministic QWAs $\B = (\Sup,\hat{f},\T_\B)$.

	This source problem is \PSpaceH: the ${\geq}$-case is standard~\cite{DBLP:journals/tocl/ChatterjeeDH10,DBLP:conf/vmcai/KupfermanL07}, and the ${>}$-case polynomial-time interreduces with the ${\geq}$-case because every word value belongs to the finite set of transition weights.
	Moreover, unrestricted and lasso-word universality coincide by the same $\omega$-regularity argument as above.

	Now, let $\A' = \hat{\B} = (\Inf,f,\hat{\T}_\B)$, so that $\A'(w) = -\B(w)$ for every word $w$.
	Then, $\B$ is lasso-word ${\hat{\compare}}$-universal for $k$ iff $\A'$ is not lasso-word ${\compare}$-nonempty for $-k$.
	Since, $\PSpace = \coPSpace$, lasso-word ${\compare}$-nonemptiness of universal QWAs of the form $(\Inf,f,\T)$ is \PSpaceH.
	Finally, for $h \in \{\Inf,\Sup,\E\}$, let $\AA' = (h,\A')$.
	Since $\top_{\A'}$ is achievable by a lasso word, \cref{bigreduction}(b)(iii) yields that $\AA'$ is ${\compare}$-nonempty for $-k$ iff it is finite-state ${\compare}$-nonempty for $-k$ iff $\A'$ is lasso-word ${\compare}$-nonempty for $-k$, so both problems are \PSpaceH.
\end{proof}

\paragraph{QLAs with Universal Limit-Average QWAs.}
Next, we focus on QLAs with the limit-average run aggregators.
We first show that the bottom value of a nondeterministic $\LimSupAvg$ QWA can be approximated arbitrarily well by lasso words; dually, the same holds for the top value of a universal $\LimInfAvg$ QWA.
This is in sharp contrast with nondeterministic $\LimInfAvg$ QWAs: such an automaton may be lasso-word universal for a threshold $k$, while still admitting a non-lasso word of value below $k$ \cite[Lem.~4]{DBLP:journals/tocl/ChatterjeeDH10}.
The key point is that for $\LimSupAvg$, if a word has value below some threshold $\theta$, then sufficiently long prefixes cannot keep witnessing averages above $\theta$; otherwise one could extract a run whose $\LimSupAvg$ value is at least $\theta$.
This makes it possible to isolate a long factor between two occurrences of the same reachable-state set and pump that factor, yielding an ultimately periodic word whose value stays below the threshold.

\begin{lem}\label{limsupavg-approximability}
	Consider a QWA $\A = (\Sup, \LimSupAvg, \T)$.
	For every $\varepsilon > 0$ there exists a lasso word $z$ such that $\A(z) < \bot_{\A} + \varepsilon$.
\end{lem}
\begin{proof}
	Let $\varepsilon > 0$ and $\theta = \bot_{\A} + \varepsilon$.
	Choose a word $w \in \Sigma^\omega$ such that $\A(w) < \theta$.

	For each $n \in \NN$, let $v_n$ be the prefix of $w$ of length $n$, and let $P(n) = \delta(q_I, v_n) \subseteq Q$ be the set of states reachable after reading $v_n$.
	Since $Q$ is finite, there exist a set $S \subseteq Q$ and indices $n_0 < n_1 < n_2 < \cdots$ such that $P(n_t) = S$ for every $t \in \NN$.

	For $m < n$, let $u(m,n) \in \Sigma^+$ be the unique word satisfying $v_n = v_m u(m,n)$.
	Because $P(m) = P(n) = S$, we have $\delta(S, u(m,n)) = S$.

	For a finite run fragment $\pi$, write $\operatorname{wt}(\pi)$ for its total weight.

	Now, consider a nonempty word $u \in \Sigma^+$ with $\delta(S,u) = S$.
	We associate with $u$ a finite weighted graph $H(u)$:
	its vertex set is $S$, and for $p, q \in S$ there is an edge $p \to q$ iff there exists a run fragment over $u$ from $p$ to $q$.
	The weight of this edge is $\eta_u(p,q) = \max\{\operatorname{wt}(\pi) \mid \pi \text{ is a run fragment over } u \text{ from } p \text{ to } q\}$.
	Since $\delta(S,u) = S$, every vertex of $H(u)$ has at least one outgoing edge, hence $H(u)$ contains a directed cycle.
	Let $\operatorname{mcm}(H(u))$ denote its maximum cycle mean.
	Notice that cycle means in $H(u)$ are measured per copy of $u$; dividing by $|u|$ converts them into averages per input letter.

	We will now prove two claims that together yield the existence of a lasso word $z$ such that $\A(z) < \theta$.

	\textbf{Claim 1.}
	Let $u \in \Sigma^+$ satisfy $\delta(S,u) = S$, and let $v \in \Sigma^*$ satisfy $\delta(q_I,v) = S$.
	Then, $\A(vu^\omega) = \operatorname{mcm}(H(u)) / |u|$.

	\emph{Proof of Claim 1.}
	Let $\mu_u = \operatorname{mcm}(H(u))$.
	We first show $\A(vu^\omega) \leq \mu_u / |u|$.
	Consider a run $\rho$ of $\A$ on $vu^\omega$.
	The finite prefix over $v$ does not affect $\LimSupAvg$, so it suffices to analyze the successive copies of $u$.

	For each $k \in \NN$, let $q_k \in S$ be the state of $\rho$ after reading $vu^k$, and let $y_k$ be the total weight collected while reading the $(k+1)$st copy of $u$.
	Then, $y_k \leq \eta_u(q_k, q_{k+1})$ for every $k$.
	Hence, for every $N \geq 1$, the block sum $Y_N = \sum_{k=0}^{N-1} y_k$ is bounded by the weight of the length-$N$ path $q_0 \to q_1 \to \cdots \to q_N$ in $H(u)$.

	Since $H(u)$ is finite, there exists a constant $C_u$ such that every path of length $N$ in $H(u)$ has total weight at most $\mu_u N + C_u$.
	Indeed, delete cycles one by one until only a simple path remains: each deleted cycle has mean at most $\mu_u$, and the remaining simple path has bounded length because $H(u)$ is finite.

	Therefore, $Y_N \leq \mu_u N + C_u$ for all $N \geq 1$.
	Since transition weights are bounded, the total weight collected between any prefix length and the nearest block boundary is bounded by a constant independent of the prefix length, so the difference between the two averages vanishes as the prefix grows.
	The $\limsup$ of prefix averages is therefore the same whether taken over all prefix lengths or only over block boundaries.
	Thus, $\LimSupAvg(\gamma(\rho)) = \frac{1}{|u|} \limsup_{N \to \infty} \frac{Y_N}{N} \leq \frac{\mu_u}{|u|}$.
	Since $\rho$ was arbitrary, this yields $\A(vu^\omega) \leq \mu_u / |u|$.

	For the reverse inequality, consider a directed cycle $D = r_0 \to r_1 \to \cdots \to r_{t-1} \to r_0$ in $H(u)$ whose mean is exactly $\mu_u$.
	Because $r_0 \in S = \delta(q_I,v)$, there exists a run prefix over $v$ ending in $r_0$.
	For each edge $r_a \to r_{a+1}$ of $D$, take a run fragment over $u$ from $r_a$ to $r_{a+1}$ whose weight is exactly $\eta_u(r_a, r_{a+1})$.
	Repeating these fragments periodically yields a run on $vu^\omega$ whose block sums are eventually periodic with mean $\mu_u$.
	Hence, that run has $\LimSupAvg$ value $\mu_u / |u|$, so $\A(vu^\omega) \geq \mu_u / |u|$.

	Therefore $\A(vu^\omega) = \operatorname{mcm}(H(u)) / |u|$, proving Claim~1.

	\textbf{Claim 2.}
	There exist indices $i < j$ such that $\operatorname{mcm}(H(u(n_i, n_j))) < \theta \cdot |u(n_i, n_j)|$.

	\emph{Proof of Claim 2.}
	Suppose toward contradiction that $\operatorname{mcm}(H(u(n_i, n_j))) \geq \theta \cdot |u(n_i, n_j)|$ for all $i < j$.
	For each pair $i < j$, choose a simple directed cycle $C_{i,j}$ in $H(u(n_i, n_j))$ whose mean is at least $\theta \cdot |u(n_i, n_j)|$.
	This is possible because every cycle decomposes into simple cycles, and one of those simple cycles has mean at least that of the original cycle.

	There are only finitely many simple directed cycles on the finite vertex set $S$.
	We may therefore color each pair $\{i,j\}$, with $i < j$, by the chosen cycle $C_{i,j}$ and apply the infinite Ramsey theorem.
	Thus, we obtain a subsequence $m_0 < m_1 < m_2 < \cdots$ of $n_0 < n_1 < n_2 < \cdots$ and a fixed simple cycle $C = s_0 \to s_1 \to \cdots \to s_{r-1} \to s_0$ such that, for all $p < q$, the cycle $C$ appears in $H(u(m_p, m_q))$ and has mean at least $\theta \cdot |u(m_p, m_q)|$.

	For each $t \in \NN$, define $b_t = u(m_t, m_{t+1})$ and $\ell_t = |b_t|$.
	Then, $C$ appears in $H(b_t)$ for every $t$, and its mean there is at least $\theta \ell_t$.

	For each $a \in \{0, \dots, r-1\}$, let $x_t(a) = \eta_{b_t}(s_a, s_{a+1})$, where indices are taken modulo $r$.
	Since the mean of $C$ in $H(b_t)$ is at least $\theta \ell_t$, we have for every $t \in \NN$
	\begin{equation}\label{eq:limsupavg-claim2}
		\sum_{a=0}^{r-1} x_t(a) \geq r \theta \ell_t.
	\end{equation}
	
	For $c \in \{0, \dots, r-1\}$ and $N \geq 1$, define $P_c(N) = \sum_{t=0}^{N-1} x_t((t+c) \bmod r)$.
	The point of this definition is that the chosen edges concatenate: in block $b_t$ we use the edge $s_{(t+c) \bmod r} \to s_{(t+1+c) \bmod r}$, so the target state selected in block $t$ is exactly the source state needed in block $t+1$.

	Averaging over all offsets $c$ gives, by \Cref{eq:limsupavg-claim2},
	\begin{align*}
		\frac{1}{r} \sum_{c=0}^{r-1} P_c(N)
		&= \frac{1}{r} \sum_{t=0}^{N-1} \sum_{a=0}^{r-1} x_t(a) \\
		&\geq \theta \sum_{t=0}^{N-1} \ell_t.
	\end{align*}
	Hence, for every $N \geq 1$, there exists some $c_N \in \{0, \dots, r-1\}$ such that $P_{c_N}(N) \geq \theta \sum_{t=0}^{N-1} \ell_t$.
	Because there are only finitely many offsets, one fixed $c \in \{0, \dots, r-1\}$ satisfies $P_c(N) \geq \theta \sum_{t=0}^{N-1} \ell_t$ for infinitely many values of $N$.

	Since $s_c \in S = \delta(q_I, v_{m_0})$, choose a run prefix $\pi_0$ over $v_{m_0}$ ending in $s_c$.
	For each $t \in \NN$, choose a run fragment $\sigma_t$ over $b_t$ from $s_{(t+c) \bmod r}$ to $s_{(t+1+c) \bmod r}$ whose weight is exactly $x_t((t+c) \bmod r)$.
	Because the original word decomposes as $w = v_{m_0} b_0 b_1 b_2 \cdots$, the concatenation $\rho = \pi_0 \sigma_0 \sigma_1 \sigma_2 \cdots$ is a run of $\A$ on $w$.

	For each $N \geq 1$, let $\rho^{(N)}$ be the prefix of $\rho$ over $v_{m_N}$.
	Then $\operatorname{wt}(\rho^{(N)}) = \operatorname{wt}(\pi_0) + P_c(N)$ and $m_N = m_0 + \sum_{t=0}^{N-1} \ell_t$.
	So, for infinitely many $N$,
	\[
		\frac{\operatorname{wt}(\rho^{(N)})}{m_N}
		= \frac{\operatorname{wt}(\pi_0) + P_c(N)}{m_0 + \sum_{t=0}^{N-1} \ell_t}
		\geq \frac{\operatorname{wt}(\pi_0) + \theta \sum_{t=0}^{N-1} \ell_t}{m_0 + \sum_{t=0}^{N-1} \ell_t}.
	\]
	The right-hand side tends to $\theta$ because $\sum_{t=0}^{N-1} \ell_t \to \infty$.
	Hence, along infinitely many prefixes of $\rho$, the average weight is at least a quantity converging to $\theta$.
	Therefore $\LimSupAvg(\gamma(\rho)) \geq \theta$.
	This contradicts $\A(w) < \theta$, because $\rho$ is a run of $\A$ on $w$.

	This proves Claim~2.

	Finally, choose indices $i < j$ as in Claim~2 and define $z = v_{n_i} u(n_i, n_j)^\omega$.
	By Claim~1, $\A(z) = \operatorname{mcm}(H(u(n_i, n_j))) / |u(n_i, n_j)| < \theta = \bot_{\A} + \varepsilon$.
	So, $z$ is the required lasso word.
\end{proof}

For universal limit-average QWAs with $f \in \{\LimInfAvg, \LimSupAvg\}$, the $>$-nonemptiness problem is undecidable~\cite{DBLP:conf/csl/DegorreDGRT10}.
Dually, the $\geq$-universality problem is undecidable for the corresponding nondeterministic automata.
In contrast, the decidability of $\geq$-nonemptiness over arbitrary words, and dually of $>$-universality for nondeterministic automata, is open.
As far as we know, the decidability of approximate-nonemptiness for universal limit-average QWAs over arbitrary words is also open.
For lasso words, the picture is sharper: both $>$-nonemptiness and $\geq$-nonemptiness are undecidable.

\begin{thmC}[\cite{DBLP:conf/csl/DegorreDGRT10}]\label{undecidability:limavg}
	Consider a QWA $\A = (\Inf, f, \T)$ with $f\in\{\LimInfAvg$, $\LimSupAvg\}$.
	The $>$-nonemptiness of $\A$ is undecidable.
	When we consider only lasso words, both $>$-nonemptiness and $\geq$-nonemptiness are undecidable.
\end{thmC}
\begin{proof}
	The undecidability of $>$-nonemptiness for $\LimInfAvg$ and $\LimSupAvg$ was proved in~\cite[Thm. 4]{DBLP:conf/csl/DegorreDGRT10}.
	The undecidability of lasso-word $\geq$-nonemptiness for $\LimInfAvg$ and $\LimSupAvg$ was proved in~\cite[Thm. 6]{DBLP:conf/csl/DegorreDGRT10}.
	The lasso-word $>$-nonemptiness for $\LimInfAvg$ is undecidable since the problem is equivalent to $>$-nonemptiness for $\LimInfAvg$ by the dual version of~\cref{limsupavg-approximability}.
	The corresponding problems for $\LimSupAvg$ are equivalent to those for $\LimInfAvg$ since for all lasso words the $\LimSupAvg$ and $\LimInfAvg$ values coincide.
\end{proof}

Using the undecidability results in \Cref{undecidability:limavg}, we obtain the following.

\begin{thm}\label{emptiness:avg}
	\label{emptiness:avg:stochastic}
	Consider a QLA $\AA = (h,(\Inf,f,\T))$ with $f\in\{\LimInfAvg$, $\LimSupAvg\}$ and $h \in \{\Inf,\Sup,\E\}$.
	The $>$-nonemptiness of $\AA$ is undecidable.
	For the finite-state restriction, the ${\compare}$-nonemptiness is undecidable for these cases:
	(i) $h = \Inf$ and ${\compare}\in\{{>}, {\geq}\}$;
	(ii) $h = \Sup$ and ${\compare} = {>}$;
	(iii) $h = \E$, $f = \LimInfAvg$, and ${\compare} = {>}$. 
\end{thm}
\begin{proof}
	Let $\A$ be the underlying QWA of $\AA$.
	For the unrestricted cases and ${\compare} \in \{{>}, {\geq}\}$, we use~\cref{bigreduction}(a)(i-ii) to reduce the ${\compare}$-nonemptiness or approximate-nonemptiness of $\A$ to the ${\compare}$-nonemptiness of $\AA$. Thanks to \Cref{undecidability:limavg}, the case ${\compare} = {>}$ is undecidable.

	For the finite-state restrictions, we have the following.
	If $h = \Inf$, we use~\cref{bigreduction}(b)(ii) to reduce the lasso-word ${\compare}$-nonemptiness of $\A$, which is undecidable by \Cref{undecidability:limavg}, to the finite-state ${\compare}$-nonemptiness of $\AA$.
	If $h = \Sup$, we use~\cref{bigreduction}(b)(i) to reduce the unrestricted ${\compare}$-nonemptiness of $\AA$, which we proved undecidable above for ${\compare} = {>}$, to the finite-state ${\compare}$-nonemptiness of $\AA$.
	If $h = \E$ and $f = \LimInfAvg$, by~\cref{approximability-reduction} and the dual version of~\cref{limsupavg-approximability} (namely, for every QWA $\A = (\Inf, \LimInfAvg, \T)$ and every $\varepsilon > 0$ there is a lasso word $w$ such that $\A(w) > \top_{\A} - \varepsilon$), the finite-state $>$-nonemptiness of $\AA$ is equivalent to the unrestricted $>$-nonemptiness of $\AA$, which we proved undecidable above.
	We leave the remaining cases open and we conjecture them to be undecidable.
\end{proof}

\paragraph{QLAs with Universal Discounted-Sum QWAs.}
We now consider QLAs with the discounted-sum run aggregator.
We first show that the bottom value of a nondeterministic $\DSum$ QWA is achievable.
By duality, it follows that the top value of a universal $\DSum$ QWA is achievable.
The key idea is to build a word incrementally by extending a finite prefix with a letter that preserves the current infimum over all continuation values.
Since discounted-sum automata are co-safe in the quantitative sense~\cite{DBLP:conf/fossacs/HenzingerMS23,DBLP:conf/concur/BokerHMS23,DBLP:journals/lmcs/BokerHMS25}, the value of the limit word equals the limit of its prefix infima, which converges to the bottom value.
We use this realizability property below to obtain transfer hardness results for nonemptiness in the universal discounted-sum setting.

\begin{prop}\label{realizability:dsum}
	Consider a QWA $\A = (\Sup, \DSum, \T)$.
	There is a word $w$ such that $\A(w) = \bot_{\A}$.
\end{prop}
\begin{proof}
	Let $\A$ be a QWA as in the statement.
	For every $u \in \Sigma^*$, let $P_u = \{\A(uw) \st w \in \Sigma^\omega\}$.
	We claim that for every $u \in \Sigma^*$, there exists $\sigma \in \Sigma$ such that $\inf P_u = \inf P_{u\sigma}$.
	To prove, let $u$ be a finite word and observe the following:
	$P_u = \bigcup_{\sigma \in \Sigma} P_{u\sigma}$.
	Then, since $\inf(\bigcup_{i=1}^n X_i) = \min_{1 \leq i \leq n} \inf X_i$ for every finite collection $X_1, \ldots, X_n$ of sets and $\Sigma$ is finite, we have the following: $\inf P_u = \min_{\sigma \in \Sigma} (\inf P_{u\sigma})$.
	Finally, since some $\hat{\sigma} \in \Sigma$ achieves this minimum by definition, we have $\inf P_u = \inf P_{u\hat{\sigma}}$.

	Now, let $u_0 = \epsilon$ be the empty word.
	For each $i \geq 0$, choose a letter $\sigma_i \in \Sigma$ and set $u_{i+1} = u_i \sigma_i$.
	Then, define $w$ as the limit of this sequence, i.e., $w = \sigma_0 \sigma_1 \ldots$.
	By the definition of $w$, we have $\lim_{i \to \infty} \inf P_{u_i} = \bot_{\A}$.
	Since the sequence of infima is nondecreasing, we can equivalently write $\sup_{u \prec w} \inf P_u = \bot_{\A}$.
	By~\cite[Definition 24]{DBLP:conf/fossacs/HenzingerMS23} the left-hand side expresses the co-safety closure of $\A$.
	Moreover, since every nondeterministic discounted-sum word automaton is co-safe~\cite[Theorems 4.15 and 4.16]{DBLP:conf/concur/BokerHMS23}, the left-hand side equals $\A(w)$.
	Therefore, $\A(w) = \bot_{\A}$.
\end{proof}

Next, we relate finite-state Markov-chain values to lasso-word attainability for discounted-sum QLAs with language aggregator $\E$.
We show that if no lasso word attains $\top_\A$ (resp.\ $\bot_\A$), then no finite-state Markov chain attains $\top_\A$ (resp.\ $\bot_\A$) at the QLA level.
The proof is by contradiction: from a reachable cycle of the Markov chain we extract a lasso word, and then use the discounted-sum dependence on long common prefixes to show that words generated with positive probability force the expected value to stay strictly away from the extremal value.

\begin{prop}\label{convexity:dsum}
	Consider a QLA $\AA = (\E, \A)$ where $\A = (g, \DSum, \T)$ and $g \in \{\Sup, \Inf, \E\}$. If there is no lasso word $w \in \Sigma^{\omega}$ such that $\A(w) = \top_\A$ ($\A(w) = \bot_\A$) then there is no finite Markov chain $\mu$ such that $\AA(\mu) = \top_\A$ ($\AA(\mu) = \bot_\A$).
\end{prop}
\begin{proof}
	Let us consider a finite Markov chain $\mu$ such that $\AA(\mu) = \top_\A$ and that there is no lasso word $w \in \Sigma^{\omega}$ such that $\A(w) = \top_\A$. We arrive at a contradiction. We know that this is not possible for $g = \Sup$ as the top value is always achievable by a lasso word. Hence, we only consider $g = \Inf$ and $g = \E$. The dual statement is proved dually.

	First, we take some $u, v \in \Sigma^*$ such that $\mu(uv^n\Sigma^{\omega}) > 0$ for every $n \in \mathbb{N}$.
	One can take any reachable cycle in the Markov chain.
	The prefix part $u$ corresponds to reaching the cycle and $v$ corresponds to the cycle itself.

	We prove that there exists a lasso word $w \in \Sigma^{\omega}$ such that $\AA(\mu) \leq \A(w)$.

	Let $S$ be the set of weights of $\T$ and $\lambda$ the discount factor.
	Let $W = \max(\{x ~|~ x \in S\} \cup \{-x ~|~ x \in S\})$.
	Take some $n \in \mathbb{N}$, words $w_1, w_2 \in \Sigma^{\omega}$ and a finite run $\rho$ over $uv^n$.
	Then, for any two run continuations $\rho_1, \rho_2$ of the run $\rho$ over words $w_1, w_2$ in $\A$, respectively, we have that $|\Val(\rho \rho_1) - \Val(\rho\rho_2)| \leq \lambda^{|uv^n|}\frac{2W}{1 - \lambda}$ since the value of the suffixes $w_1, w_2$ are discounted $|uv^n|$ number of times and an absolute value of an infinite run in $\A$ is at most $\frac{W}{1 - \lambda}$.

	Since there is no lasso word achieving the top value, we have that $\A(uv^\omega) < \top_\A$.
	Let us take $n$ such that $\lambda^{|uv^n|}\frac{2W}{1 - \lambda} < \top_\A - \A(uv^\omega).$
	For both $g = \Inf$ and $g = \E$, we prove that for every $w \in \Sigma^{\omega}$, we have $\A(uv^nw) < \top_{\A}$.

	\begin{enumerate}[(i)]
		\item Case $g = \Inf$.
		Let $\varepsilon > 0$ and take a run $\rho\rho_1$ over the word $uv^\omega$ such that $\rho$ is a run over the finite word $uv^n$ and $\Val(\rho\rho_1) \leq \A(uv^\omega) + \varepsilon$.
		Thanks to the property proved above, for any run continuation $\rho_2$ of the run $\rho$ over the word $w$, we have that $|\Val(\rho\rho_1) - \Val(\rho\rho_2)| \leq \lambda^{|uv^n|}\frac{2W}{1 - \lambda}.$
		Hence, there exists a run over the word $uv^nw$ that achieves a value at most $\A(uv^\omega) + \lambda^{|uv^n|}\frac{2W}{1 - \lambda} + \varepsilon$.
		As this holds for every $\varepsilon > 0$, this implies $\A(uv^nw) \leq \A(uv^\omega) + \lambda^{|uv^n|}\frac{2W}{1 - \lambda} < \top_\A.$
		
		\item Case $g = \E$.
		Let us take a run $\rho$ over the finite word $uv^n$.
		Let $\A(uv^nw ~|~ \rho) := \E\left[\Val(\rho') ~|~ \rho' \text{ a run over } uv^nw \text{ that starts with } \rho\right]$ denote the value of the word $uv^nw$ in $\A$ conditioned on the fact that the run starts with $\rho$, and define $\A(uv^\omega ~|~ \rho)$ analogously.
		Thanks to the property proved above, we have that for any run continuation $\rho_1$ over the word $v^\omega$ and any run continuation $\rho_2$ over the word $w$, we have that $|\Val(\rho\rho_1) - \Val(\rho\rho_2)| \leq \lambda^{|uv^n|}\frac{2W}{1 - \lambda}.$
		Hence, we have that $\A(uv^nw~|~\rho) \leq \A(uv^\omega~|~\rho) + \lambda^{|uv^n|}\frac{2W}{1 - \lambda}.$
		Since this holds for any run $\rho$ over $uv^n$, we obtain that
		\begin{align*}
			\A(uv^nw) &= \underset{\rho \text{ over }uv^n}{\E}\A(uv^nw ~|~ \rho) \\
			&\leq \underset{\rho \text{ over }uv^n}{\E} \A(uv^\omega ~|~ \rho) + \lambda^{|uv^n|}\frac{2W}{1 - \lambda} \\
			&= \A(uv^\omega) + \lambda^{|uv^n|}\frac{2W}{1 - \lambda} \\
			&< \top_\A.
		\end{align*}
	\end{enumerate}

	Now, since $\mu(uv^n\Sigma^{\omega}) > 0$, $\A(uv^nw) < \top_\A$ for any $w \in \Sigma^{\omega}$, and $\A(x) \leq \top_\A$ for every $x \in \Sigma^\omega$, we get that the expected value $\AA(\mu) < \top_\A$, which is a contradiction.
\end{proof}

The next proposition handles the finite-state case for discounted-sum QLAs with outer aggregator $\E$.
It shows that finite-state nonemptiness is equivalent to lasso-word nonemptiness of the underlying QWA.

\begin{prop}\label{finite-state:dsum}
	Consider a QLA $\AA = (\E, \A)$ where $\A = (g, \DSum, \T)$, $g \in \{\Inf, \Sup, \E\}$.
	Let ${\compare} \in \{>, \geq\}$ and $k \in \QQ$.
	Then, $\AA$ is finite-state ${\compare}$-nonempty for $k$ iff $\A$ is lasso-word ${\compare}$-nonempty for $k$.
\end{prop}
\begin{proof}
	If $\AA$ is finite-state ${\compare}$-nonempty for $k$, then there exists a finite-state Markov chain $\mu$ such that $\AA(\mu) \compare k$.
	If $\AA(\mu) = \top_{\A}$ then there exists a lasso word $v$ such that $\A(v) = \top_\A$ due to \Cref{convexity:dsum}. If $\AA(\mu) < \top_{\A}$ then we can use the approximability of the top value by lasso words to claim the existence of a lasso word $v$ such that $\AA(\mu) < \A(v)$.
	Therefore, $\A(v) \compare k$ for some lasso word $v$.
	
	If $\A$ is lasso word ${\compare}$-nonempty for $k$, there exists a lasso word $w$ such that $\A(w) \compare k$
	We can take a Markov chain $\mu$ which generates $w$ with probability 1 and we get that $\AA(\mu) \compare k$.
\end{proof}

By combining the discounted-sum realizability and transfer results above, in particular \cref{bigreduction} and \cref{finite-state:dsum}, we show that the nonemptiness problem for QLAs whose underlying QWAs are universal $\DSum$ QWAs is at least as hard as the universality problem for nondeterministic $\DSum$ QWAs, which is a long-standing open problem~\cite{DBLP:journals/tocl/ChatterjeeDH10,DBLP:conf/lics/BokerHO15}.

\begin{thm}\label{emptiness:dsum} 
	Consider a QLA $\AA = (h,(\Inf,\DSum,\T))$ with $h \in \{\Inf,\Sup,\E\}$.
	Let ${\compare}\in\{{>}, {\geq}\}$ and let ${\hat{\compare}} \in\{{>}, {\geq}\} \setminus \{{\compare}\}$ be the dual of ${\compare}$.
	If the (finite-state) ${\compare}$-nonemptiness of $\AA$ is decidable, then the (lasso-word) ${\hat{\compare}}$-universality of nondeterministic discounted-sum word automata is decidable.
\end{thm}
\begin{proof}
	Consider a nondeterministic discounted-sum QWA $\B = (\Sup,\DSum,\T_\B)$ and a threshold $k \in \QQ$.
	Let $\A = \hat{\B} = (\Inf,\DSum,\hat{\T}_\B)$ and let $\AA = (h,\A)$.
	Since discounted sum is self-dual, for every word $w$ we have $\A(w) = -\B(w)$.
	Hence, $\B$ is ${\hat{\compare}}$-universal for $k$ iff $\A$ is not ${\compare}$-nonempty for $-k$, and the same equivalence holds when quantification is restricted to lasso words.

	For the unrestricted case, if $h \neq \Sup$ or ${\compare} = {>}$, then by~\cref{bigreduction}(a)(i), $\AA$ is ${\compare}$-nonempty for $-k$ iff $\A$ is ${\compare}$-nonempty for $-k$.
	If $h = \Sup$ and ${\compare} = {\geq}$, then by~\cref{bigreduction}(a)(ii),then $\AA$ is ${\geq}$-nonempty for $-k$ iff $\A$ is approximate-nonempty for $-k$.
	Moreover, $\top_{\A}$ is achievable by the dual of~\cref{realizability:dsum}, so approximate-nonemptiness and ${\geq}$-nonemptiness coincide for $\A$.
	Therefore, decidability of ${\compare}$-nonemptiness for $\AA$ yields decidability of ${\hat{\compare}}$-universality for $\B$.

	For the finite-state case, if $h \in \{\Inf,\E\}$, then by~\cref{finite-state:dsum} and by~\cref{bigreduction}(b)(ii), lasso-word ${\compare}$-nonemptiness of $\A$ for $-k$ reduces to finite-state ${\compare}$-nonemptiness of $\AA$ for $-k$.
	Combining this with the equivalence above between lasso-word ${\hat{\compare}}$-universality of $\B$ for $k$ and lasso-word ${\compare}$-nonemptiness of $\A$ for $-k$, we obtain a reduction from lasso-word ${\hat{\compare}}$-universality of $\B$ to finite-state ${\compare}$-nonemptiness of $\AA$.
	If $h = \Sup$, then by~\cref{bigreduction}(b)(i), unrestricted ${\compare}$-nonemptiness of $\AA$ reduces to finite-state ${\compare}$-nonemptiness of $\AA$.
	Combined with the unrestricted equivalence above, this gives a reduction from unrestricted ${\hat{\compare}}$-universality of $\B$ to finite-state ${\compare}$-nonemptiness of $\AA$.
	In particular, this also covers lasso-word ${\hat{\compare}}$-universality.
\end{proof}

\paragraph{QLAs with Probabilistic QWAs.}
We finally consider QLAs whose underlying QWA is probabilistic, that is, with $g = \E$.
In this setting, undecidability already arises at the word-automaton level: several variants of nonemptiness for probabilistic QWAs are undecidable.
We first recall the relevant word-level result~\cite[Cor. 3.4]{DBLP:journals/ai/MadaniHC03} and then use it to derive undecidability for the corresponding QLA nonemptiness problems, including the finite-state cases.

\begin{thmC}[\cite{DBLP:journals/ai/MadaniHC03}]\label{inapproximability}
	For any $\varepsilon$ such that $0 < \varepsilon < \frac{1}{2}$, the following problem is undecidable:
	Let $\A = (\E, \Sup, \T)$ be a probabilistic word automaton with weights over $\{0, 1\}$ and accepting states, i.e., a subset of states $S \subseteq Q$ such that all transitions leading to $S$ have weight 1 and every other transition has weight 0. Given that either of the two cases hold:
	\begin{enumerate}[(i)]
		\item There exists a word $w \in \Sigma^{\omega}$ such that $\A(w) > 1 - \varepsilon$.
		\item For all words $w \in \Sigma^{\omega}$, we have $\A(w) < \varepsilon$.
	\end{enumerate}
	Decide whether (i) holds.
\end{thmC}

We now use \cref{inapproximability} to derive undecidability of several nonemptiness variants for probabilistic QWAs.
These word-level results will then be lifted to the corresponding QLA problems.
We treat discounted-sum separately as some cases remain open.

\begin{thm}\label{stochastic:undecidability:rest}
	Consider a QWA $\A = (\E, f, \T)$ with $f \in \{\Inf$, $\Sup$, $\LimInf,$ $\LimSup$, $\LimInfAvg$, $\LimSupAvg\}$.
	Let ${\compare}\in\{{>}, {\geq}\}$.
	The ${\compare}$-nonemptiness of $\A$ is undecidable, whether we consider all words from $\Sigma^{\omega}$ or only lasso words.
	Moreover, its approximate-nonemptiness is also undecidable.
\end{thm}
\begin{proof}
	The proof is based on the proofs of Theorems 4.2., 4.3., and 4.7. of \cite{DBLP:journals/ai/MadaniHC03}.

	We prove the undecidability for both lasso words and all words. Given a probabilistic automaton $\B = (\E, \Sup, \T_\B)$, we construct a QWA $\A' = (\E, f, \T')$ where $\T'$ is the same as $\T_\B$ but we change all outgoing transitions of accepting states to be a loop with value 1.
	\Cref{inapproximability} gives us that either (i) or (ii) is true.
	If (i) holds for some $w \in \Sigma^{\omega}$ then there exists a long enough prefix $u \in \Sigma^*$ of $w$ such that $\B(u v^\omega) > \frac{1}{2}$ for any $v \in \Sigma^*$ since $\B(u v^{\omega})$ is at least the probability of getting to an accepting state after reading $u$. Directly from the construction, we obtain that $\A'(uv^\omega) > \frac 12$.
	If (ii) holds, then for all words $w \in \Sigma^{\omega}$ we have that $\A'(w)< \frac 12$.
	Hence, we reduce the problem of \Cref{inapproximability} to the nonemptiness problem of whether $\A'(w)\ {\compare}\ \frac 12$ for some (lasso-word) $w \in \Sigma^{\omega}$.
	For $f = \Inf$, the reduction is dual as for $f = \Sup$. We swap all weights in $\T'$ from 0 to 1 and vice-versa. If there is a word $w \in \Sigma^{\omega}$ such that $\B(w) > \frac 12$ then we have $\A'(uv^\omega) < \frac 12$ for some prefix $u\in \Sigma^\omega$ of $w$ and $v \in \Sigma^*$. Furthermore, if $\B(w) < \frac 12$ for all words $w \in \Sigma^{\omega}$, then $\A'(w) > \frac 12$ for all words $w \in \Sigma^{\omega}$.

	Regarding the approximate-nonemptiness problem, the above construction proves undecidability of approximate-nonemptiness thanks to \Cref{inapproximability}
	as $\varepsilon < \frac{1}{2}$.
\end{proof}

The decision questions for discounted-sum probabilistic QWAs are generally undecidable. We leave open the case of strict nonemptiness for the unrestricted variant.

\begin{thm}\label{stochastic:undecidability:dsum}
	Consider a QWA $\A = (\E, \DSum, \T)$.
	The ${>}$-nonemptiness of $\A$ is undecidable. When we consider only lasso words, then ${>}$-nonemptiness of $\A$ and ${\geq}$-nonemptiness of $\A$ are undecidable.
\end{thm}
\begin{proof}
	The undecidability of ${>}$-nonemptiness for all words follows from \cite[Thm. 4.4]{DBLP:journals/ai/MadaniHC03}.

	For the lasso-word variants, since the top value can be arbitrarily approximated by lasso words, then the lasso-word $>$-nonemptiness is equivalent to the unrestricted $>$-nonemptiness.
	For $\geq$-nonemptiness, we use \cite[Thm. 4.8]{DBLP:journals/ai/MadaniHC03}: it is undecidable to determine whether the top value (which is known to be $\frac{1}{2}$) of a particular automaton is achieved by a lasso word.
	Since this problem can be encoded as a lasso-word $\geq$-nonemptiness (with threshold $\frac{1}{2}$), this nonemptiness problem is undecidable.
\end{proof}

Using the word-level undecidability results above and the transfer arguments developed earlier, we now lift undecidability to the nonemptiness problem for QLAs with probabilistic underlying QWAs, including the finite-state restriction.
We leave open the cases of $>$-nonemptiness for unrestricted discounted-sum and the case of finite-state $>$-nonemptiness for discounted-sum with supremum as the language aggregator.

\begin{thm}\label{emptiness:mixe}
	Consider a QLA $\AA = (h, (\E, f, \T))$ with $f \in \{\Inf$, $\Sup$, $\LimInf,$ $\LimSup$, $\LimInfAvg$, $\LimSupAvg$, $\DSum\}$ and $h \in \{\Inf,\Sup,\E\}$.
	Let ${\compare}\in\{{>}, {\geq}\}$.
	The ${\compare}$-nonemptiness of $\AA$ is undecidable when $f \neq \DSum$ or ${\compare} = >$.
	The finite-state ${\compare}$-nonemptiness of $\AA$ is undecidable when $h \neq \Sup$ or $f \neq \DSum$ or ${\compare} = >$.
\end{thm}
\begin{proof}
	Let $\A = (\E, f, \T)$ be the underlying QWA of $\AA$.
	For the unrestricted cases, we use~\cref{bigreduction}(a)(i-ii), depending on the case, to reduce the ${\compare}$-nonemptiness or approximate-nonemptiness of $\A$, both of which are undecidable (for $f=\DSum$, only $>$-nonemptiness applies) by~\cref{stochastic:undecidability:rest,stochastic:undecidability:rest}, to the ${\compare}$-nonemptiness of $\AA$.

	Let us now prove the finite-state case.

	For $f = \DSum$ and $h = \Sup$, we combine the unrestricted $>$-nonemptiness undecidability proved above with~\cref{bigreduction}(b)(i).
	For $f = \DSum$ and $h \in \{\Inf, \E\}$, we use~\cref{finite-state:dsum} and~\cref{bigreduction}(b)(ii) to reduce the lasso-word ${\compare}$-nonemptiness of $\A$ to the finite-state ${\compare}$-nonemptiness of $\AA$.

	For the remaining run aggregators, the proof follows the same lines as in the proof of \Cref{stochastic:undecidability:rest}.
	Let $\B = (\E, \Sup, \T_B)$ be the source probabilistic automaton from \Cref{inapproximability}, let $\A'$ be the QWA constructed from $\B$ in the proof of \Cref{stochastic:undecidability:rest} for the chosen run aggregator, and let $\AA' = (h, \A')$.
	If case~(i) of \Cref{inapproximability} holds for some $w \in \Sigma^{\omega}$, then there exists a long enough prefix $u \in \Sigma^*$ of $w$ such that $\B(uv) > \frac 12$ for every $v \in \Sigma^{\omega}$.
	By the construction of $\A'$, we also have $\A'(uv) > \frac 12$ for every $v \in \Sigma^{\omega}$.
	Let $v_0 \in \Sigma^+$ and $w_0 = uv_0^\omega$.
	Then, $\A'(w_0) > \frac 12$.
	Hence, for $h \in \{\Inf, \Sup\}$, the singleton language $\{w_0\}$ witnesses finite-state ${\compare}$-nonemptiness of $\AA'$, and for $h = \E$, the finite-state Markov chain generating $w_0$ witnesses finite-state ${\compare}$-nonemptiness of $\AA'$.
	If case~(ii) holds, then there exists a fixed $\varepsilon < \frac 12$ such that $\A'(w) < \varepsilon$ for all words $w \in \Sigma^{\omega}$.
	Therefore, for $h \in \{\Inf, \Sup\}$ we have $\AA'(L) < \frac 12$ for every nonempty $\omega$-regular language $L$, and for $h = \E$ we have $\AA'(\mu) < \frac 12$ for every finite-state Markov chain $\mu$.
\end{proof}

\section{Language Automata with Limit Aggregators}\label{sec:limit}

We now consider QLAs in which all three aggregators come from $\{\Inf,\Sup,\LimInf,\LimSup\}$ and at least one word or language aggregator is a limit aggregator.
The central difficulty is that $\LimInf$ and $\LimSup$ depend on infinite support: one must determine not merely which values are achieved, but which are achieved infinitely often, both across runs on a fixed word and across words in the language.

The section proceeds as follows.
We first develop combinatorial criteria for infinite achievability, including a finite pattern characterizing states visited infinitely often over infinitely many words, and automata constructions isolating exact-value runs.
These tools support the subsequent analysis: we characterize the expressive power of limit aggregators, establish a uniform \PSpace upper bound for evaluation via reductions to emptiness and infiniteness tests, and handle comparison, nonemptiness, and universality through a finite-profile argument.
We further show that nonemptiness drops to \PTime when the word aggregator is $\Sup$.

\subsection{Infiniteness Patterns and Exact-Value Languages}

To analyze QLAs with a word or language aggregator in $\{\LimInf,\LimSup\}$, we must determine which values occur infinitely often.
This gives rise to two tasks. 
At the language level, we identify a finite pattern witnessing infinitely many distinct words with the same qualitative behavior (see \Cref{fig:pattern:words:fixed}).
At the run level, for a fixed word and value $x$, we determine whether infinitely many runs achieve value $x$, handled not by a structural characterization but by automata: we isolate exact-value runs and recognize the words on which such runs are infinite.
Throughout, we restrict to run aggregators $f \in \{\Inf,\Sup,\LimInf,\LimSup\}$.

\begin{figure}[t]
	\begin{tikzpicture}
		\node (constuction0) {};
		\node[state, label=center:$q_I$, shift = (180:1cm), initial] (q1) at (constuction0) {};
		\node[shift = (300:1cm)] (q5) at (constuction0) {};
		\node[shift = (60:1cm)] (q4) at (constuction0) {};
		\node[state, label=center:$q_1$, right = of q1] (q2){};
		\node[state, label=center:$q_2$, right = of q2] (q3) {};
		\path[transition]
		(q1) edge node[above] {$u_0$} (q2)
		(q2) edge[loop above] node {$v_1$} (q2)
		(q2) edge node[above] {$u_1u_2$} (q3)
		(q3) edge[loop above] node {$v_2$} (q3)
		;
	\end{tikzpicture}
	\caption{
		The pattern used in \Cref{characterize:words:fixed} witnessing word-level infiniteness.
		From $q_I$, the automaton reaches $q_1$, may iterate the loop labeled $v_1$, then moves to $q_2$ via $u_1u_2$, and may iterate the loop labeled $v_2$ at $q_2$, where $u_0,u_1,u_2,v_1,v_2 \in \Sigma^+$, $|u_1| = |v_1|$, and $u_1 \neq v_1$.
		Consequently, the words $u_0(v_1)^i u_1u_2(v_2)^\omega$, for $i \in \NN$, are pairwise distinct, and each admits a run that visits $q_2$ infinitely often.
	}
	\label{fig:pattern:words:fixed}
\end{figure}

We begin with the language-level task.
The next lemma characterizes when a state is visited infinitely often along runs over infinitely many distinct words.
The forward direction extracts a finite pattern from an infinite family of witnesses; the backward direction shows that this pattern generates infinitely many pairwise distinct lasso words.
The argument is closely related to classical ambiguity results for finite automata~\cite{DBLP:journals/tcs/WeberS91,DBLP:conf/dlt/LodingP18}.

\begin{lem}\label{characterize:words:fixed}
	Let $\T$ be a labeled transition system with initial state $q_I$.
	For every state $q_2$ of $\T$, the following are equivalent:
	\begin{enumerate}[(i)]
		\item There exist infinitely many words $w \in \Sigma^\omega$ with a run from $q_I$ visiting $q_2$ infinitely often.
		\item There exist $u_0,u_1,u_2,v_1,v_2\in\Sigma^+$, a state $q_1$ of $\T$, and run fragments
		\[
		\rho_1 : q_I \xrightarrow{u_0} q_1,\qquad
		\ell_1 : q_1 \xrightarrow{v_1} q_1,\qquad
		\rho_2 : q_1 \xrightarrow{u_1u_2} q_2,\qquad
		\ell_2 : q_2 \xrightarrow{v_2} q_2
		\]
		such that $|u_1| = |v_1|$ and $u_1 \neq v_1$.
	\end{enumerate}
	In particular, Item~(ii) is exactly the pattern of \Cref{fig:pattern:words:fixed} parameterized by $q_2$.
\end{lem}
\begin{proof}
	We start with the backward implication.
	Given the state $q_2$, we assume that $\T$ complies with the pattern in \Cref{fig:pattern:words:fixed} exhibiting four finite runs of the form $\rho_1 \colon q_I \xrightarrow{u_0} q_1$, $\ell_1 \colon q_1 \xrightarrow{v_1} q_1$, $\rho_2 \colon q_1 \xrightarrow{u_1u_2} q_2$, $\ell_2 \colon q_2 \xrightarrow{v_2} q_2$.
	Additionally we have that $|u_1|=|v_1|$ and $u_1\neq v_1$.
	For all $i\in\NN$, we define the run $\pi_i = \rho_1^{}(\ell_1^{})^{i}\rho_2^{}(\ell_2^{})^\omega$ from $q_I$ over $u_0(v_1)^{i}u_1u_2(v_2)^\omega$.
	Since $|u_1|=|v_1|$ and $u_1\neq v_1$, we have that $\pi_{i}$ and $\pi_{j}$ read distinct words for all $i \neq j$.
	In particular, the set $\{u_0(v_1)^iu_1u_2(v_2)^\omega \st i \in \NN\}$ of ultimately periodic words with a run visiting $q_2$ infinitely often is infinite.
	
	Next, we prove the forward implication.
	We say that a state $p$ is infinitely productive if infinitely many words admit a run visiting $q_2$ infinitely often starting from $p$.
	By assumption, the initial state $q_I$ is infinitely productive.
	Since every state has finitely many successors, at least one successor of every infinitely productive state must be infinitely productive.
	Iterating this argument starting from $q_I$ gives an arbitrarily long path of infinitely productive states.
	As the state space is finite, some state $q_1$ eventually repeats, yielding a run $\rho_1 \colon q_I \xrightarrow{u_0} q_1$ and a loop $\ell'_1 \colon q_1 \xrightarrow{v_0} q_1$ with $v_0 \neq \varepsilon$.
	By construction, $q_1$ is infinitely productive, and thus we can pick a witness word $w \neq v_0^\omega$ admitting a run that visits $q_2$ infinitely often starting from $q_1$.
	The word $w$ is of the form $w = w_1w_2\ldots$ where each $|w_i| = |v_0|$ for all $i\in\NN$.
	Let $m \geq 1$ be the least index such that $w_m \neq v_0$.
	We define $u_1 = w_1 \ldots w_m$.
	Let $v_1 = v_0^m$ and $\ell_1 \colon q_1 \xrightarrow{v_1} q_1$.
	Observe that $|u_1| = |v_1|$ and $u_1 \neq v_1$.
	Consider a run $\pi$ from $q_1$ on $w$ that visits $q_2$ infinitely often.
	Since $|u_1|$ is finite and $\pi$ visits $q_2$ infinitely often, there exist two positions $i<j$ with $i > |u_1|$ such that $\pi$ is in state $q_2$ after reading the prefix of length $i$ and again after reading the prefix of length $j$.
	Let $u_2$ be the factor of $w$ between positions $|u_1|$ and $i$, and let $v_2$ be the factor of $w$ between positions $i$ and $j$.
	Then, $u_2 \neq \varepsilon$ and $v_2 \neq \varepsilon$.
	Let $\rho_2 \colon q_1 \xrightarrow{u_1u_2} q_2$ be the prefix of $\pi$ ending at position $i$, and let $\ell_2 \colon q_2 \xrightarrow{v_2} q_2$ be the segment of $\pi$ between positions $i$ and $j$.	
	Finally, if $u_0 = \varepsilon$, then necessarily $q_1 = q_I$, and we can replace $\rho_1$ by $\rho_1 \ell_1$, whose label is $u_0 v_1 = v_1 \neq \varepsilon$.
	Hence, we may assume that $u_0 \neq \varepsilon$.
    Therefore, $\T$ complies with the pattern in \Cref{fig:pattern:words:fixed}.
\end{proof}

The previous lemma reduces infiniteness of an $\omega$-regular language to the existence of a finite pattern in the corresponding B\"uchi automaton.
As a first application, we use it to decide whether the difference of two B\"uchi languages is infinite.
The proof keeps the complemented automaton implicit and checks the components of the pattern on the fly in polynomial space.

\begin{thm}\label{main:words}
	Let $\B_1$ and $\B_2$ be two B\"uchi automata.
	We can decide in \PSpace whether $L(\B_1) \setminus L(\B_2)$ is infinite.
\end{thm}
\begin{proof}[Proof sketch]
	Let $n = |\B_1| + |\B_2|$ and let $m = |Q_{\B_2}| \leq n$.
	We use a standard rank-based complementation of $\B_2$~\cite{DBLP:journals/tocl/KupfermanV01,DBLP:conf/stacs/Schewe09}.
	In this construction, a state of the complement is a pair $(g,O)$, where $g \colon Q_{\B_2} \to \{\bot, 0, \dots, 2m\}$ is a level ranking and $O \subseteq Q_{\B_2}$ is an obligation set.
	Hence, the complement may have exponentially many states, but each state is polynomially representable: encoding $g$ and $O$ uses $O(m \log m)$ bits.
	Intersecting this complement with $\B_1$ adds only the current state of $\B_1$ and a constant-size flag for the B\"uchi product.
	Therefore, we obtain a B\"uchi automaton $\D$ with $L(\D) = L(\B_1) \setminus L(\B_2)$, with at most $2^{p(n)}$ states for some polynomial $p$, such that every state of $\D$ has an encoding of length $p(n)$ and both its acceptance condition and its transition relation are decidable from these encodings in polynomial space: given encoded states $q,q'$ and a letter $a$, one only checks the corresponding transition of $\B_1$, that the next ranking is compatible with the current one on $a$, that the obligation set is updated correctly, and that the product flag is updated correctly.

	By \Cref{characterize:words:fixed}, $L(\D)$ is infinite iff there exist states $q_1, q_2$ of $\D$ with $q_2$ accepting, words $u_0, u_1, u_2, v_1, v_2 \in \Sigma^+$ with $|u_1| = |v_1|$ and $u_1 \neq v_1$, and run fragments $q_I \xrightarrow{u_0} q_1$, $q_1 \xrightarrow{v_1} q_1$, $q_1 \xrightarrow{u_1 u_2} q_2$, $q_2 \xrightarrow{v_2} q_2$ in $\D$.
	We verify each condition in polynomial space.
	For ordinary reachability (e.g., the existence of $q_1$ and $u_0$ with $q_I \xrightarrow{u_0} q_1$), it suffices to guess a path of length at most $|Q_\D| \leq 2^{p(n)}$, storing only the current encoded state and a counter of $O(p(n))$ bits, and to verify each guessed transition on the fly.
	Hence, reachability in $\D$ is in $\NPSpace = \PSpace$ by Savitch's theorem.
	The self-loop conditions ($q_1 \xrightarrow{v_1} q_1$ and $q_2 \xrightarrow{v_2} q_2$) and the connecting path ($q_1 \xrightarrow{u_1 u_2} q_2$) reduce to the same kind of reachability check.
	For the remaining condition $|u_1| = |v_1|$ with $u_1 \neq v_1$, we use an auxiliary graph $Q_\D \times Q_\D \times \{0,1\}$, where the extra bit records whether a letter mismatch between $u_1$ and $v_1$ has occurred.
	This graph is again only exponentially large with polynomially representable vertices and an adjacency relation decidable in polynomial space, so the same argument applies.
	Thus, every condition in \Cref{characterize:words:fixed} can be verified in $\PSpace$, and the theorem follows.
\end{proof}

We now turn to the run-level task.
For limit aggregators, knowing that a value is realizable is not enough; we must also control its multiplicity.
The first step is to isolate, for a fixed value $x$, exactly those runs of value $x$.
The next construction produces a B\"uchi automaton with a stronger guarantee than language equivalence: over every word, its accepting runs are in bijection with the runs of value $x$ in the original automaton.

\begin{lem}\label{exact:value:construction}
	Consider a QWA $\A = (g,f,\T)$ over the alphabet $\Sigma$, with arbitrary word aggregator $g$ and $f \in \{\Inf$, $\Sup$, $\LimInf$, $\LimSup\}$.
	Given $x \in \QQ$, we can construct in polynomial time a B\"uchi automaton $\B_x$ such that for every word $w \in \Sigma^\omega$, there is a bijection between the accepting runs of $\B_x$ over $w$ and the runs $\rho \in R(\T,w)$ with $f(\gamma(\rho)) = x$.
	In particular, for every word $w$, the automaton $\B_x$ has infinitely many accepting runs over $w$ iff $\A$ has infinitely many runs of value $x$ over $w$.
\end{lem}
\begin{proof}
	A transition $\trans{p}{a:y}{q}$ of $\T$ is \emph{$x$-exact} if $y = x$, and it is \emph{$x$-safe} if $y \leq x$ when $f \in \{\Sup,\LimSup\}$, and $y \geq x$ when $f \in \{\Inf,\LimInf\}$; moreover, when $f \in \{\LimSup,\LimInf\}$, it is \emph{$x$-bad} if it is not $x$-safe.

	First, we describe the construction when $f \in \{\Sup,\Inf\}$.
	A run $\rho$ of $\T$ has value $x$ iff every transition of $\rho$ is $x$-safe and at least one transition of $\rho$ is $x$-exact.

	We construct a B\"uchi automaton $\B_x$ with state set $\{s_I\} \cup (Q \times \{0,1\})$, initial state $s_I$, and accepting states $Q \times \{1\}$.
	The bit records whether an $x$-exact transition has already been seen.
	The transitions are defined as follows:
	\begin{itemize}
		\item from $s_I$, for every $x$-safe transition $\trans{q_I}{a:y}{q}$ of $\T$, add one transition on $a$ to $(q,b)$, where $b=1$ iff $y=x$;
		\item from $(p,b)$, for every $x$-safe transition $\trans{p}{a:y}{q}$ of $\T$, add one transition on $a$ to $(q,b')$, where $b'=1$ iff $b=1$ or $y=x$.
	\end{itemize}
	There are no other transitions.

	Second, we describe the construction when $f \in \{\LimSup,\LimInf\}$.
	Since $\T$ has only finitely many transition weights, a run $\rho$ of $\T$ has value $x$ iff $\rho$ contains only finitely many $x$-bad transitions and infinitely many $x$-exact transitions.

	We construct a B\"uchi automaton $\B_x$ with state set $\{s_I\} \cup \bigl(Q \times \{\mathrm{pre},\mathrm{post}_0,\mathrm{post}_1\}\bigr)$, initial state $s_I$, and accepting states $Q \times \{\mathrm{post}_1\}$.
	Intuitively, mode $\mathrm{pre}$ means that we have not yet committed to being past the last $x$-bad transition; mode $\mathrm{post}_0$ means that we have committed and the most recent transition is not $x$-exact; mode $\mathrm{post}_1$ means that we have committed and the most recent transition is $x$-exact.

	The transitions are defined as follows:
	\begin{itemize}
		\item from $s_I$, for every transition $\trans{q_I}{a:y}{q}$ of $\T$, add a transition on $a$ to $(q,\mathrm{pre})$; additionally, add a transition on $a$ to $(q,\mathrm{post}_1)$ if $y = x$, and to $(q,\mathrm{post}_0)$ if $y \neq x$;
		\item from $(p,\mathrm{pre})$, for every transition $\trans{p}{a:y}{q}$ of $\T$, add a transition on $a$ to $(q,\mathrm{pre})$; additionally, if the transition is $x$-bad, add a transition on $a$ to $(q,\mathrm{post}_0)$;
		\item from $(p,m)$ with $m \in \{\mathrm{post}_0,\mathrm{post}_1\}$, for every $x$-safe transition $\trans{p}{a:y}{q}$ of $\T$, add a transition on $a$ to $(q,\mathrm{post}_1)$ if $y = x$, and to $(q,\mathrm{post}_0)$ otherwise.
	\end{itemize}
	There are no other transitions.

	In both cases, $\B_x$ is constructible in polynomial time and has size $O(|Q|)$.

	We now prove correctness.
	Since $\T$ has no parallel transitions with the same source, letter, and target, once we know two consecutive $Q$-components in a run of $\B_x$ and the input letter, the corresponding transition of $\T$ is uniquely determined.
	Hence, every run of $\B_x$ over a word $w$ has a well-defined projection to a run of $\T$ over $w$, obtained by erasing the extra mode or the bit component.

	First, let $\pi$ be an accepting run of $\B_x$, and let $\rho_\pi$ be its projection.
	If $f \in \{\Sup,\Inf\}$, every transition of $\pi$ follows an $x$-safe transition of $\T$, and since the accepting bit $1$ is absorbing, acceptance means that some projected transition is $x$-exact.
	Thus, $f(\gamma(\rho_\pi)) = x$.
	If $f \in \{\LimSup,\LimInf\}$, the run $\pi$ cannot stay in mode $\mathrm{pre}$ forever, so from some point on it remains in post modes.
	By construction, every transition read in a post mode is $x$-safe; hence $\rho_\pi$ has only finitely many $x$-bad transitions.
	Moreover, $\pi$ visits $\mathrm{post}_1$ infinitely often, so $\rho_\pi$ has infinitely many $x$-exact transitions.
	Thus again $f(\gamma(\rho_\pi)) = x$.

	Conversely, let $\rho$ be a run of $\T$ over $w$ with value $x$.
	If $f \in \{\Sup,\Inf\}$, $\rho$ uses only $x$-safe transitions and contains at least one $x$-exact transition.
	There is therefore exactly one run of $\B_x$ compatible with $\rho$: its bit is $0$ exactly before the first $x$-exact transition and $1$ from then on.
	This run is accepting.
	If $f \in \{\LimSup,\LimInf\}$, and if $\rho$ has no $x$-bad transition, then there is exactly one compatible accepting run of $\B_x$, namely the one that leaves $s_I$ directly to the appropriate post mode and then follows the unique compatible post transitions.
	If $\rho$ has a last $x$-bad transition at position $k$, then there is again exactly one compatible accepting run: it stays in mode $\mathrm{pre}$ up to position $k-1$, switches to $\mathrm{post}_0$ while reading the $k$th transition, and afterwards follows the unique compatible post transitions.
	Since $\rho$ has infinitely many $x$-exact transitions, this run visits $\mathrm{post}_1$ infinitely often and is therefore accepting.

	Thus, for every word $w$, projection yields a bijection between the accepting runs of $\B_x$ over $w$ and the runs $\rho \in R(\T,w)$ with $f(\gamma(\rho))=x$.
\end{proof}

Once exact-value runs are encoded by a B\"uchi automaton, the remaining question is automata-theoretic: for which words does a nondeterministic B\"uchi automaton admit infinitely many accepting runs?
The next lemma answers this by constructing an alternating B\"uchi automaton recognizing exactly those words.
The construction uses a designated ``spine'' branch together with repeatedly spawned side branches that certify the existence of infinitely many distinct accepting runs.

\begin{lem}\label{inf:acc:runs}
	Let $\B = (Q, \Sigma, q_I, \delta, F)$ be a nondeterministic Büchi automaton.
	We can construct in polynomial time an alternating Büchi automaton
	$\mathsf{InfAcc}(\B) = (Q', \Sigma, q_I', \delta', F')$ such that, for every word $w \in \Sigma^\omega$, we have $w \in L(\mathsf{InfAcc}(\B))$ iff $\B$ has infinitely many accepting runs on $w$.
\end{lem}
\begin{proof}
	Let $Q^x = \{q^x : q \in Q\}$ for $x \in \{\mathrm{sp}, \mathrm{mk}, \mathrm{acc}\}$, and let $F^{\mathrm{acc}} = \{q^{\mathrm{acc}} : q \in F\}$.
	We define $Q' = Q^{\mathrm{sp}} \uplus Q^{\mathrm{mk}} \uplus Q^{\mathrm{acc}}$, $q_I' = q_I^{\mathrm{sp}}$, and $F' = Q^{\mathrm{mk}} \cup F^{\mathrm{acc}}$.
	For $q \in Q$ and $a \in \Sigma$, set
	\[
		\delta'(q^{\mathrm{acc}}, a) = \bigvee_{r \in \delta(q,a)} r^{\mathrm{acc}} \quad\text{and}\quad
		\delta'(q^{\mathrm{sp}}, a) = \delta'(q^{\mathrm{mk}}, a) =
		\left(\bigvee_{p \in \delta(q,a)} p^{\mathrm{sp}}\right)
		\vee
		\left(\bigvee_{\substack{p,r \in \delta(q,a) \\ p \neq r}} (p^{\mathrm{mk}} \wedge r^{\mathrm{acc}})\right).
		\]
	Thus, $\mathsf{InfAcc}(\B)$ has $3|Q|$ states, and its transition formulas have polynomial size.

	Intuitively, the automaton maintains a \emph{spine} run in $Q^{\mathrm{sp}} \cup Q^{\mathrm{mk}}$, and uses $Q^{\mathrm{acc}}$ to spawn side threads that certify accepting runs branching off the spine.
	At each spine step, the transition either advances to a single successor $p^{\mathrm{sp}}$ (no split), or simultaneously advances to $p^{\mathrm{mk}}$ and spawns a side thread from a \emph{different} successor $r^{\mathrm{acc}}$ (a split).
	Since $Q^{\mathrm{mk}} \subseteq F'$ and $Q^{\mathrm{sp}} \cap F' = \varnothing$, the Büchi condition forces the spine to split infinitely often.
	Each side thread lives in $Q^{\mathrm{acc}}$, where transitions are pure disjunctions, so the thread follows a single run of $\B$; the condition $F^{\mathrm{acc}} \subseteq F'$ then forces that run to be accepting.

	Consider a word $w = a_0 a_1 a_2 \cdots \in \Sigma^\omega$.
	We prove below $w \in L(\mathsf{InfAcc}(\B))$ iff $\B$ has infinitely many accepting runs on $w$.

	\smallskip\noindent\emph{($\Rightarrow$)}
	Assume that $\mathsf{InfAcc}(\B)$ has an accepting run tree on $w$.
	Since all transition formulas are positive boolean formulas, any set of children satisfying a formula has an inclusion-minimal subset that still satisfies it.
	We may therefore prune children at each node to such a minimal set; since this only removes branches entirely without altering the states on surviving branches, Büchi acceptance is preserved.
	We thus obtain an accepting run tree $T$ that is inclusion-minimal at every node.

	Now, every node of $T$ labeled in $Q^{\mathrm{acc}}$ has exactly one child, again in $Q^{\mathrm{acc}}$: the formula $\delta'(q^{\mathrm{acc}}, a)$ is a disjunction of single states, so a minimal satisfying set is a singleton.
	Likewise, every node labeled in $Q^{\mathrm{sp}} \cup Q^{\mathrm{mk}}$ has exactly one child in $Q^{\mathrm{sp}} \cup Q^{\mathrm{mk}}$, and possibly one additional child in $Q^{\mathrm{acc}}$: a minimal satisfying set for
	$\left(\bigvee_{p \in \delta(q,a)} p^{\mathrm{sp}}\right) \vee \left(\bigvee_{\substack{p,r \in \delta(q,a) \\ p \neq r}} (p^{\mathrm{mk}} \wedge r^{\mathrm{acc}})\right)$
	is either $\{p^{\mathrm{sp}}\}$ or $\{p^{\mathrm{mk}}, r^{\mathrm{acc}}\}$.

	Hence, there is a unique infinite branch of $T$ that stays inside $Q^{\mathrm{sp}} \cup Q^{\mathrm{mk}}$; we call it the \emph{spine}.
	Projecting it to $Q$ yields a run $\rho = q_0 \xrightarrow{a_0} q_1 \xrightarrow{a_1} q_2 \xrightarrow{a_2} \cdots$ of $\B$ on $w$.

	The spine never enters $Q^{\mathrm{acc}}$, and $Q^{\mathrm{sp}} \cap F' = \varnothing$, so for $T$ to be accepting the spine must visit $Q^{\mathrm{mk}}$ infinitely often.
	So, for infinitely many $i$, the transition chosen at depth $i$ on the spine is of the form $q_{i+1}^{\mathrm{mk}} \wedge r_{i+1}^{\mathrm{acc}}$ with $r_{i+1} \in \delta(q_i, a_i) \setminus \{q_{i+1}\}$.

	Consider such an $i$.
	The branch starting from $r_{i+1}^{\mathrm{acc}}$ stays forever in $Q^{\mathrm{acc}}$, and because $T$ is accepting, it visits $F^{\mathrm{acc}}$ infinitely often.
	Its projection is therefore an accepting run of $\B$ on the suffix $a_{i+1} a_{i+2} \cdots$, starting from $r_{i+1}$.
	Consequently, $q_0 \xrightarrow{a_0} \cdots \xrightarrow{a_{i-1}} q_i \xrightarrow{a_i} r_{i+1} \xrightarrow{a_{i+1}} \cdots$ is an accepting run of $\B$ on $w$.

	If $i < j$ are two such positions, the run obtained from $i$ diverges from $\rho$ at step $i+1$, while the run obtained from $j$ agrees with $\rho$ through step $j$, so the two runs are distinct.
	Since the spine visits $Q^{\mathrm{mk}}$ infinitely often, $\B$ has infinitely many accepting runs on~$w$.

	\smallskip\noindent\emph{($\Leftarrow$)}
	Assume that $\B$ has infinitely many accepting runs on $w$.
	A finite run prefix $q_0 \xrightarrow{a_0} \cdots \xrightarrow{a_{n-1}} q_n$ is \emph{prolific} if it has infinitely many accepting extensions on $w$.
	Note that the empty prefix is prolific.
	Since $\B$ is finitely branching, every prolific prefix has a prolific one-step extension: otherwise each immediate successor would admit only finitely many accepting extensions, and hence so would the prefix itself.

	By recursion, we obtain a run $\rho = q_0 \xrightarrow{a_0} q_1 \xrightarrow{a_1} q_2 \xrightarrow{a_2} \cdots$ such that every finite prefix of $\rho$ is prolific.
	Note that $\rho$ itself need not be accepting.

	A position $i$ is a \emph{split position} if there exists $r_{i+1} \in \delta(q_i, a_i) \setminus \{q_{i+1}\}$ such that the prefix $q_0 \xrightarrow{a_0} \cdots \xrightarrow{a_{i-1}} q_i \xrightarrow{a_i} r_{i+1}$ has an accepting extension on the suffix $a_{i+1} a_{i+2} \cdots$.
	We claim that there are infinitely many split positions.
	Suppose towards contradiction, and choose a position $N$ larger than all of them.
	Then, every accepting extension of $q_0 \xrightarrow{a_0} \cdots \xrightarrow{a_{N-1}} q_N$ must continue with $q_{N+1}$, since otherwise $N$ would be a split position.
	The same argument applies at $N+1, N+2, \ldots$: every accepting extension is forced to follow $\rho$ forever.
	Hence, that prefix has at most one accepting extension, contradicting its prolificness.

	For each split position $i$, fix one witness $r_{i+1} \in \delta(q_i, a_i) \setminus \{q_{i+1}\}$ and fix one accepting run of $\B$ on the suffix $a_{i+1} a_{i+2} \cdots$ starting from $r_{i+1}$.
	We now define a run tree of $\mathsf{InfAcc}(\B)$ on $w$ whose spine branch follows $\rho$.
	At a non-split position $i$, we choose the disjunct $q_{i+1}^{\mathrm{sp}}$.
	At a split position $i$, we choose the conjunct $q_{i+1}^{\mathrm{mk}} \wedge r_{i+1}^{\mathrm{acc}}$, and the spawned $Q^{\mathrm{acc}}$-branch follows the fixed accepting run from $r_{i+1}$.

	This is a valid run tree.
	Its spine branch stays in $Q^{\mathrm{sp}} \cup Q^{\mathrm{mk}}$ and visits $Q^{\mathrm{mk}}$ infinitely often because there are infinitely many split positions.
	Every other infinite branch eventually stays in $Q^{\mathrm{acc}}$ and follows one of the fixed accepting runs, hence visits $F^{\mathrm{acc}}$ infinitely often.
	Therefore, every infinite branch is accepting, so the run tree is accepting, and thus $w \in L(\mathsf{InfAcc}(\B))$.
\end{proof}

Combining the previous two lemmas yields the exact-value languages of the original QWA.
For each $x$, we can recognize both the words admitting at least one run of value $x$ and those admitting infinitely many.
The four word aggregators are then expressed as finite boolean combinations of these languages, yielding polynomial-size alternating parity automata for exact-value languages.
This in turn gives polynomial-space emptiness and infiniteness procedures for their intersections with a given alternating parity automaton.

\begin{lem}\label{limit:exact:value:languages}
	Consider a QWA $\A = (g, f, \T)$ over $\Sigma$ with $f, g \in \{\Inf$, $\Sup$, $\LimInf$, $\LimSup\}$.
	Let $X_\A$ be the set of transition weights of $\A$.
	For each $x \in X_\A$, define $L_x^\A = \{w \in \Sigma^\omega \mid \A(w) = x\}$.
	Then, the following hold.
	\begin{enumerate}[(i)]
		\item For every word $w \in \Sigma^\omega$, we have $\A(w) \in X_\A$.
		\item For every $x \in X_\A$, there exists an alternating parity automaton $\mathcal{P}_x^\A$ of size polynomial in $|\A|$ such that $L(\mathcal{P}_x^\A) = L_x^\A$.
		\item Given $x \in X_\A$ and an alternating parity automaton $\C$ over $\Sigma$, one can decide in space polynomial in $|\A| + |\C|$ whether $L(\C) \cap L_x^\A$ is empty, and whether it is infinite.
	\end{enumerate}
\end{lem}
\begin{proof}
	For $w \in \Sigma^\omega$, write $M_w = M_{w,\T,f}$ for the multiset of values obtained by applying $f$ on the runs of $\T$ over $w$.
	For each $x \in X_\A$, let
	\[
		E_x = \{w \in \Sigma^\omega \mid x \in \textit{Supp}(M_w)\}
		\qquad\text{and}\qquad
		I_x = \{w \in \Sigma^\omega \mid x \in \textit{InfSupp}(M_w)\}.
	\]
	By \Cref{exact:value:construction}, for every $x \in X_\A$ there is a polynomial-size Büchi automaton $\mathcal{R}_x$ with $L(\mathcal{R}_x) = E_x$,
	whose accepting runs over $w$ are in bijection with the runs of $\T$ over $w$ of $f$-value $x$.
	Applying \Cref{inf:acc:runs} to $\mathcal{R}_x$, we obtain a polynomial-size alternating Büchi automaton $\mathcal{S}_x$ with $L(\mathcal{S}_x) = I_x$.

	\smallskip\noindent\emph{(i)}
	Every run of $\T$ uses only weights from $X_\A$.
	Since $X_\A$ is finite, for every run $\rho$ the value $f(\gamma(\rho))$ is one of the weights occurring along $\rho$.
	Hence, for every word $w$ we have $\textit{Supp}(M_w) \subseteq X_\A$ and $\textit{InfSupp}(M_w) \subseteq X_\A$.

	Since $\T$ is complete, every word admits at least one run, so $\textit{Supp}(M_w) \neq \emptyset$.
	If $g \in \{\Inf, \Sup\}$, then $\A(w)$ is the minimum or maximum of the nonempty set $\textit{Supp}(M_w)$, hence $\A(w) \in X_\A$.
	If $g = \LimSup$, then either $\textit{InfSupp}(M_w) \neq \emptyset$, in which case $\A(w) = \sup \textit{InfSupp}(M_w) \in X_\A$,
	or $\textit{InfSupp}(M_w) = \emptyset$, in which case $\A(w) = \beta_\A := \bot_{\A^{(g \gets \Sup)}}$.
	The range of $\A^{(g \gets \Sup)}$ is a nonempty finite subset of $X_\A$, so $\beta_\A \in X_\A$.
	The case $g = \LimInf$ is dual: if $\textit{InfSupp}(M_w) = \emptyset$, then $\A(w) = \tau_\A := \top_{\A^{(g \gets \Inf)}}$,
	and again $\tau_\A \in X_\A$.
	Thus $\A(w) \in X_\A$ for every $w \in \Sigma^\omega$.

	\smallskip\noindent\emph{(ii)}
	By Item~(i), the default values $\beta_\A$ and $\tau_\A$ belong to $X_\A$.
	By the definitions of the four word aggregators, the exact-value language of $\A$ is given by the following finite boolean combinations of the languages $E_x$ and $I_x$:
	\[
		L_x^\A =
		\begin{cases}
			E_x \cap \bigcap_{\substack{y \in X_\A \\ y > x}} E_y^\complement,
			& \text{if } g = \Sup, \\[2mm]
			E_x \cap \bigcap_{\substack{y \in X_\A \\ y < x}} E_y^\complement,
			& \text{if } g = \Inf, \\[2mm]
			I_x \cap \bigcap_{\substack{y \in X_\A \\ y > x}} I_y^\complement,
			& \text{if } g = \LimSup \text{ and } x \neq \beta_\A, \\[2mm]
			\left(I_x \cap \bigcap_{\substack{y \in X_\A \\ y > x}} I_y^\complement\right)
			\cup
			\bigcap_{y \in X_\A} I_y^\complement,
			& \text{if } g = \LimSup \text{ and } x = \beta_\A, \\[2mm]
			I_x \cap \bigcap_{\substack{y \in X_\A \\ y < x}} I_y^\complement,
			& \text{if } g = \LimInf \text{ and } x \neq \tau_\A, \\[2mm]
			\left(I_x \cap \bigcap_{\substack{y \in X_\A \\ y < x}} I_y^\complement\right)
			\cup
			\bigcap_{y \in X_\A} I_y^\complement,
			& \text{if } g = \LimInf \text{ and } x = \tau_\A.
		\end{cases}
	\]
	For $g = \Sup$ and $g = \Inf$, this says exactly that $x$ is realized and no strictly better value is realized.
	For $g = \LimSup$ and $g = \LimInf$, the same characterization applies to $\textit{InfSupp}(M_w)$;
	the additional disjunct covers the case $\textit{InfSupp}(M_w) = \emptyset$, using that $\textit{InfSupp}(M_w) \subseteq X_\A$.

	Each language $E_x$ is recognized by a Büchi automaton, and each language $I_x$ by an alternating Büchi automaton.
	Since alternating parity automata are closed under finite boolean operations, and $|X_\A|$ is polynomial in $|\A|$,
	every displayed language is recognized by an alternating parity automaton of size polynomial in $|\A|$, which we denote by $\mathcal{P}_x^\A$.

	\smallskip\noindent\emph{(iii)}
	Let $x \in X_\A$ and consider an alternating parity automaton $\C$ over $\Sigma$.
	In the limit cases, the formulas above depend on whether $x$ is the default value, which we can determine in \PSpace as follows.

	If $g = \LimSup$, define $K_y^{\sup} = E_y \cap \bigcap_{\substack{z \in X_\A \\ z > y}} E_z^\complement$ for each $y \in X_\A$.
	Then, $x = \beta_\A$ iff $K_x^{\sup} \neq \emptyset$ and $K_y^{\sup} = \emptyset$ for every $y < x$.
	Dually, if $g = \LimInf$, define $K_y^{\inf} = E_y \cap \bigcap_{\substack{z \in X_\A \\ z < y}} E_z^\complement$ for each $y \in X_\A$.
	Then, $x = \tau_\A$ iff $K_x^{\inf} \neq \emptyset$ and $K_y^{\inf} = \emptyset$ for every $y > x$.
	Each $K_y^{\sup}$ and $K_y^{\inf}$ is a finite boolean combination of the Büchi languages $E_z$, hence recognized by a polynomial-size alternating parity automaton, whose emptiness is decidable in \PSpace by the argument below.

	Once the correct case is fixed, we intersect the corresponding automaton for $L_x^\A$ with $\C$	and obtain an alternating parity automaton $\D$ of size polynomial in $|\A| + |\C|$ with $L(\D) = L(\C) \cap L_x^\A$.
	By the standard alternation-removal construction~\cite{DBLP:journals/tcs/MiyanoH84,DBLP:journals/tcs/MullerS95}, $\D$ can be translated on the fly into an equivalent nondeterministic Büchi automaton $\N$ of exponential size.
	Crucially, each state of $\N$ is representable by polynomially many bits,	and its transition relation and acceptance condition are computable in space polynomial in $|\A| + |\C|$.
	Hence, emptiness of $L(\D)$ is decidable in \PSpace by an on-the-fly emptiness check on $\N$, and infiniteness of $L(\D)$ is decidable in \PSpace by the corresponding on-the-fly infiniteness test for Büchi automata as in the proof of \Cref{main:words}.
	Therefore, one can decide in space polynomial in $|\A| + |\C|$ whether $L(\C) \cap L_x^\A$ is empty, and whether it is infinite.
\end{proof}

\subsection{Expressive Power of Limit Aggregators}

We begin with a preprocessing observation on run aggregators.
The classical QWA constructions replacing $\Inf$ or $\Sup$ by a limit aggregator~\cite{DBLP:journals/tocl/ChatterjeeDH10,DBLP:conf/concur/BokerHMS23,DBLP:journals/lmcs/BokerHMS25} lift to QLAs, as they preserve not only word values but also run multiplicities.
This stronger preservation is essential when the semantics depends on infinite support.
We then compare expressive power, first at the word level and then at the language level.

\begin{prop}\label{all:inf:sup}
	Consider a QLA $\AA = (h,(g,f,\T))$ with $f\in\{\Inf$, $\Sup\}$, $g \in \{\Inf$, $\Sup$, $\LimInf$, $\LimSup\}$, and $h$ any language aggregator function.
	We can construct in \PTime a QLA $\BB = (h,(g,f',\T'))$ with $f' \in \{\LimInf$, $\LimSup\}$ such that $\AA(S) = \BB(S)$ for all $S\subseteq \Sigma^\omega$.
\end{prop}
\begin{proof}
	A careful observation shows that the constructions of~\cite{DBLP:journals/tocl/ChatterjeeDH10,DBLP:conf/concur/BokerHMS23,DBLP:journals/lmcs/BokerHMS25} allowing to convert a word automaton to a $\LimSup$ one extends to language automata.
	This proof provides the correctness of the construction from~\cite{DBLP:conf/concur/BokerHMS23,DBLP:journals/lmcs/BokerHMS25} in the context of language automata.

	Consider a language automaton $\AA = (h,(g,\Sup,\T))$ where $\T = (\Sigma, Q, q_I, \delta)$ with $\gamma$ the weight function of $\T$, the word aggregator function $g\in\{\Inf$, $\Sup$, $\LimInf$, $\LimSup\}$, and $h$ can be any language aggregator function.
	Let $\A = (g,\Sup,\T)$.
	The idea is to construct an equivalent word automaton $\A' = (g,f',\T')$ where $f' \in \{\LimInf, \LimSup\}$ and transition system $\T'$ memorizes the maximal visited weight, and thus be used to construct a language automaton where runs are aggregated indifferently with $\Sup$, $\LimInf$, or $\LimSup$.
	Let $X$ be the set of weights of $\A$.
	Since $|X| < \infty$, we can fix the minimal weight $X_0 = \min X$.
	We construct $\T' = (\Sigma, Q \times X, (q_I, X_0), \delta')$ where $\trans{(q_1,v)}{a:x}{(q_2,x)}$ is a transition in $\T'$ iff $\trans{q_1}{a:v'}{q_2}$ in $\T$ where $x = \max\{v, v'\}$.
	The construction of $\T'$ can be done in \PTime in the size of $|\T|$.
	Observe that, by construction, every run $\pi$ of $\T'$ yields a non-decreasing weight sequence for which there exists $i \in \NN$ such that for all $j \geq i$ we have $\gamma(\pi[i]) = \gamma(\pi[j]) = \sup(\gamma(\pi))$.
	Therefore, $\A$ and $\A'$ are equivalent regardless of the choice of $f' \in \{\LimInf, \LimSup\}$.
	Moreover, again by construction, there is a bijection between their runs, implying that for all $w\in\Sigma^\omega$ the number of runs over $w$ is the same in $\A$ and $\A'$.
	Now, let $\AA' = (h,\A')$ and observe that since $\A$ and $\A'$ are equivalent, so are $\AA$ and $\AA'$.
	The construction for a given automaton with $f=\Inf$ is dual as it consists in memorizing the minimal visited weight, therefore the weight sequences are non-increasing.
\end{proof}

\begin{rem}
	The preceding proposition relies on a multiplicity-preserving compilation: for each word, runs of the original and compiled automata are in bijection and share the same value.
	This is why the construction carries over from QWAs to QLAs even when the outer semantics depends on infinite support.
	This contrasts with the classical translation of nondeterministic $\LimInf$ QWAs into nondeterministic $\LimSup$ QWAs, and dually of universal $\LimSup$ into universal $\LimInf$ QWAs~\cite{DBLP:journals/tocl/ChatterjeeDH10}.
	Those translations preserve the pointwise word value but need not preserve run multiplicities.
	Hence they extend to QLAs with word aggregator $\Sup$ (dually, $\Inf$), but not to those with word aggregator $\LimSup$ (dually, $\LimInf$), where infinite multiplicity is semantically significant.
	This is consistent with the expressive incomparability of deterministic $\LimInf$ and $\LimSup$ QWAs~\cite{DBLP:journals/tocl/ChatterjeeDH10}.
\end{rem}

We now compare the word aggregators, beginning by showing that the limit variants can be simulated by the ordinary ones.
For each value $x$ in the finite range of the QWA, we construct an automaton outputting $x$ exactly on words of value $x$ and a uniform fallback otherwise.
Taking the disjoint union of these automata yields a single QWA whose $\Sup$ word aggregator recovers the original $\LimSup$ semantics.
The $\LimInf$ case follows by duality.

\begin{lem}\label{limit:reduction:g}
	Consider a QLA $\AA = (h,(g,f,\T))$ with $f \in \{\Inf$, $\Sup$, $\LimInf$, $\LimSup\}$, $g = \LimSup$ (resp.\ $g = \LimInf$), and $h$ any language aggregator.
	Then, we can effectively construct a QLA $\BB = (h,(g',f',\T'))$ with $g' = \Sup$ and $f' = \LimSup$ (resp.\ $g' = \Inf$ and $f' = \LimInf$) such that $\AA(S) = \BB(S)$ for every $S \subseteq \Sigma^\omega$.
\end{lem}
\begin{proof}
	It suffices to construct a pointwise equivalent QWA.

	Suppose $g = \LimSup$.
	Let $\A = (\LimSup, f, \T)$ be the QWA underlying $\AA$.
	Let $X_\A$ be the set of weights of $\A$ and $L_x^\A = \{ w \in \Sigma^\omega \mid \A(w) = x \}$ for $x \in X_\A$ be as in \Cref{limit:exact:value:languages}.
	Then, $X_\A$ is finite, every value of $\A$ belongs to $X_\A$, and the family $\{L_x^\A\}_{x \in X_\A}$ is pairwise disjoint and covers $\Sigma^\omega$.
	Let $m = \min X_\A$.

	For each $x \in X_\A$, the language $L_x^\A$ is $\omega$-regular, so let $\N_x = (Q_x, \Sigma, q_x, \delta_x, F_x)$ be a complete nondeterministic Büchi automaton recognizing $L_x^\A$.
	From $\N_x$, we build a QWA $\C_x = (\Sup, \LimSup, \T_x)$ as follows:
	$\T_x$ is the transition graph of $\N_x$, every transition entering a state of $F_x$ receives weight $x$, and every other transition receives weight $m$.
	Then, for every $w \in \Sigma^\omega$ we have $\C_x(w) = x$ if $w \in L_x^\A$, and $\C_x(w) = m$ if $w \notin L_x^\A$.
	
	Now, construct a QWA $\B = (\Sup, \LimSup, \T')$ by taking the disjoint union of all $\T_x$ ($x \in X_\A$) and adding a fresh initial state $\hat{q}$.
	For each $a \in \Sigma$, the outgoing $a$-transitions of $\hat{q}$ are the union of the outgoing $a$-transitions of the initial states $q_x$, with the same weights and targets.
	Every run of $\B$ therefore selects one component on the first step and remains inside it, reproducing exactly the weight sequence of the corresponding run in that component.
	Consequently, $\B(w) = \max_{x \in X_\A} \C_x(w)$ for every $w \in \Sigma^\omega$.
	For any $w \in \Sigma^\omega$, let $x_w \in X_\A$ be the unique index with $w \in L_{x_w}^\A$.
	Then, $\C_{x_w}(w) = x_w$, while $\C_x(w) = m \leq x_w$ for every $x \neq x_w$, so $\B(w) = \max_{x \in X_\A} \C_x(w) = x_w = \A(w)$.
	Hence, $\A$ and $\B$ are pointwise equivalent.

	Now, suppose $g = \LimInf$.
	Let $\A = (\LimInf, f, \T)$ and consider its dual QWA $\hat{\A}$, which has word aggregator $\LimSup$.
	By the previous case there exists a QWA $\C = (\Sup, \LimSup, \T'')$ with $\C(w) = \hat{\A}(w)$ for every $w \in \Sigma^\omega$.
	Set $\B = \hat{\C}$; then $\B$ has word aggregator $\Inf$ and run aggregator $\LimInf$, and for every $w \in \Sigma^\omega$, we have $\B(w) = {-}\C(w) = {-}\hat{\A}(w) = \A(w)$.
	So, $\A$ and $\B$ are pointwise equivalent.
	
	Finally, set $\BB = (h, \B)$.
	Since the underlying QWAs of $\AA$ and $\BB$ agree on every word, the two QLAs agree on every language, and the construction is effective.
\end{proof}

For the converse, starting from a QWA with word aggregator $\Sup$ (dually, $\Inf$), we duplicate the transition system and allow each run to switch once from the first copy to the second.
Every original run thereby induces infinitely many distinct runs with the same weight sequence, one per switch point.
Hence every realizable value becomes infinitely realizable, so replacing $\Sup$ by $\LimSup$ (dually, $\Inf$ by $\LimInf$) preserves the value of every word and thus every induced QLA.

\begin{lem}\label{classic:reduction:g}
	Let $\AA = (h,(g,f,\T))$ be a QLA with $f$ any run aggregator function, $g = \Sup$ (resp. $g = \Inf$), and $h$ any language aggregator function.
	We can construct in \PTime a QLA $\BB = (h,(g',f,\T'))$ where $g' = \LimSup$ (resp. $g' = \LimInf$)
	such that $\AA(S) = \BB(S)$ for all $S\subseteq \Sigma^\omega$.
\end{lem}
\begin{proof}
	The transition system $\T'$ consists of two copies of $\T$ that are connected by nondeterministic transitions, allowing for every run in $\AA$ infinitely many runs in $\BB$.
	In particular, for every transition $\trans{q}{\sigma:x}{p}$ in $\T$, there are three transitions in $\T'$ as follows: $\trans{q}{\sigma:x}{p}$ copying the original transition, $\trans{q}{\sigma:x}{p'}$ allowing to jump from the first copy of the transition system to the second, and $\trans{q'}{\sigma:x}{p'}$ allowing to stay in the second copy.
	Evidently, for every run $\rho$ in $\T$ and every $i \in \NN$, there is a run $\rho_i'$ in $\T'$ such that $\rho_i'$ imitates $\rho$ in the first copy in the first $i$ transitions and in the second copy afterwards.
	Moreover, the values of $\rho_i'$ and $\rho$ coincide as they yield the same weight sequences.
	Since the word automata $(g,f,\T)$ and $(g',f,\T')$ are equivalent, so are $\AA$ and $\BB$.
\end{proof}

The previous two lemmas show that, for the run-aggregator fragment $\{\Inf,\Sup,\LimInf,\LimSup\}$, the non-limit and limit word aggregators are equally expressive in each polarity.

\begin{cor}\label{thm:word:equiexpressive}
	For QLAs with run aggregators $f \in \{ \Inf$, $\Sup$, $\LimInf$, $\LimSup \}$, the word aggregators $\Inf$ and $\LimInf$ (resp. $\Sup$ and $\LimSup$) are equally expressive.
\end{cor}

We now turn to language aggregators, where the situation is different: $\LimInf$ and $\LimSup$ depend on infinite support across words and collapse to a default on finite languages.
On one hand, a non-limit language aggregator distinguishes singleton languages by their unique word's value, whereas a limit aggregator assigns every singleton the default.
On the other hand, a limit aggregator can exploit a value being realized by infinitely many words, even though every singleton sees only the default.
These two features yield expressive incomparability.

\begin{prop}\label{limit:incomparable}
	QLAs with the language aggregators $\Inf$ and $\LimInf$ (resp. $\Sup$ and $\LimSup$) are expressively incomparable.
\end{prop}
\begin{proof}
	Let $\Sigma = \{a,b\}$.
	We prove each of the four separations.

	Consider a QLA $\AA = (\Inf,\A)$ over the value domain $\{0,1\}$ such that $\A(a^\omega) = 0$ and $\A(w) = 1$ for every $w\neq a^\omega$.
	Then, $\AA(\{a^\omega\}) = 0$ and $\AA(\{ba^\omega\}) = 1$.
	Assume that there exists a QLA $\BB = (\LimInf,\B)$ equivalent to $\AA$.
	For every word $w$, the language $\{w\}$ is finite, hence $\textit{InfSupp}(M_{\{w\},\B}) = \emptyset$.
	Therefore, $\BB(\{w\})$ is the default value of the language aggregator $\LimInf$, and so it is independent of $w$.
	This contradicts $\AA(\{a^\omega\}) \neq \AA(\{ba^\omega\})$.

	Consider now a QLA $\AA = (\LimInf,\A)$ over the value domain $\{0,1\}$ such that $\A(w) = 0$ iff $w \in \Sigma^* a^\omega$, and $\A(w) = 1$ otherwise.
	Let $U = \Sigma^* a^\omega$.
	Then, $U$ is infinite and $\AA(U) = 0$.
	Moreover, for every $u \in U$, the language $\{u\}$ is finite, so $\AA(\{u\})$ is the default value of the language aggregator $\LimInf$, namely $\top_{\A} = 1$.
	Assume that there exists a QLA $\BB = (\Inf,\B)$ equivalent to $\AA$.
	Then, for every $u \in U$, we have $\B(u) = \BB(\{u\}) = \AA(\{u\}) = 1$.
	Hence $\BB(U) = \inf_{u \in U} \B(u) = 1$, contradicting $\AA(U) = 0$.

	The remaining two separations ($\Sup$ versus $\LimSup$) follow by symmetric arguments: swap values $0$ and $1$, replace $\Inf$ by $\Sup$ and $\LimInf$ by $\LimSup$, and note that the default of $\LimSup$ on singletons is $\bot_{\A}$ rather than $\top_{\A}$.
\end{proof}

\subsection{Evaluation with Limit Aggregators}

We begin with the evaluation problem for QLAs $\AA = (h, (g, f, \T))$ whose run, word, and language aggregators all belong to $\{\Inf, \Sup, \LimInf, \LimSup\}$ and such that at least one of $g$ and $h$ is a limit aggregator.
The key observation is that every value of the underlying QWA $\A = (g, f, \T)$ lies in the finite set of its transition weights.
Hence, evaluation reduces to determining, for each transition weight $x$, whether the input $\omega$-regular language contains no words, finitely many words, or infinitely many words of value exactly $x$.
This is precisely what \Cref{limit:exact:value:languages} provides in polynomial space: for each $x$, it yields an automaton for the exact-value language $L_x^\A = \{w \in \Sigma^\omega \mid \A(w) = x\}$ together with polynomial-space procedures for testing emptiness and infiniteness after intersection with the input language.
Scanning the finitely many candidates then suffices to reconstruct the outer aggregation, including the default cases when the relevant support is empty.
This yields a uniform \PSpace upper bound for the whole fragment, complementing the non-limit cases $g, h \in \{\Inf, \Sup\}$ already settled in \Cref{evaluation:classic}.

\begin{thm}\label{classic:membership:reduction}
	Consider a QLA $\AA = (h,\A)$ with $\A = (g,f,\T)$ and $f,g,h \in \{\Inf$, $\Sup$, $\LimInf$, $\LimSup\}$, where at least one of $g$ and $h$ belongs to $\{\LimInf,\LimSup\}$.
	The evaluation of $\AA$ is in \PSpace.
\end{thm}
\begin{proof}
	Let $S$ be an $\omega$-regular language given as a B\"uchi automaton $\B_S$.
	Let $X_\A$ be the finite set of weights of $\A$, and note that every value of $\A$ belongs to $X_\A$.
	For each $x \in X_\A$, let $\mathcal{P}_x^\A$ be as in \Cref{limit:exact:value:languages}, satisfying $L(\mathcal{P}_x^\A) = L_x^\A$, defined as $\{w \in \Sigma^\omega \mid \A(w) = x\}$.

	For each $x \in X_\A$, let $S_x = S \cap L_x^\A$.
	By \Cref{limit:exact:value:languages}(iii), applied with $\C = \B_S$ viewed as an alternating parity automaton, we can decide in \PSpace whether $S_x$ is empty and whether $S_x$ is infinite.

	Now, let $\beta = \min \{x \in X_\A \mid L_x^\A \neq \emptyset\}$ and $\tau = \max \{x \in X_\A \mid L_x^\A \neq \emptyset\}$.
	Applying the same emptiness test with $\C$ equal to the one-state B\"uchi automaton for $\Sigma^\omega$ computes $\beta$ and $\tau$ in \PSpace.
	Hence, $\beta$ and $\tau$ are exactly the canonical defaults for the outer aggregators $\Sup,\LimSup$ and $\Inf,\LimInf$, respectively.
	
	Since the range of $\A$ is a nonempty finite subset of $X_\A$, we have $\beta = \bot_\A$ and $\tau = \top_\A$.
	For $h \in \{\LimSup,\LimInf\}$ these are exactly the canonical default values of the outer limit aggregator.
	For $h \in \{\Sup,\Inf\}$ these defaults are irrelevant.
	Therefore,
	\[
		\AA(S)=
		\begin{cases}
			\max\bigl(\textit{Supp}(M_{S,\A})\cup\{\beta\}\bigr) & \text{if } h=\Sup,\\[1mm]
			\min\bigl(\textit{Supp}(M_{S,\A})\cup\{\tau\}\bigr) & \text{if } h=\Inf,\\[1mm]
			\max\bigl(\textit{InfSupp}(M_{S,\A})\cup\{\beta\}\bigr) & \text{if } h=\LimSup,\\[1mm]
			\min\bigl(\textit{InfSupp}(M_{S,\A})\cup\{\tau\}\bigr) & \text{if } h=\LimInf.
		\end{cases}
	\]
	The extra singleton handles exactly the case in which the relevant support is empty.

	Since $X_\A$ is finite and polynomial in the size of $\A$, we can scan all values in $X_\A$ one by one and reuse the same polynomial workspace.
	While scanning, it suffices to maintain the current maximum or minimum over the relevant set, together with $\beta$ and $\tau$.
	Thus, the exact value $\AA(S)$ can be computed using polynomial space.
	The threshold problem follows immediately.
\end{proof}

We now turn to lower bounds.
For evaluation, hardness is witnessed already for very simple fixed input languages.
We first handle the mixed cases, where the word and language aggregators have opposite polarity (e.g., $h = \LimInf$ and $g = \LimSup$).
The starting point is the \PSpaceH universality problem for QWAs.
The reduction duplicates the alphabet so that every word witnessing the bottom value of the source automaton gives rise to infinitely many distinct words sharing that value.
Evaluating the resulting QLA on all words over the enlarged alphabet then lets an outer $\Inf$ or $\LimInf$ recover exactly that bottom value, while the inner supremum-type aggregator preserves each individual word's value.
The remaining mixed cases follow by the translations between $\Sup$ and $\LimSup$ and by duality.

\begin{lem}\label{lem:PSpace:swap}
	Consider a QLA $\AA = (h,(g,f,\T))$ over the alphabet $\Sigma$ with $f \in \{\Inf$, $\Sup$, $\LimInf$, $\LimSup\}$, at least one of $g$ and $h$ in $\{\LimInf,\LimSup\}$, and
	either (i) $g \in \{\Inf,\LimInf\}$ and $h \in \{\Sup,\LimSup\}$,
	or (ii) $g \in \{\Sup,\LimSup\}$ and $h \in \{\Inf,\LimInf\}$.
	Deciding whether $\AA(\Sigma^\omega) \geq k$ for a given threshold $k \in \QQ$ is \PSpaceH.
\end{lem}
\begin{proof}
	We prove the stronger statement for $g = \Sup$ and $h \in \{\Inf, \LimInf\}$.
	The corresponding cases with $g \in \{\Sup,\LimSup\}$ then follow from \Cref{classic:reduction:g}.
	The cases with $g \in \{\Inf,\LimInf\}$ and $h \in \{\Sup,\LimSup\}$ then follow by the duality property of \Cref{all:dual}, applied to the corresponding construction for the dual run aggregator $\hat{f}$.

	The proof goes by reduction from deciding whether $\bot_{\C} \geq 1$ for a QWA $\C=(\Sup, f ,\T)$ over weights $\{0, 1\}$.
	Since $\bot_{\C} \in \{0,1\}$, this is exactly the ${\geq}$-universality problem for threshold $1$, which is \PSpaceH for $f \in \{\Inf$, $\Sup$, $\LimInf$, $\LimSup\}$~\cite{DBLP:journals/tocl/ChatterjeeDH10,DBLP:conf/vmcai/KupfermanL07}.

	Let $\Sigma' = \{a' \st a \in \Sigma\}$ be a copy of $\Sigma$, and let $\Gamma = \Sigma \cup \Sigma'$.
	We transform $\C$ into $\C' = (\Sup, f, \T')$ by duplicating every transition label:
	for every transition $(p, a, x, q)$ in $\C$, the automaton $\C'$ contains both $(p, a, x, q)$ and $(p, a', x, q)$.
	For every word $w \in \Gamma^\omega$, let $\pi(w) \in \Sigma^\omega$ be obtained by replacing every letter $a'$ by $a$.
	By construction, there is a surjective mapping from the runs of $\C'$ over $w$ to runs of $\C$ over $\pi(w)$, and corresponding runs have the same weight sequence.
	Hence, $\C'(w) = \C(\pi(w))$ for all $w \in \Gamma^\omega$.
	In particular, $\bot_{\C'} = \bot_{\C}$.
	Moreover, if $\C$ maps some word to $0$, then $\C'$ maps infinitely many words to $0$ as well (e.g., by priming only the $i$-th letter, for arbitrarily large $i$), and if $\C$ maps no word to $0$, then $\C'$ maps no word to $0$ either.

	Now, define $\AA_{\Inf} = (\Inf,\C')$ and $\AA_{\LimInf} = (\LimInf,\C')$.
	If $\bot_{\C} = 1$, then every word of $\Gamma^\omega$ is mapped by $\C'$ to $1$, and therefore $\AA_{\Inf}(\Gamma^\omega) = \AA_{\LimInf}(\Gamma^\omega) = 1$.
	If $\bot_{\C} = 0$, then some word of $\Sigma^\omega$ is mapped by $\C$ to $0$, hence infinitely many words of $\Gamma^\omega$ are mapped by $\C'$ to $0$.
	Consequently, $\AA_{\Inf}(\Gamma^\omega) = 0$, and also $0 \in \textit{InfSupp}(M_{\Gamma^\omega,\C'})$, so $\AA_{\LimInf}(\Gamma^\omega) = 0$.
	Thus, for $h \in \{\Inf,\LimInf\}$, the value of the QLA $(h,\C')$ on $\Gamma^\omega$ is exactly $\bot_{\C}$.
	In particular, $(h,\C')(\Gamma^\omega) \geq 1$ iff $\bot_{\C} \geq 1$.
	Since deciding the latter is \PSpaceH, so is the evaluation problem in the statement.
	Since the alphabet in the statement is arbitrary, we may rename $\Gamma$ back to $\Sigma$.
\end{proof}

The aligned cases require a different construction.
When the word and language aggregators point in the same direction, producing infinitely many copies of a witness word no longer forces the evaluation to match the bottom value of the source automaton.
Instead, we add a fresh letter sending every run to a designated weight-$1$ sink and evaluate the resulting automaton on a singleton language.
Depending on whether the limit aggregator appears at the word or language level, the value on this singleton is determined by the corresponding default case, which is arranged to coincide with the bottom value of the original QWA.
This yields \PSpace hardness for the remaining supremum-type cases, and the infimum-type cases follow by duality.

\begin{lem}\label{lem:PSpace:same}
	Consider a QLA $\AA = (h,(g,f,\T))$ with $f \in \{\Inf,\Sup,\LimInf,\LimSup\}$, at least one of $g$ and $h$ in $\{\LimInf,\LimSup\}$, and either $g,h \in \{\Sup,\LimSup\}$ or $g,h \in \{\Inf,\LimInf\}$.
	The evaluation of $\AA$ is \PSpaceH.
\end{lem}
\begin{proof}
	Without loss of generality, we show only the cases $(h,g) = (\Sup,\LimSup)$ and $(h,g) = (\LimSup,\Sup)$.
	The case $(\LimSup,\LimSup)$ follows from \Cref{classic:reduction:g}.
	By duality (\Cref{all:dual}), the cases $(\Inf,\LimInf)$, $(\LimInf,\Inf)$, and $(\LimInf,\LimInf)$ follow from $(\Sup,\LimSup)$, $(\LimSup,\Sup)$, and $(\LimSup,\LimSup)$, respectively.

	Consider a QWA $\C = (\Sup, f, \T)$ over weights $\{0, 1\}$ and alphabet $\Gamma$.
	The proof goes by reduction from deciding whether $\bot_{\C} \geq 1$, which is \PSpaceH for $f \in \{\Inf$, $\Sup$, $\LimInf$, $\LimSup\}$~\cite{DBLP:journals/tocl/ChatterjeeDH10,DBLP:conf/vmcai/KupfermanL07}.
	Let $\lozenge \notin \Gamma$ be a fresh symbol, and let $\Sigma = \Gamma \cup \{\lozenge\}$.
	The graph $\T'$ is a copy of $\T$ with an additional state that is reachable from all states via $\lozenge$-labeled transitions and has a self loop over all letters in $\Sigma$.
	All transitions entering the additional state, as well as all self loops on it, have weight 1.

	We define $\AA_1 = (\Sup, (\LimSup, f, \T'))$ and $\AA_2 = (\LimSup, (\Sup, f, \T'))$, and let $S = \{\lozenge^\omega\}$, which is $\omega$-regular.
	The word $\lozenge^\omega$ admits finitely many runs, and thus $\AA_1(S) = \bot_{(\Sup,f,\T')}$ by the default value chosen for the inner $\LimSup$.
	Similarly, $\AA_2(S) = \bot_{(\Sup,f,\T')}$ since $S$ is finite, by the default value chosen for the outer $\LimSup$ and \Cref{all:top:bot}.

	We now show that $\bot_{(\Sup,f,\T')} = \bot_{\C}$.
	Since all weights are boolean, it is enough to show that $\bot_{(\Sup,f,\T')} = 1$ iff $\bot_{\C} = 1$.
	If $\bot_{\C} = 0$, then $\bot_{(\Sup,f,\T')} = 0$ because $(\Sup,f,\T')$ and $\C$ coincide on $\Gamma^\omega$.
	Conversely, assume that $\bot_{\C} = 1$, and let $w \in \Sigma^\omega$.
	If $w \in \Gamma^\omega$, then $(\Sup,f,\T')(w) = \C(w) = 1$.
	Otherwise, let $w = u \lozenge v$ with $u \in \Gamma^*$ and $v \in \Sigma^\omega$.
	If $f \in \{\Sup,\LimInf,\LimSup\}$, then the run that follows any run of $\T$ over $u$, jumps to the additional state on $\lozenge$, and then stays there has value 1.
	If $f = \Inf$, then every finite word $u \in \Gamma^*$ admits a run in $\T$ over $u$ whose transition weights are all 1; otherwise every extension of $u$ would be mapped to 0 by $\C$, contradicting $\bot_{\C} = 1$.
	Hence, also for $f = \Inf$, the run that follows such a prefix run, jumps to the additional state on $\lozenge$, and then stays there has value 1.
	Therefore, every word of $\Sigma^\omega$ is mapped to 1 by $(\Sup,f,\T')$, and so $\bot_{(\Sup,f,\T')} = 1$.

	Thus, $\AA_1(S) = \AA_2(S) = \bot_{\C}$.
	It follows that evaluation is \PSpaceH for the two displayed cases, and the remaining cases follow as explained above.
\end{proof}

Together, \Cref{classic:membership:reduction,lem:PSpace:swap,lem:PSpace:same} cover all combinations in which at least one of $g$ and $h$ is a limit aggregator, yielding a complete \PSpace classification for evaluation in this fragment.

\begin{cor}\label{cor:PSpace:complete:evaluation:limit}
	Consider a QLA $\AA = (h,(g,f,\T))$ with $f,g,h \in \{\Inf$, $\Sup$, $\LimInf$, $\LimSup\}$ and at least one of $g$ and $h$ in $\{\LimInf,\LimSup\}$.
	The evaluation of $\AA$ is \PSpaceC.
\end{cor}

\subsection{Comparison, Nonemptiness, and Universality with Limit Aggregators}

We now turn to comparison, nonemptiness, and universality for the fragment where all three aggregators belong to $\{\Inf, \Sup, \LimInf, \LimSup\}$.
In this setting, every run value is one of the finitely many transition weights, so every word value and language value also belongs to a finite set.
Consequently, the extremal values $\top_{\AA}$ and $\bot_{\AA}$ are actual maxima and minima of the image of $\AA$ over nonempty inputs.
We exploit this finite-valued behavior twice: first to reduce nonemptiness and universality to threshold questions about $\top_{\AA}$ and $\bot_{\AA}$, and then to support a finite profile argument for comparing two QLAs.

\begin{prop} \label{main:top}
	Consider a QLA $\AA = (h,(g,f,\T))$ with $f, g, h \in\{\Inf$, $\Sup$, $\LimInf$, $\LimSup\}$.
	Let ${\compare}\in\{{>}, {\geq}\}$.
	Then,
	(i) $\top_{\AA} \compare k$ iff $\AA$ is ${\compare}$-nonempty for $k$, and
	(ii) $\bot_{\AA} \compare k$ iff $\AA$ is ${\compare}$-universal for $k$.
\end{prop}
\begin{proof}
	For (i), notice that there are only finitely many values to which a nonempty language $L$ can be mapped by $\AA$, and the value $\top_{\AA}$ is the maximum among these values.
	So, if $\AA(L) {\not}\compare k$ for all nonempty $L \subseteq \Sigma^\omega$, then $\top_{\AA} {\not}\compare k$ as well.
	Conversely, if $\AA(L) \compare k$ for some nonempty $L \subseteq \Sigma^\omega$, then $\top_{\AA} \geq \AA(L)$ by definition, and thus $\top_{\AA} \compare k$.

	For (ii), simply observe that $\bot_{\AA} \leq \AA(L)$ by definition for every nonempty language $L \subseteq \Sigma^\omega$.
	Hence, if $\bot_{\AA} \compare k$, then $\AA(L) \compare k$ for every nonempty language $L \subseteq \Sigma^\omega$, i.e., $\AA$ is ${\compare}$-universal for $k$.
	Conversely, there are only finitely many values to which a nonempty language $L$ can be mapped by $\AA$, and $\bot_{\AA}$ is the minimum among these values.
	Therefore, if $\AA(L) \compare k$ for all nonempty languages $L \subseteq \Sigma^\omega$, then $\bot_{\AA} \compare k$ as well.
\end{proof}

For the upper bounds, we study a more general comparison problem.
Rather than analyzing candidate languages directly, the proof uses the exact-value languages from \Cref{limit:exact:value:languages} to partition $\Sigma^\omega$ by the pair of values induced by the two automata.
Each language is then summarized by a finite profile recording, for each cell of this partition, whether it contributes no words, finitely many words, or infinitely many words.
Since the outer aggregators depend only on support and infinite support, they are determined entirely by this profile.
Moreover, every realizable profile can be witnessed by an $\omega$-regular language, taking one lasso representative per cell marked $1$ and the whole cell per entry marked $\infty$.
This yields both the reduction to $\omega$-regular languages and the \PSpace decision procedure.

\begin{thm}\label{inclusion:limit}
	Consider two QLAs $\AA = (h, (g, f, \T))$ and $\BB = (h', (g', f', \T'))$ over the same alphabet $\Sigma$ such that all aggregators of $\AA$ and $\BB$ belong to $\{\Inf, \Sup, \LimInf, \LimSup\}$.
	Let ${\compare} \in \{{>}, {\geq}\}$.
	The following are equivalent:
	\begin{enumerate}[(i)]
		\item $\AA(S) \compare \BB(S)$ for every language $S \subseteq \Sigma^\omega$.
		\item $\AA(S) \compare \BB(S)$ for every $\omega$-regular language $S \subseteq \Sigma^\omega$.
	\end{enumerate}
	Moreover, these equivalent conditions are decidable in \PSpace.
\end{thm}
\begin{proof}
	Let $\A$ and $\B$ the underlying QWAs of $\AA$ and $\BB$, respectively.
	Let $X_\A$ and $X_\B$ be the finite weight sets from \Cref{limit:exact:value:languages}.
	For $x \in X_\A$ and $y \in X_\B$, let $L_x^\A = \{w \in \Sigma^\omega \mid \A(w) = x\}$ and $L_y^\B = \{w \in \Sigma^\omega \mid \B(w) = y\}$.
	By \Cref{limit:exact:value:languages}, each $L_x^\A$ and $L_y^\B$ is recognized by a polynomial-size alternating parity automaton.
	Hence, for every pair $(x, y) \in X_\A \times X_\B$, the language $K_{x,y} = L_x^\A \cap L_y^\B$ is also recognized by a polynomial-size alternating parity automaton, and emptiness and infiniteness of $K_{x,y}$ are decidable in \PSpace.
	Moreover, the languages $K_{x,y}$ form a partition of $\Sigma^\omega$.

	For a language $S \subseteq \Sigma^\omega$, define its \emph{profile} $\pi_S \colon X_\A \times X_\B \to \{0, 1, \infty\}$ by
	\[
		\pi_S(x, y) =
		\begin{cases}
			0 & \text{if } S \cap K_{x,y} = \emptyset, \\
			1 & \text{if } S \cap K_{x,y} \text{ is finite and nonempty}, \\
			\infty & \text{if } S \cap K_{x,y} \text{ is infinite}.
		\end{cases}
	\]
	Intuitively, $\pi_S(x, y)$ records whether $S$ contributes no words, finitely many words, or infinitely many words whose values are simultaneously $x$ in $\A$ and $y$ in $\B$.

	A function $\pi \colon X_\A \times X_\B \to \{0, 1, \infty\}$ \emph{realizable} if for every $(x, y) \in X_\A \times X_\B$ we have that $\pi(x, y) = 1$ implies $K_{x,y} \neq \emptyset$, and $\pi(x, y) = \infty$ implies $K_{x,y}$ is infinite.
	Since the $K_{x,y}$ are pairwise disjoint, this is equivalent to saying that $\pi = \pi_S$ for some language $S \subseteq \Sigma^\omega$.

	To recover the values of $\AA$ and $\BB$ from a profile $\pi$, define $\textit{Supp}_\A(\pi) = \{x \in X_\A \mid \exists y \in X_\B \colon \pi(x, y) \neq 0\}$ and $\textit{InfSupp}_\A(\pi) = \{x \in X_\A \mid \exists y \in X_\B \colon \pi(x, y) = \infty\}$, and symmetrically $\textit{Supp}_\B(\pi) = \{y \in X_\B \mid \exists x \in X_\A \colon \pi(x, y) \neq 0\}$ and $\textit{InfSupp}_\B(\pi) = \{y \in X_\B \mid \exists x \in X_\A \colon \pi(x, y) = \infty\}$.
	Further, let $\ell_\A = \min\{x \in X_\A \mid L_x^\A \neq \emptyset\}$ and $u_\A = \max\{x \in X_\A \mid L_x^\A \neq \emptyset\}$, and define $\ell_\B, u_\B$ analogously.
	These constants are computable in \PSpace from \Cref{limit:exact:value:languages} by checking emptiness of the exact-value languages.
	Moreover, they are exactly the default values that arise when the relevant support or infinite support is empty.

	Now, let
	\[
		\operatorname{val}_\AA(\pi) =
		\begin{cases}
			\min \textit{Supp}_\A(\pi) & \text{if } h = \Inf \text{ and } \textit{Supp}_\A(\pi) \neq \emptyset, \\
			u_\A & \text{if } h = \Inf \text{ and } \textit{Supp}_\A(\pi) = \emptyset, \\
			\max \textit{Supp}_\A(\pi) & \text{if } h = \Sup \text{ and } \textit{Supp}_\A(\pi) \neq \emptyset, \\
			\ell_\A & \text{if } h = \Sup \text{ and } \textit{Supp}_\A(\pi) = \emptyset, \\
			\min \textit{InfSupp}_\A(\pi) & \text{if } h = \LimInf \text{ and } \textit{InfSupp}_\A(\pi) \neq \emptyset, \\
			u_\A & \text{if } h = \LimInf \text{ and } \textit{InfSupp}_\A(\pi) = \emptyset, \\
			\max \textit{InfSupp}_\A(\pi) & \text{if } h = \LimSup \text{ and } \textit{InfSupp}_\A(\pi) \neq \emptyset, \\
			\ell_\A & \text{if } h = \LimSup \text{ and } \textit{InfSupp}_\A(\pi) = \emptyset,
		\end{cases}
	\]
	and define $\operatorname{val}_\BB(\pi)$ symmetrically.

	We claim that $\AA(S) = \operatorname{val}_\AA(\pi_S)$ and $\BB(S) = \operatorname{val}_\BB(\pi_S)$ for every language $S \subseteq \Sigma^\omega$.
	We prove the first equality; the second is symmetric.
	Fix $x \in X_\A$.
	Since the languages $K_{x,y}$ partition $L_x^\A$, we have the disjoint union $S \cap L_x^\A = \bigcup_{y \in X_\B} (S \cap K_{x,y})$.
	Because $X_\B$ is finite, the multiplicity of $x$ in the multiset $M_{S,\A}$ is zero iff $\pi_S(x, y) = 0$ for all $y$, finite and nonzero iff $\pi_S(x, y) \in \{0, 1\}$ for all $y$ and $\pi_S(x, y) = 1$ for at least one $y$, and infinite iff $\pi_S(x, y) = \infty$ for some $y$.
	Hence the support and infinite support of $M_{S,\A}$ are exactly $\textit{Supp}_\A(\pi_S)$ and $\textit{InfSupp}_\A(\pi_S)$.
	Applying the outer aggregator $h$ yields $\AA(S) = \operatorname{val}_\AA(\pi_S)$, as claimed.

	Conversely, every realizable profile is witnessed by an $\omega$-regular language.
	Let $\pi$ be realizable.
	For every $(x, y)$ with $\pi(x, y) = 1$, choose one ultimately periodic word $w_{x,y} \in K_{x,y}$; this is possible because every nonempty $\omega$-regular language contains an ultimately periodic word.
	Define
	\[
		S_\pi = \bigcup_{\pi(x,y) = 1} \{w_{x,y}\} \cup \bigcup_{\pi(x,y) = \infty} K_{x,y}.
	\]
	Since the languages $K_{x,y}$ are pairwise disjoint, the profile of $S_\pi$ is exactly $\pi$.
	Moreover, $S_\pi$ is $\omega$-regular.

	We have thus shown: there exists a language $S$ with $\AA(S) \not\compare \BB(S)$ iff there exists a realizable profile $\pi$ such that $\operatorname{val}_\AA(\pi) \not\compare \operatorname{val}_\BB(\pi)$.
	Moreover, whenever such a profile exists, it is realized by an $\omega$-regular language.
	This proves the equivalence of (i) and (ii).

	For the complexity bound, it therefore suffices to decide whether there exists a realizable profile $\pi$ with $\operatorname{val}_\AA(\pi) \not\compare \operatorname{val}_\BB(\pi)$.
	A nondeterministic polynomial-space procedure guesses the polynomial-size function $\pi \colon X_\A \times X_\B \to \{0, 1, \infty\}$, checks realizability entry by entry using the emptiness and infiniteness procedures from \Cref{limit:exact:value:languages}, computes $\operatorname{val}_\AA(\pi)$ and $\operatorname{val}_\BB(\pi)$, and accepts iff the comparison fails.
	Hence, the complement of the comparison problem is in $\NPSpace = \PSpace$, and since \PSpace is closed under complement, the comparison problem itself is in \PSpace.
\end{proof}

As a consequence, nonemptiness and universality admit a uniform \PSpace upper bound in the limit-aggregator setting.
Both reduce to the comparison problem above: universality is a direct instance, while nonemptiness is the complement of the reverse comparison, each against a constant automaton returning threshold $k$.
The non-limit cases were already handled in \Cref{ptime:nonemptiness,emptiness:classic}; the next corollary covers the remaining combinations.
The rest of the subsection sharpens this bound according to the word aggregator.

\begin{cor}\label{emptiness:limit}
	Consider a QLA $\AA = (h,(g,f,\T))$ with $f, g, h \in\{\Inf$, $\Sup$, $\LimInf$, $\LimSup\}$ and at least one $g$ and $h$ in $\{\LimInf$, $\LimSup\}$.
	Let ${\compare}\in\{{>}, {\geq}\}$.
	The ${\compare}$-nonemptiness (resp. ${\compare}$-universality) of $\AA$ is in \PSpace.
	The statement holds also for the finite-state restriction.
\end{cor}
\begin{proof}
	Let $\D_k = (\Inf,(\Sup,\Sup,\T_k))$, where $\T_k$ consists of a single state with a self loop of weight $k$ on every letter.
	Then $\D_k(S)=k$ for every language $S \subseteq \Sigma^\omega$.

	Moreover, the proof of \Cref{inclusion:limit} yields the same \PSpace upper bound when the universal quantification ranges only over nonempty languages (and likewise only over nonempty $\omega$-regular languages), because every counterexample language constructed there is nonempty whenever the original witness is nonempty.

	For universality, $\AA$ is ${\compare}$-universal for $k$ iff $\AA(S) \compare \D_k(S)$ for every nonempty language $S$.
	Hence the claim follows directly from \Cref{inclusion:limit}.

	For nonemptiness, if ${\compare} = {\geq}$, then $\AA$ is ${\geq}$-nonempty for $k$ iff it is not the case that $\D_k(S) > \AA(S)$ for every nonempty language $S$.
	If ${\compare} = {>}$, then $\AA$ is ${>}$-nonempty for $k$ iff it is not the case that $\D_k(S) \geq \AA(S)$ for every nonempty language $S$.
	By \Cref{inclusion:limit}, the universal comparison is decidable in \PSpace, and closure of \PSpace under complement yields the result.
	The same argument applies to the finite-state restriction.
\end{proof}

We first consider the case where the word aggregator is $\LimSup$.
The lower bound uses a finite-ambiguity construction: the reduction builds a QWA in which every word has only finitely many runs, so the inner $\LimSup$ always falls back to its default value.
That default is arranged to coincide with the bottom value of a companion nondeterministic automaton encoding a polynomial-space computation.
Combined with \Cref{emptiness:limit}, this yields \PSpaceC, even under the finite-state restriction.

\begin{thm}\label{top:PSpace}
	Consider a QLA $\AA = (h,(\LimSup,f,\T))$ with $f, h \in \{\Inf$, $\Sup$, $\LimInf$, $\LimSup\}$.
	Let ${\compare} \in \{{>}, {\geq}\}$.
	The ${\compare}$-nonemptiness of $\AA$ is \PSpaceC.
	The statement also holds for the finite-state restriction.
\end{thm}
\begin{proof}
	Membership in \PSpace follows from \Cref{emptiness:limit}, also under the finite-state restriction.
	For hardness, fix $f \in \{\Inf, \Sup, \LimInf, \LimSup\}$ and use only boolean weights.

	We reduce from the following standard \PSpaceC problem: given a deterministic Turing machine $M_0$, an input $x$, and a space bound $n$ in unary, decide whether $M_0$ accepts $x$ without ever leaving the first $n$ tape cells.
	Let $Q_0$ and $\Gamma_0$ be the state set and tape alphabet of $M_0$, respectively, and let $N = |Q_0| \cdot n \cdot |\Gamma_0|^n$ be the number of configurations of $M_0$ on those $n$ cells.
	We transform $(M_0, x, n)$ in polynomial time into an equivalent instance $(M, x, m)$ as follows: $M$ simulates $M_0$ on the first $n$ cells, stores a binary step counter up to $N + 1$ on $O(\log N)$ additional cells, rejects if the simulation tries to leave the first $n$ cells, and also rejects if the counter reaches $N + 1$ before the simulation halts.
	Then, $M$ is deterministic, halts on $x$, and can be viewed as a machine whose tape consists exactly of the cells $0, \dots, m-1$, where $m = n + O(\log N)$, with any attempted move outside these cells goes immediately to rejection.
	Moreover, $M$ accepts $x$ iff $M_0$ accepts $x$ within the first $n$ cells.

	Let $M = (Q, \Gamma, \delta, q_{\mathsf{init}}, Q_{\mathsf{acc}}, Q_{\mathsf{rej}})$.
	Define the configuration alphabet $\Gamma_{\mathsf{cfg}} = \Gamma \cup (Q \times \Gamma)$, where symbols from $Q \times \Gamma$ are ``tagged'' and hence distinct from the plain tape symbols in $\Gamma$.
	A word $D = d_0 \cdots d_{m-1} \in \Gamma_{\mathsf{cfg}}^m$ is \emph{well formed} if exactly one $d_i$ lies in $Q \times \Gamma$.
	Such a block encodes a unique configuration: if $d_i = (q, a)$, then the head is at position $i$, the control state is $q$, the tape symbol at cell $i$ is $a$, and for $j \neq i$ the symbol at position $j$ is $d_j \in \Gamma$.
	Let $C_0, C_1, \dots, C_t$ be the unique computation of $M$ on $x$, encoded in this way.
	Let $\# \notin \Gamma_{\mathsf{cfg}}$ and $\Sigma = \Gamma_{\mathsf{cfg}} \cup \{\#\}$.
	Define $H_{\mathsf{rej}}(M, x) = \{C_0 \# C_1 \# \cdots \# C_t \# u \mid u \in \Sigma^\omega, C_t \text{ is rejecting}\}$.
	Then, $H_{\mathsf{rej}}(M, x) = \emptyset$ iff $M$ accepts $x$.

	For each cell position $i \in \{0, \dots, m-1\}$, the successor monitor $S_i$ checks only what happens at cell $i$.
	For a well-formed block $D = d_0 \cdots d_{m-1}$, let $N_i(D) := (e_{i-1}, e_i, e_{i+1})$, where $e_j = d_j$ for $0 \leq j \leq m-1$, $e_{-1} = \vdash$, and $e_m = \dashv$.
	Thus $N_i(D)$ is the radius-$1$ neighbourhood of position $i$, with boundary markers at the tape ends.
	Since $M$ is deterministic and the head moves by at most one cell in one step, this local neighbourhood uniquely determines what symbol must appear at position $i$ in the successor configuration.
	We denote that symbol by $g_i(N_i(D))$.
	Hence, whenever $D'$ is the successor configuration of a non-halting well-formed block $D$, one has $D'[i] = g_i(N_i(D))$.

	We now construct a complete weighted transition system $\T_f$ over $\Sigma$.
	It consists of the following deterministic monitors, together with two common absorbing sinks $q_0$ and $q_1$: $I$, $F$, $S_0, \dots, S_{m-1}$, and $H$.
	The monitors are as follows.
	\begin{enumerate}[(1)]
		\item \emph{Initial monitor $I$.}
		It checks that the first block is exactly $C_0$ and that it is followed by a separator after exactly $m$ symbols.
		Any mismatch sends it to $q_1$.
		After the first separator, it no longer checks the computation history; it only keeps a one-bit flag saying whether the current block is halting, and as soon as it reads the separator after the first halting block it moves to $q_0$.

		\item \emph{Format monitor $F$.}
		It keeps a counter modulo $m + 1$ and a flag $\eta \in \{0, 1, \geq 2\}$ counting how many tagged symbols from $Q \times \Gamma$ have appeared in the current block.
		It checks that positions $0, \dots, m-1$ of every block lie in $\Gamma_{\mathsf{cfg}}$, that position $m$ is $\#$, and that at each separator one has $\eta = 1$.
		Thus it verifies that the word begins with blocks $D_0 \# D_1 \# D_2 \# \cdots$ where every $D_k \in \Gamma_{\mathsf{cfg}}^m$ is well formed.
		Any violation sends it to $q_1$.
		It also keeps track of whether the current block is halting; after verifying $\eta = 1$ at a separator, it moves to $q_0$ if that block is the first halting block.

		\item \emph{Successor monitors $S_i$ for $i = 0, \dots, m-1$.}
		Monitor $S_i$ checks the update rule for cell $i$ between consecutive blocks.
		While reading a block $D$, it stores the triple $N_i(D)$ and one bit saying whether $D$ is halting.
		If $D$ is halting, then at the following separator it moves to $q_0$.
		Otherwise it scans the next block $D'$ and, when it reaches position $i$ of $D'$, it compares the current symbol with $g_i(N_i(D))$.
		If they differ, it moves to $q_1$.
		By the end of $D'$, it has stored $N_i(D')$, and then repeats the same test for the pair $(D', D'')$.

		\item \emph{Halting monitor $H$.}
		It scans the blocks and records whether the current block is halting, and if so whether it is accepting or rejecting.
		When it reaches the separator after the first halting block, it moves to $q_0$ if that block is rejecting and to $q_1$ if it is accepting.
	\end{enumerate}
	Every transition not explicitly specified above is defined to go directly to the common sink $q_1$; so, each monitor is a total deterministic transition system, and every previously unspecified behaviour is treated as a defect.
	From a fresh initial state, on the first input symbol we nondeterministically enter one of these monitors: for each letter $a \in \Sigma$, the fresh initial state has one $a$-transition into the unique $a$-successor of the start state of each monitor.
	Thus, every word has exactly $m + 3$ runs in $\T_f$, one per monitor.
	The initial and format monitors have $O(m)$ states, each successor monitor has $O(m \cdot |\Gamma_{\mathsf{cfg}}|^3)$ states, and the halting monitor has $O(m)$ states.
	Therefore, $\T_f$ has size polynomial in $|M| + |x| + m$.

	We first claim the following:
	if $w \in H_{\mathsf{rej}}(M, x)$, then every run of $\T_f$ on $w$ reaches $q_0$;
	if $w \notin H_{\mathsf{rej}}(M, x)$, then some run of $\T_f$ on $w$ reaches $q_1$.
	First, let $w \in H_{\mathsf{rej}}(M, x)$.
	Then the prefix of $w$ up to the first halting block is exactly $C_0 \# C_1 \# \cdots \# C_t \#$, where $C_t$ is rejecting.
	Along this prefix, every monitor follows only explicitly described transitions, so none of the default transitions to $q_1$ is taken.
	Hence $I$ sees the correct initial block, $F$ sees only well-formed blocks, and for every $i$ the monitor $S_i$ sees the correct local update at every step.
	Thus none of these monitors reaches $q_1$, and each of them moves to $q_0$ at the separator after $C_t$.
	The halting monitor $H$ also moves to $q_0$, because $C_t$ is rejecting.
	So every run of $\T_f$ on $w$ reaches $q_0$.
	Conversely, let $w \notin H_{\mathsf{rej}}(M, x)$.
	If the first block is not $C_0$, then $I$ reaches $q_1$.
	So assume from now on that the first block is $C_0$.
	If the block structure is wrong or some block is not well formed, then $F$ reaches $q_1$.
	So assume from now on that the scanned prefix consists of well-formed blocks $D_0 \# D_1 \# D_2 \# \cdots$ with $D_0 = C_0$.
	If for some $k$ and some $i$ the pair $(D_k, D_{k+1})$ violates the transition rule at cell $i$, i.e., $D_{k+1}[i] \neq g_i(N_i(D_k))$, then $S_i$ reaches $q_1$.
	So assume finally that no successor monitor reaches $q_1$.
	Then for every $k$ before the first halting block and for every $i \in \{0, \dots, m-1\}$ we have $D_{k+1}[i] = g_i(N_i(D_k))$, hence $D_{k+1}$ is exactly the successor configuration of $D_k$.
	Since $D_0 = C_0$, it follows by induction that $D_k = C_k$ for all $k$ up to the first halting block.
	Because $M$ halts on $x$, that first halting block is $C_t$.
	Since $w \notin H_{\mathsf{rej}}(M, x)$, this halting block is accepting.
	The halting monitor $H$ then reaches $q_1$ at the separator after $C_t$.
	Thus, every word outside $H_{\mathsf{rej}}(M, x)$ has some run that reaches $q_1$.

	We now choose the weights.
	If $f \in \{\Sup, \LimSup\}$, every transition outside the sinks $q_0, q_1$ has weight $0$.
	If $f \in \{\Inf, \LimInf\}$, every transition outside the sinks $q_0, q_1$ has weight $1$.
	Every transition into $q_b$ and every self-loop on $q_b$ has weight $b$ for $b \in \{0, 1\}$.
	Hence, under $f$, every run that reaches $q_0$ has value $0$, and every run that reaches $q_1$ has value $1$.
	Let $\B_f = (\Sup, f, \T_f)$.
	Together with the earlier claim, this yields $\B_f(w) = 0$ if $w \in H_{\mathsf{rej}}(M, x)$, and $\B_f(w) = 1$ if $w \notin H_{\mathsf{rej}}(M, x)$.
	Hence,$\bot_{\B_f} = 0$ if $M$ rejects $x$ (i.e., $H_{\mathsf{rej}}(M, x) \neq \emptyset$), and  $\bot_{\B_f} = 1$ if $M$ accepts $x$ (i.e., $H_{\mathsf{rej}}(M, x) = \emptyset$).

	Now, let $\A_f = (\LimSup, f, \T_f)$.
	Every word has exactly $m + 3$ runs in $\T_f$, so every run value has finite multiplicity in $M_{w,\T_f,f}$, hence $\textit{InfSupp}(M_{w,\T_f,f}) = \emptyset$ for every $w \in \Sigma^\omega$.
	By definition, we obtain $\A_f(w) = \bot_{(\Sup,f,\T_f)} = \bot_{\B_f}$ for every $w$.
	Thus, $\A_f$ is the constant function with value $c = \bot_{\B_f} \in \{0, 1\}$.

	Let $\AA_f = (h, \A_f)$.
	Since $\A_f$ is constant, every nonempty language $S \subseteq \Sigma^\omega$ satisfies $\AA_f(S) = c$.
	For $h \in \{\Inf, \Sup\}$ this is immediate.
	For $h \in \{\LimInf, \LimSup\}$, if $S$ is infinite, then $\textit{InfSupp}(M_{S,\A_f}) = \{c\}$, hence $\AA_f(S) = c$; if $S$ is finite, then $\textit{InfSupp}(M_{S,\A_f}) = \emptyset$, and the default value is again $c$, because for $h = \LimInf$ it is $\top_{(\Inf,\A_f)} = c$, while for $h = \LimSup$ it is $\bot_{(\Sup,\A_f)} = c$.
	Therefore, $M$ accepts $x$ iff $c = 1$ iff $\AA_f$ is $\geq$-nonempty for threshold $1$.
	Since $c \in \{0, 1\}$, this is also equivalent to $\AA_f$ being $>$-nonempty for threshold $0$.

	Notice that the reduction works under the finite-state restriction.
	If $c = 1$, then any nonempty $\omega$-regular language, e.g., $\Sigma^\omega$, is a witness.
	If $c = 0$, then no witness exists.
\end{proof}

We next treat the infimum side, where $g \in \{\Inf,\LimInf\}$.
The argument takes a nondeterministic QWA whose bottom value is hard to test, dualizes to a language automaton turning it into a top-value question, and selects a witness language depending on the language aggregator.
Singleton witnesses suffice for $\hat{h} \in \{\Inf,\Sup,\LimSup\}$, while $\hat{h} = \LimInf$ requires an infinite $\omega$-regular family of alphabet-duplicated lasso witnesses so that the relevant value appears in infinite support.
The $\Inf$-to-$\LimInf$ translation from the previous subsection then transfers hardness to both word aggregators.

\begin{thm} \label{thm:limit:emptiness:hard}
	Consider a QLA $\AA = (h,(g,f,\T))$ with $f, h \in \{\Inf$, $\Sup$, $\LimInf$, $\LimSup\}$ and $g \in \{\Inf, \LimInf\}$ with at least one of $g$ and $h$ in $\{\LimInf, \LimSup\}$.
	Let ${\compare} \in \{{>}, {\geq}\}$.
	The ${\compare}$-nonemptiness of $\AA$ is \PSpaceC.
	The statement also holds for the finite-state restriction.
\end{thm}
\begin{proof}
	The \PSpace membership, both for unrestricted and finite-state versions, follows from \Cref{emptiness:limit}.
	We therefore only prove \PSpace-hardness.

	We first prove a slightly stronger hardness statement for $g = \Inf$ and $h \in \{\Inf$, $\Sup$, $\LimInf$, $\LimSup\}$.
	This covers the theorem when $g = \Inf$ (where necessarily $h \in \{\LimInf,\LimSup\}$), and the cases $g = \LimInf$ then follow from \Cref{classic:reduction:g}, which preserves the value on every language and hence also under the finite-state restriction.

	Thanks to \Cref{main:top}, we have that $\AA$ is ${\compare}$-nonempty for $k$ iff $\top_{\AA} \compare k$.
	Furthermore, $\top_{\AA} = -\bot_{\hat{\AA}}$, where $\hat{\AA} = (\hat{h}, (\hat{g}, \hat{f}, \hat{\T}))$ is the dual of $\AA$.
	Hence, we construct $\hat{\AA}$ first and then let $\AA$ be its dual.
	In the sequel, we focus on the hardness of deciding whether $\bot_{\hat{\AA}} = 0$.

	The proof goes by reduction from deciding whether $\bot_{\B} = 0$ for a QWA $\B = (\Sup,\hat{f},\T)$ over weights $\{0,1\}$ and alphabet $\Sigma$.
	Since $\bot_{\B} \in \{0,1\}$, this is the complement of the $\geq$-universality problem of $\B$ for threshold $1$, which is \PSpace-complete for $\hat{f} \in \{\Inf$, $\Sup$, $\LimInf$, $\LimSup\}$~\cite{DBLP:journals/tocl/ChatterjeeDH10,DBLP:conf/vmcai/KupfermanL07}.
	As \PSpace is closed under complement, deciding whether $\bot_{\B} = 0$ is \PSpaceH.

	If $\hat{h} \in \{\Inf,\Sup,\LimSup\}$, we simply let $\hat{\A} = \B$.
	If $\hat{h} = \LimInf$, let $\Sigma' = \{a' \mid a \in \Sigma\}$ be a disjoint copy of $\Sigma$, and let $\hat{\A} = (\Sup,\hat{f},\T')$ over $\Sigma \cup \Sigma'$ be obtained by duplicating every transition:
	for each transition $(q,a,x,q')$ of $\T$, the transition system $\T'$ contains both $(q,a,x,q')$ and $(q,a',x,q')$.
	Let $\pi : \Sigma \cup \Sigma' \to \Sigma$ be given by $\pi(a) = a$ and $\pi(a') = a$ for every $a \in \Sigma$, and extend $\pi$ letterwise to finite and infinite words.
	Then, $\hat{\A}(w) = \B(\pi(w))$ for every $w \in (\Sigma \cup \Sigma')^\omega$, hence $\bot_{\hat{\A}} = \bot_{\B}$.

	Consider a lasso word $u = p v^\omega \in \Sigma^\omega$ with $\B(u) = \bot_{\B}$; such a word exists by the lasso-achievability argument used in the proof of \Cref{emptiness:classic}.
	In either case, we have $\hat{\A}(u) = \bot_{\hat{\A}}$.

	Now, let $\hat{\AA} = (\hat{h},\hat{\A})$, and let $\AA$ be its dual.
	We emphasize that $\hat{g} = \Sup$, and that $\hat{\AA}(S) \geq \bot_{\hat{\A}}$ for all input languages $S$.

	If $\hat{h} \in \{\Inf,\Sup\}$, let $S = \{u\}$.
	Then, $\hat{\AA}(S) = \bot_{\hat{\A}}$.

	If $\hat{h} = \LimSup$, let again $S = \{u\}$.
	Then $\textit{InfSupp}(M_{S,\hat{\A}}) = \emptyset$, so by the definition of the default value of $\LimSup$ we obtain $\hat{\AA}(S) = \bot_{\hat{\AA}^{(h \gets \Sup)}} = \bot_{\hat{\A}}$, where the last equality follows from \Cref{all:top:bot}.

	If $\hat{h} = \LimInf$, let $P = \{ p' \in (\Sigma \cup \Sigma')^* \mid \pi(p') = p \}$ and $V = \{ v' \in (\Sigma \cup \Sigma')^+ \mid \pi(v') = v \}$.
	Consider $S = P V^\omega = \pi^{-1}(\{u\})$.
	The language $S$ is exactly the set of all primed versions of the lasso word $u$, so it is $\omega$-regular and infinite.
	Moreover, every word $w \in S$ satisfies $\pi(w) = u$, hence $\hat{\A}(w) = \B(\pi(w)) = \B(u) = \hat{\A}(u) = \bot_{\hat{\A}}$.
	Therefore, $\textit{InfSupp}(M_{S,\hat{\A}})=\{\bot_{\hat{\A}}\}$, and thus $\hat{\AA}(S) = \bot_{\hat{\A}}$.

	Hence, in all cases, $\bot_{\hat{\AA}} = \bot_{\hat{\A}} = \bot_{\B}$, and the witnessing language $S$ above is $\omega$-regular.
	By \Cref{all:dual}, the same language satisfies $\AA(S) = -\hat{\AA}(S) = -\bot_{\B}$, while every language $S'$ satisfies $\AA(S') = -\hat{\AA}(S') \leq -\bot_{\B}$.
	Since $\bot_{\B} \in \{0,1\}$, we obtain $-\bot_{\B} \in \{0,-1\}$.
	Therefore, $\bot_{\B} = 0$ iff $\AA$ is $\geq$-nonempty for $0$, and also iff $\AA$ is $>$-nonempty for $-1$.
	The same equivalence holds for the finite-state restriction because the language $S$ above is $\omega$-regular.
	Hence, the ${\compare}$-nonemptiness of $\AA$ is \PSpaceH, both in the unrestricted case and the finite-state.
	Together with the upper bound, this yields \PSpace-completeness in both cases.
\end{proof}

It remains to examine the case $g = \Sup$, which is simpler than those above.
For $h \in \{\Inf,\Sup,\LimInf\}$, nonemptiness reduces to computing the top value of the underlying QWA.
For $h = \LimSup$, the only additional quantity needed is the largest word value realized by infinitely many words.
The next proposition computes exactly this for nondeterministic QWAs with standard run aggregators, by scanning the finitely many candidate weights and testing infiniteness of the corresponding threshold languages.

\begin{prop} \label{thm:finiteness:easy}
	Consider a QWA $\A = (\Sup, f, \T)$ with $f \in \{\Inf, \Sup, \LimInf, \LimSup\}$ and let $X$ be its finite set of rational weights.
	The largest weight $x \in X$ for which infinitely many words are mapped to $x$ by $\A$ can be computed in time polynomial in the size of $\A$.
\end{prop}
\begin{proof}
	Note that infiniteness of a B\"uchi automaton is decidable in~\NLogSpace~\cite{DBLP:journals/ipl/Tao06}.
	Alternatively, by \Cref{characterize:words:fixed}, a B\"uchi automaton~$\B$ accepts infinitely many words if and only if some accepting state~$q_2$ of~$\B$ allows compliance with the pattern in \Cref{fig:pattern:words:fixed}, and for each candidate~$q_2$, compliance can be checked in \NLogSpace~\cite{DBLP:conf/dlt/FiliotMR18}, yielding a polynomial-time procedure overall.

	The algorithm iterates over the weights $x \in X$ in decreasing order.
	For each~$x$, it constructs a B\"uchi automaton~$\B_x$ recognizing $L(\A_{{\geq}x}) = \{w \in \Sigma^\omega \mid \A(w) \geq x\}$ in linear time~\cite{DBLP:journals/tocl/ChatterjeeDH10}, and tests whether $L(\B_x)$ is infinite using the above procedure.
	It returns the first~$x$ for which the test succeeds.

	To see that the algorithm terminates, observe that since $\T$ is complete and $f \in \{\Inf, \Sup, \LimInf, \LimSup\}$, every run value lies in $X$, and hence $\A(w) \in X$ for every $w \in \Sigma^\omega$.
	Therefore $\Sigma^\omega = \bigcup_{x \in X} L(\A_{{=}x})$, and since $\Sigma^\omega$ is infinite and $X$ is finite, some $L(\A_{{=}x})$ must be infinite.
	For correctness, since $L(\A_{{\geq}x}) = \bigcup_{x' \in X,\, x' \geq x} L(\A_{{=}x'})$ is a finite union, $L(\A_{{\geq}x})$ is infinite iff $L(\A_{{=}x'})$ is infinite for some $x' \geq x$.
	Therefore, the largest $x$ with $L(\A_{{\geq}x})$ infinite coincides with the largest $x$ with $L(\A_{{=}x})$ infinite.
	The overall procedure performs $|X|$ iterations, each involving a linear-time automaton construction and a polynomial-time infiniteness test, yielding polynomial time in total.
\end{proof}

With \Cref{thm:finiteness:easy} in hand, the nondeterministic case can be handled in polynomial time.
If $h \in \{\Inf,\Sup\}$, then \Cref{all:top:bot} reduces the problem directly to $\top_{\A}$.
If $h = \LimInf$, singleton languages suffice to attain the top value of the QLA, since the default value of the outer aggregator is already $\top_{\A}$.
If $h = \LimSup$, the relevant quantity is the maximum value occurring on infinitely many words, except over a unary alphabet where every nonempty language is a singleton and the default value again determines the answer.
This yields the \PTime bound stated below.

\begin{thm} \label{top:PTIME} 
	Consider a QLA $\AA = (h,(\Sup,f,\T))$ with $f, h\in\{\Inf$, $\Sup$, $\LimInf$, $\LimSup\}$.
	Let ${\compare}\in\{{>}, {\geq}\}$.
	The ${\compare}$-nonemptiness of $\AA$ is in \PTime.
	The statement also holds for the finite-state restriction.
\end{thm}
\begin{proof}
	By \Cref{all:inf:sup}, we may assume that $f \in \{\LimInf,\LimSup\}$.
	Let $\A = (\Sup,f,\T)$ be the underlying QWA of $\AA$.
	By \Cref{main:top}, it suffices to compute $\top_{\AA}$.
	For the finite-state restriction, it additionally suffices to verify that
	in each case below $\top_{\AA}$ is attained by some $\omega$-regular language.

	If $h \in \{\Inf,\Sup\}$, then $\top_{\AA} = \top_{\A}$ by \Cref{all:top:bot}, and $\top_{\A}$ is computable in polynomial time~\cite{DBLP:journals/tocl/ChatterjeeDH10}.
	Since $g = \Sup$, the value $\top_{\A}$ is achieved by a lasso word $w$~\cite{DBLP:journals/tocl/ChatterjeeDH10}, and the singleton $\{w\}$ is an $\omega$-regular witness attaining $\top_{\AA}$.

	Assume now that $h = \LimInf$.
	Let $\tau$ be the default value of the outer $\LimInf$.
	By definition $\tau = \top_{\AA^{(h \gets \Inf)}} = \top_{\A}$, where the last equality follows from \Cref{all:top:bot}.
	Consider any nonempty language $L$.
	If $\textit{InfSupp}(M_{L,\A}) \neq \emptyset$, then $\AA(L) = \inf \textit{InfSupp}(M_{L,\A}) \leq \top_{\A}$.
	Otherwise, $\AA(L) = \tau = \top_{\A}$.
	Hence, $\AA(L) \leq \top_{\A}$ for every nonempty language $L$, and therefore $\top_{\AA} \leq \top_{\A}$.
	Conversely, for every word $w \in \Sigma^\omega$, the singleton language $\{w\}$ satisfies $\textit{InfSupp}(M_{\{w\},\A}) = \emptyset$, since every value occurs only finitely many times in $M_{\{w\},\A}$.
	Thus, $\AA(\{w\}) = \tau = \top_{\A}$, which yields $\top_{\AA} \geq \top_{\A}$.
	We conclude that $\top_{\AA} = \top_{\A}$, again computable in polynomial time~\cite{DBLP:journals/tocl/ChatterjeeDH10}.
	In particular, choosing $w$ to be a lasso word achieving $\top_{\A}$ (which exists since $g = \Sup$), the $\omega$-regular singleton $\{w\}$ attains $\top_{\AA}$.

	It remains to consider $h = \LimSup$.
	Let $X$ be the finite set of weights of $\T$.

	Assume first that $|\Sigma| = 1$, and let $w = a^\omega$ be the unique infinite word.
	Then, every nonempty language is $\{w\}$, so $\top_{\AA} = \AA(\{w\})$.
	Moreover, $\textit{InfSupp}(M_{\{w\},\A}) = \emptyset$, hence the outer $\LimSup$ returns its default value $\beta = \bot_{\AA^{(h \gets \Sup)}} = \bot_{\A}$, where the last equality follows from \Cref{all:top:bot}.
	Since $w$ is the only infinite word, we also have $\bot_{\A} = \top_{\A} = \A(w)$.
	Therefore, $\top_{\AA} = \A(w) = \top_{\A}$, which is computable in polynomial time~\cite{DBLP:journals/tocl/ChatterjeeDH10}.
	The unique witness $\{a^\omega\}$ is $\omega$-regular.

	Assume now that $|\Sigma| \geq 2$.
	For every $x \in X$, let $L_x = \{w \in \Sigma^\omega \st \A(w) = x\}$, and define $I = \{x \in X \mid L_x \text{ is infinite}\}$.
	Since $\Sigma^\omega = \bigcup_{x \in X} L_x$, the set $I$ is nonempty: indeed, $\Sigma^\omega$ is infinite and $X$ is finite, so some $L_x$ must be infinite.
	Let $\hat{x} = \max I$.
	We claim that $\top_{\AA} = \hat{x}$.

	For the lower bound, consider the threshold language $L(\A_{{\geq}\hat{x}}) = \{w \in \Sigma^\omega \mid \A(w) \geq \hat{x}\}$,	which is $\omega$-regular~\cite{DBLP:journals/tocl/ChatterjeeDH10}. 
	It is infinite because it contains $L_{\hat{x}}$.
	Every value $y > \hat{x}$ satisfies $y \notin I$, so $L_y$ is finite; hence $\hat{x}$ is the only value occurring infinitely often in $M_{L(\A_{{\geq}\hat{x}}),\A}$.
	Thus, $\textit{InfSupp}(M_{L(\A_{{\geq}\hat{x}}),\A}) = \{\hat{x}\}$, and so $\AA(L(\A_{{\geq}\hat{x}})) = \hat{x}$.
	Hence, $\top_{\AA} \geq \hat{x}$.

	For the upper bound, consider any nonempty language $L$.
	If $\textit{InfSupp}(M_{L,\A}) \neq \emptyset$, then every value in $\textit{InfSupp}(M_{L,\A})$ belongs to $I$, and therefore is at most $\hat{x}$.
	Thus, $\AA(L) = \sup \textit{InfSupp}(M_{L,\A}) \leq \hat{x}$.
	If $\textit{InfSupp}(M_{L,\A}) = \emptyset$, then $\AA(L)$ equals the default value $\beta = \bot_{\AA^{(h \gets \Sup)}} = \bot_{\A}$.
	Since $L_{\hat{x}}$ is nonempty, we have $\bot_{\A} \leq \hat{x}$, and thus $\AA(L) \leq \hat{x}$ in this case as well.
	We conclude that $\top_{\AA} = \hat{x}$.

	Finally, since $L(\A_{{\geq}x}) = \bigcup_{y \in X,\, y \geq x} L_y$, the largest $x$ for which $L(\A_{{\geq}x})$ is infinite is exactly $\hat{x}$.
	By \Cref{thm:finiteness:easy}, this value can be computed in polynomial time.
\end{proof}

Together, \Cref{top:PSpace,thm:limit:emptiness:hard,top:PTIME} refine the general upper bound from \Cref{emptiness:limit}.
Within the limit-aggregator fragment, nonemptiness is in \PTime when $g = \Sup$, and \PSpaceC for all remaining word aggregators.
The corresponding universality classifications follow by duality.

\section{Conclusion}\label{sec:conclusion}

We introduced quantitative language automata (QLAs) as a uniform framework for specifying and verifying quantitative hyperproperties.
Our framework extends beyond both the traditional boolean view of system properties and the single-execution scope of quantitative trace properties, enabling reasoning about quantitative aspects of system-wide behavior such as performance and robustness.
We established a thorough foundation for QLAs by investigating the evaluation, nonemptiness, and universality problems, for which we provided a broad picture of decidability and complexity results.

Future research directions include studying QLAs with aggregator alternation.
Allowing aggregators to alternate more generally than in the current three-level setup could capture richer relational properties by combining.
Other directions include investigating decidable expressive fragments of QLAs, studying equivalent logical formalisms, and augmenting the software tool Quantitative Automata Kit (QuAK)~\cite{DBLP:conf/isola/ChalupaHMS24,DBLP:conf/tacas/ChalupaHMS25} with support for QLAs.

\bibliographystyle{alphaurl}
\bibliography{qla}

@article{DBLP:journals/pacmpl/ZhangZK024,
  author       = {Linpeng Zhang and
                  Noam Zilberstein and
                  Benjamin Lucien Kaminski and
                  Alexandra Silva},
  title        = {Quantitative Weakest Hyper Pre: Unifying Correctness and Incorrectness
                  Hyperproperties via Predicate Transformers},
  journal      = {Proc. {ACM} Program. Lang.},
  volume       = {8},
  number       = {{OOPSLA2}},
  pages        = {817--845},
  year         = {2024},
  url          = {https://doi.org/10.1145/3689740},
  doi          = {10.1145/3689740},
  bibsource    = {dblp computer science bibliography, https://dblp.org}
}

@inproceedings{DBLP:conf/concur/HenzingerKMS25,
  author       = {Thomas A. Henzinger and
                  Pavol Kebis and
                  Nicolas Mazzocchi and
                  N. Ege Sara{\c{c}}},
  editor       = {Patricia Bouyer and
                  Jaco van de Pol},
  title        = {Quantitative Language Automata},
  booktitle    = {36th International Conference on Concurrency Theory, {CONCUR} 2025,
                  Aarhus, Denmark, August 26-29, 2025},
  series       = {LIPIcs},
  pages        = {21:1--21:24},
  publisher    = {Schloss Dagstuhl - Leibniz-Zentrum f{\"{u}}r Informatik},
  year         = {2025},
  url          = {https://doi.org/10.4230/LIPIcs.CONCUR.2025.21},
  doi          = {10.4230/LIPICS.CONCUR.2025.21},
  bibsource    = {dblp computer science bibliography, https://dblp.org}
}

@inproceedings{DBLP:conf/memocode/NguyenKJDJ17,
  author       = {Luan Viet Nguyen and
                  James Kapinski and
                  Xiaoqing Jin and
                  Jyotirmoy V. Deshmukh and
                  Taylor T. Johnson},
  editor       = {Jean{-}Pierre Talpin and
                  Patricia Derler and
                  Klaus Schneider},
  title        = {Hyperproperties of real-valued signals},
  booktitle    = {Proceedings of the 15th {ACM-IEEE} International Conference on Formal
                  Methods and Models for System Design, {MEMOCODE} 2017, Vienna, Austria,
                  September 29 - October 02, 2017},
  pages        = {104--113},
  publisher    = {{ACM}},
  year         = {2017},
  url          = {https://doi.org/10.1145/3127041.3127058},
  doi          = {10.1145/3127041.3127058},
  bibsource    = {dblp computer science bibliography, https://dblp.org}
}

@article{DBLP:journals/lmcs/BokerHMS25,
  author       = {Udi Boker and
                  Thomas A. Henzinger and
                  Nicolas Mazzocchi and
                  N. Ege Sara{\c{c}}},
  title        = {Safety and Liveness of Quantitative Properties and Automata},
  journal      = {Log. Methods Comput. Sci.},
  volume       = {21},
  number       = {2},
  year         = {2025},
  url          = {https://doi.org/10.46298/lmcs-21(2:2)2025},
  doi          = {10.46298/LMCS-21(2:2)2025},
  bibsource    = {dblp computer science bibliography, https://dblp.org}
}

@inproceedings{DBLP:conf/stacs/Schewe09,
  author       = {Sven Schewe},
  editor       = {Susanne Albers and
                  Jean{-}Yves Marion},
  title        = {B{\"{u}}chi Complementation Made Tight},
  booktitle    = {Proceedings of the 26th International Symposium on Theoretical Aspects
                  of Computer Science, {STACS} 2009, Freiburg, Germany, February 26-28,
                  2009},
  series       = {LIPIcs},
  pages        = {661--672},
  publisher    = {Schloss Dagstuhl - Leibniz-Zentrum f{\"{u}}r Informatik, Germany},
  year         = {2009},
  url          = {https://doi.org/10.4230/LIPIcs.STACS.2009.1854},
  doi          = {10.4230/LIPICS.STACS.2009.1854},
  bibsource    = {dblp computer science bibliography, https://dblp.org}
}

@article{DBLP:journals/tocl/KupfermanV01,
  author       = {Orna Kupferman and
                  Moshe Y. Vardi},
  title        = {Weak alternating automata are not that weak},
  journal      = {{ACM} Trans. Comput. Log.},
  volume       = {2},
  number       = {3},
  pages        = {408--429},
  year         = {2001},
  url          = {https://doi.org/10.1145/377978.377993},
  doi          = {10.1145/377978.377993},
  bibsource    = {dblp computer science bibliography, https://dblp.org}
}

@article{DBLP:journals/tcs/MullerS95,
  author       = {David E. Muller and
                  Paul E. Schupp},
  title        = {Simulating Alternating Tree Automata by Nondeterministic Automata:
                  New Results and New Proofs of the Theorems of Rabin, McNaughton and
                  Safra},
  journal      = {Theor. Comput. Sci.},
  volume       = {141},
  number       = {1{\&}2},
  pages        = {69--107},
  year         = {1995},
  url          = {https://doi.org/10.1016/0304-3975(94)00214-4},
  doi          = {10.1016/0304-3975(94)00214-4},
  bibsource    = {dblp computer science bibliography, https://dblp.org}
}

@article{DBLP:journals/tcs/MiyanoH84,
  author       = {Satoru Miyano and
                  Takeshi Hayashi},
  title        = {Alternating Finite Automata on omega-Words},
  journal      = {Theor. Comput. Sci.},
  volume       = {32},
  pages        = {321--330},
  year         = {1984},
  url          = {https://doi.org/10.1016/0304-3975(84)90049-5},
  doi          = {10.1016/0304-3975(84)90049-5},
  bibsource    = {dblp computer science bibliography, https://dblp.org}
}

@inproceedings{DBLP:conf/dcfs/KieferW21,
  author       = {Stefan Kiefer and
                  Cas Widdershoven},
  editor       = {Yo{-}Sub Han and
                  Sang{-}Ki Ko},
  title        = {Image-Binary Automata},
  booktitle    = {Descriptional Complexity of Formal Systems - 23rd {IFIP} {WG} 1.02
                  International Conference, {DCFS} 2021, Virtual Event, September 5,
                  2021, Proceedings},
  series       = {Lecture Notes in Computer Science},
  pages        = {176--187},
  publisher    = {Springer},
  year         = {2021},
  url          = {https://doi.org/10.1007/978-3-030-93489-7\_15},
  doi          = {10.1007/978-3-030-93489-7\_15},
  bibsource    = {dblp computer science bibliography, https://dblp.org}
}

@article{DBLP:journals/ipl/Tao06,
  author       = {Yunfeng Tao},
  title        = {Infinity problems and countability problems for omega-automata},
  journal      = {Inf. Process. Lett.},
  volume       = {100},
  number       = {4},
  pages        = {151--153},
  year         = {2006},
  url          = {https://doi.org/10.1016/j.ipl.2006.06.011},
  doi          = {10.1016/J.IPL.2006.06.011},
  bibsource    = {dblp computer science bibliography, https://dblp.org}
}

@inproceedings{DBLP:conf/vmcai/KupfermanL07,
  author       = {Orna Kupferman and
                  Yoad Lustig},
  editor       = {Byron Cook and
                  Andreas Podelski},
  title        = {Lattice Automata},
  booktitle    = {Verification, Model Checking, and Abstract Interpretation, 8th International
                  Conference, {VMCAI} 2007, Nice, France, January 14-16, 2007, Proceedings},
  series       = {Lecture Notes in Computer Science},
  volume       = {4349},
  pages        = {199--213},
  publisher    = {Springer},
  year         = {2007},
  url          = {https://doi.org/10.1007/978-3-540-69738-1\_14},
  doi          = {10.1007/978-3-540-69738-1\_14},
  bibsource    = {dblp computer science bibliography, https://dblp.org}
}

@inproceedings{DBLP:conf/tacas/ChalupaHMS25,
  author       = {Marek Chalupa and
                  Thomas A. Henzinger and
                  Nicolas Mazzocchi and
                  N. Ege Sara{\c{c}}},
  editor       = {Arie Gurfinkel and
                  Marijn Heule},
  title        = {Automating the Analysis of Quantitative Automata with QuAK},
  booktitle    = {Tools and Algorithms for the Construction and Analysis of Systems
                  - 31st International Conference, {TACAS} 2025, Held as Part of the
                  International Joint Conferences on Theory and Practice of Software,
                  {ETAPS} 2025, Hamilton, ON, Canada, May 3-8, 2025, Proceedings, Part
                  {I}},
  series       = {Lecture Notes in Computer Science},
  volume       = {15696},
  pages        = {303--312},
  publisher    = {Springer},
  year         = {2025},
  url          = {https://doi.org/10.1007/978-3-031-90643-5\_16},
  doi          = {10.1007/978-3-031-90643-5\_16},
  bibsource    = {dblp computer science bibliography, https://dblp.org}
}

@inproceedings{DBLP:conf/isola/ChalupaHMS24,
  author       = {Marek Chalupa and
                  Thomas A. Henzinger and
                  Nicolas Mazzocchi and
                  N. Ege Sara{\c{c}}},
  editor       = {Tiziana Margaria and
                  Bernhard Steffen},
  title        = {QuAK: Quantitative Automata Kit},
  booktitle    = {Leveraging Applications of Formal Methods, Verification and Validation.
                  Software Engineering Methodologies - 12th International Symposium,
                  ISoLA 2024, Crete, Greece, October 27-31, 2024, Proceedings, Part
                  {IV}},
  series       = {Lecture Notes in Computer Science},
  volume       = {15222},
  pages        = {3--20},
  publisher    = {Springer},
  year         = {2024},
  url          = {https://doi.org/10.1007/978-3-031-75387-9\_1},
  doi          = {10.1007/978-3-031-75387-9\_1},
  bibsource    = {dblp computer science bibliography, https://dblp.org}
}

@article{DBLP:journals/iandc/AlmagorBK22,
  author       = {Shaull Almagor and
                  Udi Boker and
                  Orna Kupferman},
  title        = {What's decidable about weighted automata?},
  journal      = {Inf. Comput.},
  volume       = {282},
  pages        = {104651},
  year         = {2022},
  url          = {https://doi.org/10.1016/j.ic.2020.104651},
  doi          = {10.1016/J.IC.2020.104651},
  bibsource    = {dblp computer science bibliography, https://dblp.org}
}

@article{DBLP:journals/ijac/Krob94,
  author       = {Daniel Krob},
  title        = {The Equality Problem for Rational Series with Multiplicities in the
                  tropical Semiring is Undecidable},
  journal      = {Int. J. Algebra Comput.},
  volume       = {4},
  number       = {3},
  pages        = {405--426},
  year         = {1994},
  url          = {https://doi.org/10.1142/S0218196794000063},
  doi          = {10.1142/S0218196794000063},
  bibsource    = {dblp computer science bibliography, https://dblp.org}
}

@inproceedings{DBLP:conf/fossacs/HenzingerMS23,
	title = {Quantitative Safety and Liveness},
	author = {Thomas A. Henzinger and Nicolas Mazzocchi and N. Ege Sara{\c{c}}},
	year = 2023,
	booktitle = {Foundations of Software Science and Computation Structures - 26th International Conference, FoSSaCS 2023, Held as Part of the European Joint Conferences on Theory and Practice of Software, {ETAPS} 2023, Paris, France, April 22-27, 2023, Proceedings},
	publisher = {Springer},
	series = {Lecture Notes in Computer Science},
	volume = 13992,
	pages = {349--370},
	doi = {10.1007/978-3-031-30829-1\_17},
	editor = {Orna Kupferman and Pawel Sobocinski},
	bibsource = {dblp computer science bibliography, https://dblp.org}
}

@inproceedings{DBLP:conf/lics/BokerHO15,
	title = {The Target Discounted-Sum Problem},
	author = {Udi Boker and Thomas A. Henzinger and Jan Otop},
	year = 2015,
	booktitle = {30th Annual {ACM/IEEE} Symposium on Logic in Computer Science, {LICS} 2015, Kyoto, Japan, July 6-10, 2015},
	publisher = {{IEEE} Computer Society},
	pages = {750--761},
	doi = {10.1109/LICS.2015.74},
	bibsource = {dblp computer science bibliography, https://dblp.org}
}

@inproceedings{DBLP:conf/dlt/LodingP18,
  author       = {Christof L{\"{o}}ding and
                  Anton Pirogov},
  editor       = {Mizuho Hoshi and
                  Shinnosuke Seki},
  title        = {On Finitely Ambiguous {B}{\"{u}}chi Automata},
  booktitle    = {Developments in Language Theory - 22nd International Conference, {DLT}
                  2018, Tokyo, Japan, September 10-14, 2018, Proceedings},
  series       = {Lecture Notes in Computer Science},
  volume       = {11088},
  pages        = {503--515},
  publisher    = {Springer},
  year         = {2018},
  url          = {https://doi.org/10.1007/978-3-319-98654-8\_41},
  doi          = {10.1007/978-3-319-98654-8\_41},
  bibsource    = {dblp computer science bibliography, https://dblp.org}
}

@article{DBLP:journals/tse/Lamport77,
  author    = {Leslie Lamport},
  title     = {Proving the Correctness of Multiprocess Programs},
  journal   = {{IEEE} Trans. Software Eng.},
  volume    = {3},
  number    = {2},
  pages     = {125--143},
  year      = {1977},
  doi       = {10.1109/TSE.1977.229904},
  bibsource = {dblp computer science bibliography, https://dblp.org}
}

@article{DBLP:journals/ipl/AlpernS85,
  author    = {Bowen Alpern and
               Fred B. Schneider},
  title     = {Defining Liveness},
  journal   = {Inf. Process. Lett.},
  volume    = {21},
  number    = {4},
  pages     = {181--185},
  year      = {1985},
  doi       = {10.1016/0020-0190(85)90056-0},
  bibsource = {dblp computer science bibliography, https://dblp.org}
}

@article{DBLP:journals/dm/Karp78,
  author       = {Richard M. Karp},
  title        = {A characterization of the minimum cycle mean in a digraph},
  journal      = {Discret. Math.},
  volume       = {23},
  number       = {3},
  pages        = {309--311},
  year         = {1978},
  url          = {https://doi.org/10.1016/0012-365X(78)90011-0},
  doi          = {10.1016/0012-365X(78)90011-0},
  bibsource    = {dblp computer science bibliography, https://dblp.org}
}

@article{DBLP:journals/dc/AlpernS87,
  author       = {Bowen Alpern and
                  Fred B. Schneider},
  title        = {Recognizing Safety and Liveness},
  journal      = {Distributed Comput.},
  volume       = {2},
  number       = {3},
  pages        = {117--126},
  year         = {1987},
  url          = {https://doi.org/10.1007/BF01782772},
  doi          = {10.1007/BF01782772},
  bibsource    = {dblp computer science bibliography, https://dblp.org}
}

@inproceedings{esslli2006-students-And,
	author =              {Andersson, Daniel},
	title =               {An improved algorithm for discounted payoff games},
	editor =              {Huitink, Janneke and Katrenko, Sophia},
	booktitle =           {{P}roceedings of the 11th {ESSLLI} {S}tudent
	{S}ession},
	pages =               {91-98},
	year =                {2006},
	month =               aug,
}

@inproceedings{DBLP:conf/fsttcs/BokerL21,
  author       = {Udi Boker and
                  Karoliina Lehtinen},
  editor       = {Mikolaj Bojanczyk and
                  Chandra Chekuri},
  title        = {History Determinism vs. Good for Gameness in Quantitative Automata},
  booktitle    = {41st {IARCS} Annual Conference on Foundations of Software Technology
                  and Theoretical Computer Science, {FSTTCS} 2021, December 15-17, 2021,
                  Virtual Conference},
  series       = {LIPIcs},
  volume       = {213},
  pages        = {38:1--38:20},
  publisher    = {Schloss Dagstuhl - Leibniz-Zentrum f{\"{u}}r Informatik},
  year         = {2021},
  url          = {https://doi.org/10.4230/LIPIcs.FSTTCS.2021.38},
  doi          = {10.4230/LIPICS.FSTTCS.2021.38},
  bibsource    = {dblp computer science bibliography, https://dblp.org}
}

@inproceedings{DBLP:conf/concur/HenzingerO13,
  author       = {Thomas A. Henzinger and
                  Jan Otop},
  editor       = {Pedro R. D'Argenio and
                  Hern{\'{a}}n C. Melgratti},
  title        = {From Model Checking to Model Measuring},
  booktitle    = {{CONCUR} 2013 - Concurrency Theory - 24th International Conference,
                  {CONCUR} 2013, Buenos Aires, Argentina, August 27-30, 2013. Proceedings},
  series       = {Lecture Notes in Computer Science},
  volume       = {8052},
  pages        = {273--287},
  publisher    = {Springer},
  year         = {2013},
  url          = {https://doi.org/10.1007/978-3-642-40184-8\_20},
  doi          = {10.1007/978-3-642-40184-8\_20},
  bibsource    = {dblp computer science bibliography, https://dblp.org}
}

@article{DBLP:journals/jcs/ClarksonS10,
  author       = {Michael R. Clarkson and
                  Fred B. Schneider},
  title        = {Hyperproperties},
  journal      = {J. Comput. Secur.},
  volume       = {18},
  number       = {6},
  pages        = {1157--1210},
  year         = {2010},
  url          = {https://doi.org/10.3233/JCS-2009-0393},
  doi          = {10.3233/JCS-2009-0393},
  bibsource    = {dblp computer science bibliography, https://dblp.org}
}

@article{DBLP:journals/tcs/WeberS91,
	author       = {Andreas Weber and
	Helmut Seidl},
	title        = {On the Degree of Ambiguity of Finite Automata},
	journal      = {Theor. Comput. Sci.},
	volume       = {88},
	number       = {2},
	pages        = {325--349},
	year         = {1991},
	url          = {https://doi.org/10.1016/0304-3975(91)90381-B},
	doi          = {10.1016/0304-3975(91)90381-B},
	bibsource    = {dblp computer science bibliography, https://dblp.org}
}

@inproceedings{DBLP:conf/cav/SahaiS020,
  author       = {Shubham Sahai and
                  Pramod Subramanyan and
                  Rohit Sinha},
  editor       = {Shuvendu K. Lahiri and
                  Chao Wang},
  title        = {Verification of Quantitative Hyperproperties Using Trace Enumeration
                  Relations},
  booktitle    = {Computer Aided Verification - 32nd International Conference, {CAV}
                  2020, Los Angeles, CA, USA, July 21-24, 2020, Proceedings, Part {I}},
  series       = {Lecture Notes in Computer Science},
  volume       = {12224},
  pages        = {201--224},
  publisher    = {Springer},
  year         = {2020},
  url          = {https://doi.org/10.1007/978-3-030-53288-8\_11},
  doi          = {10.1007/978-3-030-53288-8\_11},
  bibsource    = {dblp computer science bibliography, https://dblp.org}
}

@inproceedings{DBLP:conf/cav/FinkbeinerHT18,
  author       = {Bernd Finkbeiner and
                  Christopher Hahn and
                  Hazem Torfah},
  editor       = {Hana Chockler and
                  Georg Weissenbacher},
  title        = {Model Checking Quantitative Hyperproperties},
  booktitle    = {Computer Aided Verification - 30th International Conference, {CAV}
                  2018, Held as Part of the Federated Logic Conference, FloC 2018, Oxford,
                  UK, July 14-17, 2018, Proceedings, Part {I}},
  series       = {Lecture Notes in Computer Science},
  volume       = {10981},
  pages        = {144--163},
  publisher    = {Springer},
  year         = {2018},
  url          = {https://doi.org/10.1007/978-3-319-96145-3\_8},
  doi          = {10.1007/978-3-319-96145-3\_8},
  bibsource    = {dblp computer science bibliography, https://dblp.org}
}

@inproceedings{DBLP:conf/fct/ChatterjeeDH09,
  author       = {Krishnendu Chatterjee and
                  Laurent Doyen and
                  Thomas A. Henzinger},
  editor       = {Miroslaw Kutylowski and
                  Witold Charatonik and
                  Maciej Gebala},
  title        = {Alternating Weighted Automata},
  booktitle    = {Fundamentals of Computation Theory, 17th International Symposium,
                  {FCT} 2009, Wroclaw, Poland, September 2-4, 2009. Proceedings},
  series       = {Lecture Notes in Computer Science},
  volume       = {5699},
  pages        = {3--13},
  publisher    = {Springer},
  year         = {2009},
  url          = {https://doi.org/10.1007/978-3-642-03409-1\_2},
  doi          = {10.1007/978-3-642-03409-1\_2},
  bibsource    = {dblp computer science bibliography, https://dblp.org}
}

@inproceedings{DBLP:conf/dlt/FiliotMR18,
	author       = {Emmanuel Filiot and
	Nicolas Mazzocchi and
	Jean{-}Fran{\c{c}}ois Raskin},
	editor       = {Mizuho Hoshi and
	Shinnosuke Seki},
	title        = {A Pattern Logic for Automata with Outputs},
	booktitle    = {Developments in Language Theory - 22nd International Conference, {DLT}
	2018, Tokyo, Japan, September 10-14, 2018, Proceedings},
	series       = {Lecture Notes in Computer Science},
	volume       = {11088},
	pages        = {304--317},
	publisher    = {Springer},
	year         = {2018},
	url          = {https://doi.org/10.1007/978-3-319-98654-8\_25},
	doi          = {10.1007/978-3-319-98654-8\_25},
	bibsource    = {dblp computer science bibliography, https://dblp.org}
}

@inproceedings{DBLP:conf/concur/BokerHMS23,
	author       = {Udi Boker and
	Thomas A. Henzinger and
	Nicolas Mazzocchi and
	N. Ege Sara{\c{c}}},
	editor       = {Guillermo A. P{\'{e}}rez and
	Jean{-}Fran{\c{c}}ois Raskin},
	title        = {Safety and Liveness of Quantitative Automata},
	booktitle    = {34th International Conference on Concurrency Theory, {CONCUR} 2023,
	September 18-23, 2023, Antwerp, Belgium},
	series       = {LIPIcs},
	volume       = {279},
	pages        = {17:1--17:18},
	publisher    = {Schloss Dagstuhl - Leibniz-Zentrum f{\"{u}}r Informatik},
	year         = {2023},
	url          = {https://doi.org/10.4230/LIPIcs.CONCUR.2023.17},
	doi          = {10.4230/LIPICS.CONCUR.2023.17},
	bibsource    = {dblp computer science bibliography, https://dblp.org}
}

@inproceedings{DBLP:conf/concur/ChatterjeeDEHR10,
	author       = {Krishnendu Chatterjee and
	Laurent Doyen and
	Herbert Edelsbrunner and
	Thomas A. Henzinger and
	Philippe Rannou},
	editor       = {Paul Gastin and
	Fran{\c{c}}ois Laroussinie},
	title        = {Mean-Payoff Automaton Expressions},
	booktitle    = {{CONCUR} 2010 - Concurrency Theory, 21th International Conference,
	{CONCUR} 2010, Paris, France, August 31-September 3, 2010. Proceedings},
	series       = {Lecture Notes in Computer Science},
	volume       = {6269},
	pages        = {269--283},
	publisher    = {Springer},
	year         = {2010},
	url          = {https://doi.org/10.1007/978-3-642-15375-4\_19},
	doi          = {10.1007/978-3-642-15375-4\_19},
	bibsource    = {dblp computer science bibliography, https://dblp.org}
}

@inproceedings{DBLP:conf/csl/DegorreDGRT10,
	author       = {Aldric Degorre and
	Laurent Doyen and
	Raffaella Gentilini and
	Jean{-}Fran{\c{c}}ois Raskin and
	Szymon Torunczyk},
	editor       = {Anuj Dawar and
	Helmut Veith},
	title        = {Energy and Mean-Payoff Games with Imperfect Information},
	booktitle    = {Computer Science Logic, 24th International Workshop, {CSL} 2010, 19th
	Annual Conference of the EACSL, Brno, Czech Republic, August 23-27,
	2010. Proceedings},
	series       = {Lecture Notes in Computer Science},
	volume       = {6247},
	pages        = {260--274},
	publisher    = {Springer},
	year         = {2010},
	url          = {https://doi.org/10.1007/978-3-642-15205-4\_22},
	doi          = {10.1007/978-3-642-15205-4\_22},
	bibsource    = {dblp computer science bibliography, https://dblp.org}
}

@article{DBLP:journals/tocl/ChatterjeeDH10,
	author       = {Krishnendu Chatterjee and
	Laurent Doyen and
	Thomas A. Henzinger},
	title        = {Quantitative languages},
	journal      = {{ACM} Trans. Comput. Log.},
	volume       = {11},
	number       = {4},
	pages        = {23:1--23:38},
	year         = {2010},
	url          = {https://doi.org/10.1145/1805950.1805953},
	doi          = {10.1145/1805950.1805953},
	bibsource    = {dblp computer science bibliography, https://dblp.org}
}

@article{DBLP:journals/jcss/MichaliszynO20,
  author       = {Jakub Michaliszyn and
                  Jan Otop},
  title        = {Non-deterministic weighted automata evaluated over {M}arkov chains},
  journal      = {J. Comput. Syst. Sci.},
  volume       = {108},
  pages        = {118--136},
  year         = {2020},
  url          = {https://doi.org/10.1016/j.jcss.2019.10.001},
  doi          = {10.1016/J.JCSS.2019.10.001},
  bibsource    = {dblp computer science bibliography, https://dblp.org}
}

@article{DBLP:journals/tcs/HunterPPR18,
  author       = {Paul Hunter and
                  Arno Pauly and
                  Guillermo A. P{\'{e}}rez and
                  Jean{-}Fran{\c{c}}ois Raskin},
  title        = {Mean-payoff games with partial observation},
  journal      = {Theor. Comput. Sci.},
  volume       = {735},
  pages        = {82--110},
  year         = {2018},
  url          = {https://doi.org/10.1016/j.tcs.2017.03.038},
  doi          = {10.1016/J.TCS.2017.03.038},
  bibsource    = {dblp computer science bibliography, https://dblp.org}
}

@article{DBLP:journals/ai/MadaniHC03,
  author       = {Omid Madani and
                  Steve Hanks and
                  Anne Condon},
  title        = {On the undecidability of probabilistic planning and related stochastic
                  optimization problems},
  journal      = {Artif. Intell.},
  volume       = {147},
  number       = {1-2},
  pages        = {5--34},
  year         = {2003},
  url          = {https://doi.org/10.1016/S0004-3702(02)00378-8},
  doi          = {10.1016/S0004-3702(02)00378-8},
  bibsource    = {dblp computer science bibliography, https://dblp.org}
}

\end{document}